\documentclass[prd,superscriptaddress,tightenlines,nofootinbib,
  eqsecnum]{revtex4}
\usepackage{amssymb,amsmath,amsfonts,graphicx,mathrsfs,color,bm}
\usepackage{empheq}
\usepackage{tikz}
\usepackage{tikz-cd}
\newcommand{\Comment}[1]{{}}
\definecolor{MyDarkBlue}{rgb}{0.15,0.15,0.45}
\usepackage[linktocpage=true]{hyperref}
\hypersetup{
colorlinks=true,
citecolor=MyDarkBlue,
linkcolor=MyDarkBlue,
urlcolor=MyDarkBlue,
pdfauthor={},
pdftitle={},
pdfsubject={}
}

\allowdisplaybreaks
\usepackage{ulem}
\newcommand{\be}{\begin{equation}}
\newcommand{\ee}{\end{equation}}
\newcommand{\sbe}{\begin{subequations}}
\newcommand{\see}{\end{subequations}}

\newcommand{\ba}{\begin{eqnarray}}
\newcommand{\ea}{\end{eqnarray}}
\newcommand{\p}{\partial}
\newcommand{\nn}{\nonumber}
\newcommand{\ud}{\mathrm{d}}

\begin{document}

\title{Dynamics of compact binary systems in scalar-tensor theories: II. Center-of-mass and conserved quantities to 3PN order}

\author{Laura Bernard}
\email{lbernard@perimeterinstitute.ca}
\affiliation{Perimeter Institute for Theoretical Physics, 31 Caroline St. N., Waterloo, ON, N2L 2Y5, Canada}
%\affiliation{CENTRA, Departamento de F\'{\i}sica, Instituto Superior T{\'e}cnico -- IST, Universidade de Lisboa -- UL, Avenida Rovisco Pais 1, 1049 Lisboa, Portugal}

\date{\today}

\begin{abstract}

The equations of motion of nonspinning compact binary systems at the third post-Newtonian~(PN) order in massless scalar-tensor theories have recently been obtained. In the present paper, we complete this work by computing at 3PN order the ten integrals of motion together with the equations of motion in the center-of-mass frame. We then perform the reduction of the center-of-mass equations to quasi-circular orbits and determine the conserved energy and angular momentum for circular orbits.

\end{abstract}

\maketitle

%%%%%%%%%%%%%%%%%%%%%%%%%%%%%%%%%%%%%%%%%%%%%%%%%%%%%%%
\section{Introduction}\label{sec:intro}
%%%%%%%%%%%%%%%%%%%%%%%%%%%%%%%%%%%%%%%%%%%%%%%%%%%%%%%

The detections of gravitational waves emitted by inspiralling compact binaries is a first step towards a better comprehension of gravitational physics~\cite{Barack:2018yly}. In the future, the next generation of high precision experiments will allow a better understanding of these systems, by constraining their abundance, origin and parameters. It will also provide information about the strong-field and highly-dynamical regime of gravitational physics, by challenging the theory of general relativity (GR) in this regime.

The detection and parameter estimation of gravitational wave events require a bank of highly accurate templates for the gravitational waveforms, to be match-filtered against the data. Current templates are constructed by matching the waveform at different stages of the coalescence, that are obtained using different approaches: post-Newtonian formalism for the inspiral~\cite{Blanchet:2013haa} and numerical relativity for the merger and ringdown~\cite{Sperhake:2014wpa}. There are currently two main series of template waveforms to be used by the LIGO-Virgo collaboration, based either on a direct matching (IMR models)~\cite{Ajith:2007qp,Khan:2015jqa} or on some resummation techniques (EOB waveforms)~\cite{Buonanno:1998gg,Bohe:2016gbl}. 

In order to perform tests of general relativity, one also has to construct such templates for the many alternative theories of gravity. There exist two different and complementary approaches to this problem: either theory-independent~\cite{Yunes:2009ke} or theory-dependent~\cite{Damour:1996ke}. Although fundamentally different, these two approaches are complementary as one needs to be able to map the constraints coming from the agnostic approach to specific theories. Here, we will focus on a specific model, namely massless scalar-tensor theory~\cite{Brans:2008zz,Fujii:2005}, one of the oldest, most popular and well-studied theory of gravity~\cite{Esposito-Farese:2011cha}. It consists in adding a single massless scalar field, coupled in a minimal way to gravity. One of the motivations for studying this theory, which usually arises as a low-energy limit of some string theory, relies on the fact that it may explain the accelerated expansion of the universe~\cite{DeFelice:2010aj}.

The present work is part of a series of articles whose attempt is to construct the full gravitational and scalar waveforms in scalar-tensor theories up to 2PN order\footnote{As usual, we refer to post-Newtonian order as $n\mathrm{PN}\equiv\mathcal{O}\left(v^{2}/c^{2}\right)^{n}$.}~\cite{Mirshekari:2013vb,Lang:2013fna,Lang:2014osa,Bernard:2018hta}. In the companion paper~\cite{Bernard:2018hta}, that will be referred to as Paper~I in the following, we computed the harmonic coordinate equations of motion in scalar-tensor theories at 3PN order. In the present paper, we complete this work by determining the center-of-mass and conserved quantities to the same 3PN order in the center-of-mass frame. As the leading contributions in the energy flux and scalar waveform start respectively at $-1$PN and $-0.5$PN with respect to the usual GR result, the present complete result for the 3PN dynamics is crucial in order to obtain these two quantites at $2$PN order only.

In the rest of the paper, we give in Sec.~\ref{sec:rappel} a summary of the results obtained in Paper I. In Sec.~\ref{sec:CoMcoord}, we derive the center of mass integral of motion and compute the acceleration in the center-of-mass frame. In Sec.~\ref{sec:invariants}, we determine the conserved energy and angular momentum in the center-of-mass frame. Finally in Sec.~\ref{sec:CircOrbit}, we reduce these quantities to the case of circular orbits. We conclude with some discussion. In Appendix~\ref{sec:App1}, one can find the full 3PN scalar-tensor conservative Lagrangian in harmonic coordinates, and we give some technical details in Appendix~\ref{sec:App2}.

\paragraph*{\textbf{Notations:}}

We use boldface letters to represent three-dimensional Euclidean vectors. We denote by $\mathbf{y}_{A}(t)$ the two ordinary coordinate trajectories in a harmonic coordinate system $\left\{t,\mathbf{x}\right\}$, by $\mathbf{v}_{A}(t)=\ud\mathbf{y}_{A}/\ud t$ the two ordinary velocities and by $\mathbf{a}_{A}(t)=\ud\mathbf{v}_{A}/\ud t$ the two ordinary accelerations. The ordinary separation vector reads $\mathbf{n}_{12}=\left(\mathbf{y}_{1}-\mathbf{y}_{2}\right)/r_{12}$, where $r_{12}=\left\vert\mathbf{y}_{1}-\mathbf{y}_{2}\right\vert$. Ordinary scalar producs are denoted, \textit{e.g.} $\left(n_{12}v_{1}\right)=\mathbf{n}_{12}\cdot\mathbf{v}_{1}$, while the two masses are indicated by $m_{1}$ and $m_{2}$. We also define the symmetric mass ratio $\nu\equiv\frac{m_{1}m_{2}}{(m_{1}+m_{2})^{2}}$, which verifies $0<\nu\leq 1/4$. We note $\hat{n}_{L}$ the symmetric trace-free (STF) product of $\ell$ spatial vectors $n_{i}$, with $L=i_{1}\cdots i_{l}$ a multi-index made of $\ell$ spatial indices.
Additionally, to express quantities in the centre-of-mass frame, we introduce the notations $\mathbf{n}=\mathbf{n}_{12}$ and $r=r_{12}$, and define the relative position $\mathbf{x}=\mathbf{y}_{1}-\mathbf{y}_{2}$, velocity $\mathbf{v}=\mathbf{v}_{1}-\mathbf{v}_{2}$ and acceleration $\mathbf{a}\equiv\frac{\ud \mathbf{v}}{\ud t}$. We then pose $v^{2}=(vv)=\mathbf{v}\cdot\mathbf{v}$ and $\dot{r}=(nv)=\mathbf{n}\cdot\mathbf{v}$. The orbital frequency $\omega$ is defined by the relation $v^{2}=\dot{r}^{2}+r^{2}\omega^{2}$, and will be used when dealing with quasi-circular orbits.
Finally, we also define the sum and difference of the various parameters that appear in Paper I and that we remind in Sec.~\ref{sec:rappel}. For the masses we have,
\be
m\equiv m_{1}+m_{2}\,, \qquad m_{-}\equiv \frac{m_{1}-m_{2}}{m}\,,
\ee 
while for the scalar-tensor parameters, generically called $\theta_{1,2}$, we define
\be
\theta_{+}\equiv \frac{\theta_{1}+\theta_{2}}{2}\,, \qquad \theta_{-}\equiv \frac{\theta_{1}-\theta_{2}}{2}\,,
\ee
and for the regularisation constants,
\be
r'_{+}\equiv \sqrt{r'_{1}r'_{2}}\,, \qquad r'_{-}\equiv r\sqrt{\frac{r'_{1}}{r'_{2}}}\,.
\ee
We will write the full 3PN quantities in the following form:
\be
Q = Q^{\mathrm{N}} + \frac{1}{c^2}Q^{1\mathrm{PN}} + \frac{1}{c^4}Q^{2\mathrm{PN}} + \frac{1}{c^6}Q^{3\mathrm{PN}} \,.
\ee
The 3PN piece is then decomposed into a local part and a non-local one,
\be
Q^{3\mathrm{PN}} = Q^{3\mathrm{PN,\,inst}} + Q^{3\mathrm{PN,\,tail}}\,,
\ee
and the local part will be, when needed, further split into its increasing power of $\tilde{G}$:
\be
Q_{3\mathrm{PN}}^{\mathrm{inst}} = \tilde{G}\,Q_{3\mathrm{PN}}^{(1)} + \tilde{G}^{2}\,Q_{3\mathrm{PN}}^{(2)} + \tilde{G}^{3}\,Q_{3\mathrm{PN}}^{(3)} + \tilde{G}^{4}\,Q_{3\mathrm{PN}}^{(4)} \,.
\ee

%%%%%%%%%%%%%%%%%%%%%%%%%%%%%%%%%%%%%%%%%%%%%%%%%%%%%%%
\section{Summary of previous results}\label{sec:rappel}
%%%%%%%%%%%%%%%%%%%%%%%%%%%%%%%%%%%%%%%%%%%%%%%%%%%%%%%

We consider a generic class of massless scalar-tensor theories composed of a single massless scalar field $\phi$ minimally coupled to the metric $g_{\mu\nu}$, and described by the action
\be\label{STactionJF}
S_{\mathrm{st}} = \frac{c^{3}}{16\pi G} \int\ud^{4}x\,\sqrt{-g}\left[\phi R - \frac{\omega(\phi)}{\phi}g^{\alpha\beta}\p_{\alpha}\phi\p_{\beta}\phi\right] +S_{\mathrm{m}}\left(\mathfrak{m},g_{\alpha\beta}\right)\,,
\ee
where $R$ and $g$ are respectively the Ricci scalar and the determinant of the metric, $\omega$ is a function of the scalar field and $\mathfrak{m}$ stands generically for the matter fields. The action for the matter $S_{\mathrm{m}}$ is a function only of the matter fields and the metric, and does not couple directly to the scalar field. In scalar-tensor theories, when we are dealing with compact, self-gravitating objects, we have to take into account the internal gravity of each body. To deal with it we follow the same approach as in Paper I, pioneered by Eardley~\cite{Eardley1975}, by considering that the total mass of each body may depend on the value of the scalar field at its location. The matter action is then given by the classical action for point particles, but with a scalar-field dependent mass $m_{A}(\phi)$, namely
\be\label{matteract}
S_{\mathrm{m}} = -\sum_{A}\int\ud t\,m_{A}(\phi)c^{2}\sqrt{-\left(g_{\alpha\beta}\right)_{A}\frac{v_{A}^{\alpha}v_{A}^{\beta}}{c^2}}\,,
\ee
where $v_{A}^{\mu}\equiv\frac{\ud y_{A}^{\mu}}{\ud t}=\left(c,\mathbf{v}_{A}\right)$ is the coordinate velocity of particle $A$, $y_{A}^{\mu}=\left(ct,\mathbf{y}_{A}\right)$ its trajectory and $\left(g_{\alpha\beta}\right)_{A}$ is the physical metric evaluated at the position of particle $A$ using the dimensional regularisation scheme~\cite{Blanchet:2003gy}.

%We note $\phi_{0}$ the value of the scalar field at spatial infinity and we assume that it is constant in time. We then define the rescaled scalar field $\varphi\equiv \frac{\phi}{\phi_{0}}$ and the conformally related metric,
%\be
%\tilde{g}_{\mu\nu}\equiv \varphi\,g_{\mu\nu}\,.
%\ee
%In terms of these new variables, the action~\eqref{STactionJF} can be rewritten as,
%\be\label{STactionEF}
%S_{\mathrm{st}} = \frac{c^{3}\phi_{0}}{16\pi G} \int\ud^{4}x\,\sqrt{-\tilde{g}}\left[ \tilde{R} - \frac{3+2\omega(\phi)}{2\varphi^{2}}\tilde{g}^{\alpha\beta}\p_{\alpha}\varphi\p_{\beta}\varphi\right] +S_{\mathrm{m}}\left(\mathfrak{m},g_{\alpha\beta}\right)\,,
%\ee
%Note that the matter fields still couple to the physical metric $g_{\mu\nu}$. As the scalar field is now minimally coupled to the metric, the action~\eqref{STactionEF} is often called the ``Einstein''-frame action.

In Paper I, we derived the equations of motion at 3PN order in harmonic coordinates. They depend on a finite number of parameters that were introduced following Ref.~\cite{Mirshekari:2013vb}. We start by defining the scalar-tensor parameters
\begin{equation}
\begin{aligned}
& \tilde{G} \equiv \frac{G(4+2\omega_{0})}{\phi_{0}(3+2\omega_{0})}\,,\qquad & \zeta \equiv \frac{1}{(4+2\omega_{0})}\,, \\
& \lambda_{1} \equiv \frac{\zeta^{2}}{(1-\zeta)}\left.\frac{\ud\omega}{\ud\varphi}\right\vert_{0}\,,\qquad & \lambda_{2} \equiv \frac{\zeta^{3}}{(1-\zeta)}\left.\frac{\ud^{2}\omega}{\ud\varphi^{2}}\right\vert_{0}\,,\qquad & \lambda_{3} \equiv \frac{\zeta^{4}}{(1-\zeta)}\left.\frac{\ud^{3}\omega}{\ud\varphi^{3}}\right\vert_{0}\,, 
\end{aligned}
\end{equation}
where $\phi_{0}$ is the value of the scalar field at spatial infinity and is assumed to be constant in time. We have also defined the rescaled scalar field $\varphi\equiv \frac{\phi}{\phi_{0}}$. We then define the zeroth and higher order sensitivities of each body with respect to the scalar field,
\begin{equation}
\begin{aligned}
& s_{A} \equiv \left.\frac{\ud\ln m_{A}(\phi)}{\ud\ln\phi}\right\vert_{0}\,,\qquad & s'_{A} \equiv \left.\frac{\ud^{2}\ln m_{A}(\phi)}{\ud\ln\phi^{2}}\right\vert_{0}\,,\qquad & s''_{A} \equiv \left.\frac{\ud^{3}\ln m_{A}(\phi)}{\ud\ln\phi^{3}}\right\vert_{0}\,,\qquad & s'''_{A} \equiv \left.\frac{\ud^{4}\ln m_{A}(\phi)}{\ud\ln\phi^{4}}\right\vert_{0}\,.
\end{aligned}
\end{equation}
Finally, the dynamics can be written using a finite number of PN parameters, namely,
\begin{subequations}
\begin{small}
\begin{align}
&\alpha\equiv 1-\zeta+\zeta\left(1-2s_{1}\right)\left(1-2s_{2}\right)\,,\\[8pt]
%%%%%%%%%%%%%%%%%%%%%%%%%%%%%%%%%%%%%%%%%%%%%%%%%%%%%%%%%%%%%%%%%
& \overline{\gamma} \equiv -\frac{2\zeta}{\alpha}\left(1-2s_{1}\right)\left(1-2s_{2}\right)\,,\qquad \overline{\beta}_{1} \equiv \frac{\zeta}{\alpha^{2}}\left(1-2s_{2}\right)^{2}\left(\lambda_{1}\left(1-2s_{1}\right)+2\zeta s'_{1}\right) \,,\qquad \overline{\beta}_{2} \equiv \frac{\zeta}{\alpha^{2}}\left(1-2s_{1}\right)^{2}\left(\lambda_{1}\left(1-2s_{2}\right)+2\zeta s'_{2}\right)\,,\\[8pt]
%%%%%%%%%%%%%%%%%%%%%%%%%%%%%%%%%%%%%%%%%%%%%%%%%%%%%%%%%%%%%%%%%
& \overline{\delta}_{1} \equiv \frac{\zeta\left(1-\zeta\right)}{\alpha^{2}}\left(1-2s_{1}\right)^{2}\,,\qquad \overline{\delta}_{2} \equiv \frac{\zeta\left(1-\zeta\right)}{\alpha^{2}}\left(1-2s_{2}\right)^{2}\,,\\
& \overline{\chi}_{1} \equiv \frac{\zeta}{\alpha^{3}}\left(1-2s_{2}\right)^{3}\left[\left(\lambda_{2}-4\lambda_{1}^{2}+\zeta\lambda_{1}\right)\left(1-2s_{1}\right)-6\zeta\lambda_{1}s'_{1}+2\zeta^{2}s''_{1}\right] \,,\\
& \overline{\chi}_{2} \equiv \frac{\zeta}{\alpha^{3}}\left(1-2s_{1}\right)^{3}\left[\left(\lambda_{2}-4\lambda_{1}^{2}+\zeta\lambda_{1}\right)\left(1-2s_{2}\right)-6\zeta\lambda_{1}s'_{2}+2\zeta^{2}s''_{2}\right]\,,\\[8pt]
%%%%%%%%%%%%%%%%%%%%%%%%%%%%%%%%%%%%%%%%%%%%%%%%%%%%%%%%%%%%%%%%%
& \overline{\kappa}_{1} \equiv \frac{\zeta}{\alpha^{4}}\left(1-2s_{2}\right)^{4}\left[\left(\lambda_{3}-13\lambda_{1}\lambda_{2}+28\lambda_{1}^{3}+\zeta\left(3\lambda_{2}-13\lambda_{1}^{2}\right)+\lambda_{1}\zeta^{2}\right)\left(1-2s_{1}\right) +2\zeta\left(19\lambda_{1}^{2}-4\lambda_{2}-4\lambda_{1}\zeta\right)s'_{1}-12\zeta^{2}\lambda_{1}s''_{1}+2\zeta^{3}s'''_{1}\right] \,,\\
& \overline{\kappa}_{2} \equiv \frac{\zeta}{\alpha^{4}}\left(1-2s_{1}\right)^{4}\left[\left(\lambda_{3}-13\lambda_{1}\lambda_{2}+28\lambda_{1}^{3}+\zeta\left(3\lambda_{2}-13\lambda_{1}^{2}\right)+\lambda_{1}\zeta^{2}\right)\left(1-2s_{2}\right)+2\zeta\left(19\lambda_{1}^{2}-4\lambda_{2}-4\lambda_{1}\zeta\right)s'_{2}-12\zeta^{2}\lambda_{1}s''_{2}+2\zeta^{3}s'''_{2}\right]\,.
\end{align}
\end{small}
\end{subequations}
Note that these parameters are not all independent as we have the relations $\alpha(2+\overline{\gamma})=2(1-\zeta)$ and $16\overline{\delta}_{1}\overline{\delta}_{2}=\overline{\gamma}^{2}(2+\overline{\gamma})^{2}$. The complete expression of the 3PN acceleration can be found in Eqs.~(5.10)-(5.12) of Paper I. One peculiar feature of the scalar-tensor result is the appearance of a tail term in the conservative dynamics starting at 3PN order, while in general relativity such a term is only present at 4PN. It reads
\begin{align}\label{a1itail}\nn
a_{1\,3\mathrm{PN,\,tail}}^{i} = & -\frac{4G^{2}M}{3c^{6}\phi_{0}}(1-2s_{1})\int_{0}^{+\infty}\ud\tau\,\ln\left(\frac{c\tau}{2r_{12}}\right)\left[I_{s\,i}^{(5)}(t-\tau)-I_{s\,i}^{(5)}(t+\tau)\right]\\
& +\frac{8G^{2}M}{3c^{6}\phi_{0}}(1-2s_{1})\left(\left[\ln r_{12}I_{s\,i}^{(2)}\right]^{(2)}-\ln r_{12}I_{s\,i}^{(4)}\right) -\frac{4G^{2}M}{3c^{6}m_{1}}(3+2\omega_{0})\frac{n^{i}_{12}}{r_{12}}\left(I_{s\,i}^{(2)}\right)^{2}\,,
\end{align}
where $M=m_{1}+m_{2}$ is the ADM mass. The instantaneous terms on the second line come from the introduction of the time-varying scale $r_{12}$, and the term on the first line is the nonlocal tail term. Such a term will have to be treated carefully in the following when we will derive the integrals of motion.

In Appendix~\ref{sec:App1}, we give for the first time the complete expression of the scalar-tensor Lagrangian at 3PN order. As it depends not only on the positions $\mathbf{y}_{A}$ and velocities $\mathbf{v}_{A}$ of the particles, but also on the accelerations $\mathbf{a}_{A}$ and their higher order derivatives, the 3PN Lagrangian in harmonic coordinates is a generalised one. Here, we have reduced it to a Lagrangian linear in the acceleration by adding total time derivatives and multi-zero terms~\cite{Damour:1985mt}. This reduction process does not affect the equations of motion, which were directly obtained in Eqs.~(5-10)-(5.12) of Paper~I from the Lagrangian~\eqref{L3PN}-\eqref{L3PNtail} by writing the generalised Euler-Lagrange equations.

%Replacing the scalar dipole moment by its explicit expression,
%\be
%I_{\mathrm{s}}^{i}(t) = -\frac{1}{\phi_{0}\left(3+2\omega_{0}\right)}\left[m_{1}\left(1-2s_{1}\right)y_{1}^{i}+m_{2}\left(1-2s_{2}\right)y_{2}^{i}\right]\,,
%\ee
%and using the ST parameters to express the instantaneous terms, we get
%\begin{align}\nn
%a_{1}^{i\,3\mathrm{PN,\,tail}} = & -\frac{4G^{2}M}{3c^{6}\phi_{0}}m_{1}(1-2s_{1})\int_{0}^{+\infty}\ud\tau\,\ln\left(\frac{c\tau}{2r_{12}}\right)\left[I_{s\,i}^{(5)}(t-\tau)-I_{s\,i}^{(5)}(t+\tau)\right]\\
%& +\frac{8\tilde{G}^{3}\alpha^{3}M m_{1}^{2}m_{2}}{3c^{6}r_{12}^{4}}\left(\overline{\delta}_{1}+\frac{\overline{\gamma}(2+\overline{\gamma})}{4}\right)\left[2(n_{12}v_{12})v_{12}^{i}-8(n_{12}v_{12})^{2}n_{12}^{i}+v_{12}^{2}n_{12}^{i}-\frac{\tilde{G}\alpha M}{r_{12}}n_{12}^{i}\right] \\
%& -\frac{16\tilde{G}^{4}\alpha^{4}M m_{1}^{2}m_{2}^{2}}{3c^{6}r_{12}^{5}}\left(\frac{\overline{\delta}_{1}}{4}+\frac{\overline{\gamma}(2+\overline{\gamma})}{2}+\frac{\overline{\delta}_{2}}{4}\right)n^{i}_{12}\,.
%\end{align}

%%%%%%%%%%%%%%%%%%%%%%%%%%%%%%%%%%%%%%%%%%%%%%%%%%%%%%%
\section{Center-of-mass equations of motion}\label{sec:CoMcoord}
%%%%%%%%%%%%%%%%%%%%%%%%%%%%%%%%%%%%%%%%%%%%%%%%%%%%%%%

%%%%%%%%%%%%%%%%%%%%%%%%%%%%%%%%%%%%%%%%%%%%%%%%%%%%%%%
\subsection{The center-of-mass integral}\label{subsec:CoMvec}
%%%%%%%%%%%%%%%%%%%%%%%%%%%%%%%%%%%%%%%%%%%%%%%%%%%%%%%

In Paper I, we have derived the dynamics of compact binary systems in scalar-tensor theories at 3PN order in harmonic coordinates. These equations of motion can be derived from the 3PN ST Lagrangian that is displayed in Appendix~\ref{sec:App1}. Then, as a result of the global Poincar\'e invariance of the Lagrangian, we derive the ten Noetherian conserved integrals of the 3PN harmonic-coordinate motion. It consists in the energy $E$, the angular momentum $\mathbf{J}$, the linear momentum $\mathbf{P}$ and the center of mass position $\mathbf{G}$ associated to the Lorentz boost invariance. As the results are quite long we only present here the center of mass integral, which will then be used to define the center-of-mass frame. The CM position vector obeys the relation,
\be
\frac{\mathrm{d}\mathbf{G}}{\mathrm{d}t} = \mathbf{P}\,,
\ee
and reads
\begin{subequations}\label{G3PN}
\begin{align}
\mathbf{G}_{\mathrm{N}}={}&m_{1} \mathbf{y}_{1} + 1 \leftrightarrow 2\,,\\
%%%%%%%%%%%%%%%%%%%%%%%%%%%%%%%%%%%%%%%%%%%%%%%%%%%%%%%%%%%%%%%%%
\mathbf{G}_{1\mathrm{PN}}={}&\mathbf{y}_{1} \bigl(- \frac{1}{2} \frac{\alpha \tilde{G} m_{1} m_{2}}{r_{12}}
 + \frac{1}{2} m_{1} v_1^{2}\bigr) + 1 \leftrightarrow 2\,,\\
%%%%%%%%%%%%%%%%%%%%%%%%%%%%%%%%%%%%%%%%%%%%%%%%%%%%%%%%%%%%%%%%%
\mathbf{G}_{2\mathrm{PN}}={}&\alpha \tilde{G} m_{1} m_{2} \mathbf{v}_{1} \Bigl(\bigl(- \frac{7}{4} -  \overline{\gamma}\bigr) (n_{12} v_1)
 + \bigl(- \frac{7}{4} -  \overline{\gamma}\bigr) (n_{12} v_2)\Bigr)\nonumber\\
& + \mathbf{y}_{1} \biggl[ \frac{\alpha^2 \tilde{G}^2}{r_{12}^2}\Bigl(\bigl(- \frac{5}{4}
 -  \overline{\gamma} + \overline{\beta}_{2}\bigr) m_{1}^2 m_{2}
 + \bigl(\frac{7}{4} + \overline{\gamma}\bigr) m_{1} m_{2}^2\Bigr)
 + \frac{\alpha \tilde{G} m_{1} m_{2}}{r_{12}} \Bigl(- \frac{1}{8} (n_{12} v_1)^2 -  \frac{1}{4} (n_{12} v_1) (n_{12} v_2) \nonumber\\
& \quad + \frac{1}{8} (n_{12} v_2)^2 + \bigl(- \frac{7}{4} -  \overline{\gamma}\bigr) (v_1 v_2) + \bigl(\frac{19}{8}
 + \frac{3}{2} \overline{\gamma}\bigr) v_1^{2} + \frac{1}{8} \bigl(-7 - 4 \overline{\gamma}\bigr) v_2^{2}\Bigr) + \frac{3}{8} m_{1} v_1^{4} \biggr] + 1 \leftrightarrow 2\,,
\end{align}
\end{subequations}
together with
\begin{subequations}\label{G4PN}
\begin{align}
\mathbf{G}_{3\mathrm{PN}}^{(0)}={}&\frac{5}{16} m_{1} \mathbf{y}_{1} v_1^{6} + 1 \leftrightarrow 2\,,\\
%%%%%%%%%%%%%%%%%%%%%%%%%%%%%%%%%%%%%%%%%%%%%%%%%%%%%%%%%%%%%%%%%
\mathbf{G}_{3\mathrm{PN}}^{(1)}={}&\alpha m_{1} m_{2} \mathbf{v}_{1} \biggl(\frac{1}{12} \bigl(5
 + 3 \overline{\gamma}\bigr) (n_{12} v_1)^3
 + \frac{1}{8} \bigl(3
 + 2 \overline{\gamma}\bigr) (n_{12} v_1)^2 (n_{12} v_2)
 + \frac{1}{12} \bigl(5 + 3 \overline{\gamma}\bigr) (n_{12} v_2)^3
 - \bigl( 1 + \frac{1}{2} \overline{\gamma}\bigr) (n_{12} v_2) v_1^{2} \nonumber\\
&\quad + (n_{12} v_2) \Bigl(\frac{1}{4} (v_1 v_2)
 - \bigl( \frac{15}{8}
 +  \overline{\gamma}\bigr) v_2^{2}\Bigr) + (n_{12} v_1) \Bigl(\frac{1}{8} \bigl(3
 + 2 \overline{\gamma}\bigr) (n_{12} v_2)^2
 + \frac{1}{4} (v_1 v_2)
 - \bigl( \frac{15}{8}
 +  \overline{\gamma}\bigr) v_1^{2}
 - \bigl( 1 + \frac{1}{2} \overline{\gamma}\bigr) v_2^{2}\Bigr)\biggr)\nonumber\\
& + \frac{\alpha m_{1} m_{2} \mathbf{y}_{1}}{r_{12}} \biggl[\frac{1}{16} (n_{12} v_1)^4
 + \frac{1}{8} (n_{12} v_1)^3 (n_{12} v_2)
 -  \frac{1}{16} (n_{12} v_2)^4
 + \bigl(\frac{17}{8}
 + \overline{\gamma}\bigr) (v_1 v_2)^2 + \frac{1}{16} \bigl(53
 + 30 \overline{\gamma}\bigr) v_1^{4} \nonumber\\
&\quad + \bigl(- \frac{15}{8}
 -  \overline{\gamma}\bigr) (v_1 v_2) v_2^{2}
 + (n_{12} v_1)^2 \Bigl(\frac{3}{16} (n_{12} v_2)^2
 + \frac{1}{8} \bigl(3 + 2 \overline{\gamma}\bigr) (v_1 v_2)
 + \frac{1}{16} \bigl(-5
 - 2 \overline{\gamma}\bigr) v_1^{2}
 + \frac{1}{8} \bigl(-2
 -  \overline{\gamma}\bigr) v_2^{2}\Bigr) \nonumber\\
&\quad + v_1^{2} \Bigl(\frac{1}{8} \bigl(-11
 - 7 \overline{\gamma}\bigr) (n_{12} v_2)^2 -  \frac{5}{2} \bigl(2
 + \overline{\gamma}\bigr) (v_1 v_2)
 + \bigl(\frac{31}{16}
 + \overline{\gamma}\bigr) v_2^{2}\Bigr)
 + (n_{12} v_2)^2 \Bigl(\frac{1}{4} \bigl(5
 + 3 \overline{\gamma}\bigr) (v_1 v_2)
 + \frac{1}{16} \bigl(5 + 2 \overline{\gamma}\bigr) v_2^{2}\Bigr) \nonumber\\
&\quad + (n_{12} v_1) \biggl(\frac{1}{4} (n_{12} v_2)^3
 + \frac{1}{4} \bigl(-2
 -  \overline{\gamma}\bigr) (n_{12} v_2) v_1^{2}
 + (n_{12} v_2) \Bigl(\frac{1}{4} \bigl(3 + 2 \overline{\gamma}\bigr) (v_1 v_2)
 + \frac{1}{8} \bigl(-5
 - 2 \overline{\gamma}\bigr) v_2^{2}\Bigr)\biggr) \nonumber\\
&\quad + \frac{1}{16} \bigl(-11
 - 6 \overline{\gamma}\bigr) v_2^{4}\biggr] + 1 \leftrightarrow 2\,,\\
%%%%%%%%%%%%%%%%%%%%%%%%%%%%%%%%%%%%%%%%%%%%%%%%%%%%%%%%%%%%%%%%%
\mathbf{G}_{3\mathrm{PN}}^{(2)}={}&\frac{\alpha^2 \mathbf{v}_{1}}{r_{12}} \biggl[m_{1}^2 m_{2} \biggl(\bigl(\frac{235}{24}
 + 10 \overline{\gamma}
 + \frac{7}{3} \overline{\gamma}^2
 -  \overline{\beta}_{2}
 + \frac{4}{3} \overline{\delta}_{1}\bigr) (n_{12} v_1)
 + \Bigl(\frac{1}{24} \bigl(-235
 - 312 \overline{\gamma} - 98 \overline{\gamma}^2\bigr)
 -  \frac{1}{3} \overline{\delta}_{1}\Bigr) (n_{12} v_2)\biggr) \nonumber\\
&\ + m_{1} m_{2}^2 \biggl(\Bigl(\frac{1}{24} \bigl(-235
 - 312 \overline{\gamma}
 - 98 \overline{\gamma}^2\bigr)
 -  \frac{1}{3} \overline{\delta}_{2}\Bigr) (n_{12} v_1) + \bigl(\frac{235}{24}
 + 10 \overline{\gamma}
 + \frac{7}{3} \overline{\gamma}^2
 -  \overline{\beta}_{1}
 + \frac{4}{3} \overline{\delta}_{2}\bigr) (n_{12} v_2)\biggr)\biggr] \nonumber\\
& + \frac{\alpha^2 \mathbf{y}_{1}}{r_{12}^2} \biggl[m_{1}^2 m_{2} \biggl(\Bigl(\frac{1}{12} \bigl(79
 + 45 \overline{\gamma} + 2 \overline{\gamma}^2\bigr)
 + \frac{2}{3} \overline{\delta}_{1}\Bigr) (n_{12} v_1)^2
 + \Bigl(\frac{1}{6} \bigl(-34
 - 21 \overline{\gamma}
 - 2 \overline{\gamma}^2\bigr)
 -  \frac{4}{3} \overline{\delta}_{1}\Bigr) (n_{12} v_1) (n_{12} v_2) \nonumber\\
&\quad + \Bigl(\frac{1}{12} \bigl(34
 + 21 \overline{\gamma}
 + 2 \overline{\gamma}^2\bigr)
 + \frac{2}{3} \overline{\delta}_{1}\Bigr) (n_{12} v_2)^2
 + \Bigl(\frac{1}{12} \bigl(160
 + 138 \overline{\gamma}
 + 29 \overline{\gamma}^2\bigr) + \frac{5}{3} \overline{\delta}_{1}\Bigr) (v_1 v_2) \nonumber\\
&\quad + \Bigl(\frac{1}{24} \bigl(-175
 - 114 \overline{\gamma}
 - 8 \overline{\gamma}^2\bigr)
 + \overline{\beta}_{2}
 -  \frac{4}{3} \overline{\delta}_{1}\Bigr) v_1^{2}
 + \Bigl(\frac{1}{12} \bigl(-80
 - 87 \overline{\gamma} - 25 \overline{\gamma}^2\bigr)
 -  \frac{1}{2} \overline{\beta}_{2}
 -  \frac{1}{3} \overline{\delta}_{1}\Bigr) v_2^{2}\biggr) \nonumber\\
&\ + m_{1} m_{2}^2 \biggl(\Bigl(\frac{1}{24} \bigl(-56
 - 30 \overline{\gamma}
 -  \overline{\gamma}^2\bigr)
 -  \frac{1}{6} \overline{\delta}_{2}\Bigr) (n_{12} v_1)^2 + \Bigl(\frac{1}{12} \bigl(29
 + 6 \overline{\gamma}
 + \overline{\gamma}^2\bigr)
 + 2 \overline{\beta}_{1}
 + \frac{1}{3} \overline{\delta}_{2}\Bigr) (n_{12} v_1) (n_{12} v_2) \nonumber\\
&\quad + \Bigl(\frac{1}{24} \bigl(16
 + 18 \overline{\gamma}
 -  \overline{\gamma}^2\bigr)
 -  \overline{\beta}_{1} -  \frac{1}{6} \overline{\delta}_{2}\Bigr) (n_{12} v_2)^2
 + \Bigl(\frac{1}{6} \bigl(-80
 - 87 \overline{\gamma}
 - 25 \overline{\gamma}^2\bigr)
 -  \overline{\beta}_{1}
 -  \frac{2}{3} \overline{\delta}_{2}\Bigr) (v_1 v_2) \nonumber\\
&\quad + \Bigl(\frac{1}{24} \bigl(202 + 234 \overline{\gamma}
 + 71 \overline{\gamma}^2\bigr)
 + \overline{\beta}_{1}
 -  \frac{1}{6} \overline{\delta}_{2}\Bigr) v_1^{2}
 + \Bigl(\frac{1}{24} \bigl(139
 + 126 \overline{\gamma}
 + 29 \overline{\gamma}^2\bigr)
 + \frac{5}{6} \overline{\delta}_{2}\Bigr) v_2^{2}\biggr)\biggr] + 1 \leftrightarrow 2\,,\\
%%%%%%%%%%%%%%%%%%%%%%%%%%%%%%%%%%%%%%%%%%%%%%%%%%%%%%%%%%%%%%%%%
\mathbf{G}_{3\mathrm{PN}}^{(3)}={}&\mathbf{y}_{1} \Biggl( \frac{\alpha^2}{r_{12}^3}\biggl(\Bigl(- \frac{11}{24} \overline{\gamma} \bigl(2
 + \overline{\gamma}\bigr)
 + \frac{\bigl(10 + \overline{\gamma}\bigr) \overline{\delta}_{1}}{12 + 6 \overline{\gamma}}\Bigr) m_{1}^3 m_{2}
 + \Bigl(\frac{11}{24} \overline{\gamma} \bigl(2
 + \overline{\gamma}\bigr) -  \frac{\bigl(10 + \overline{\gamma}\bigr) \overline{\delta}_{2}}{6 \bigl(2 + \overline{\gamma}\bigr)}\Bigr) m_{1} m_{2}^3\biggr) \nonumber\\
&\  + \frac{\alpha^3}{r_{12}^3} \biggl[\Bigl(- \frac{19}{8}
 -  \overline{\gamma}
 + \bigl(- \frac{17}{4}
 - 2 \overline{\gamma}\bigr) \overline{\beta}_{2}
 + \overline{\beta}_{1} \bigl(\frac{9}{4}
 + 2 \overline{\gamma} + \frac{4 \overline{\beta}_{2}}{\overline{\gamma}}\bigr)
 + \frac{19}{9} \overline{\delta}_{1}
 -  \frac{19}{9} \overline{\delta}_{2}\Bigr) m_{1}^2 m_{2}^2 \nonumber\\
&\  + m_{1}^3 m_{2} \biggl(\frac{1}{72} \bigl(1124
 + 1256 \overline{\gamma}
 + 405 \overline{\gamma}^2
 + 33 \overline{\gamma}^3\bigr) + \bigl(\frac{7}{4}
 + 2 \overline{\gamma}\bigr) \overline{\beta}_{2}
 + \frac{1}{6} \bigl(3
 -  \overline{\gamma}\bigr) \overline{\delta}_{1}
 + \overline{\chi}_{2} \nonumber\\
&\quad + \Bigl(\frac{11}{4} \bigl(2
 + \overline{\gamma}\bigr)^2
 -  \overline{\delta}_{1}\Bigr) \ln\bigl(r'_{1}\bigr) + \Bigl(- \frac{11}{4} \bigl(2 + \overline{\gamma}\bigr)^2
 + \overline{\delta}_{1}\Bigr) \ln\bigl(r_{12}\bigr)\biggr) \nonumber\\
&\  + m_{1} m_{2}^3 \biggl(\frac{1}{72} \bigl(-1160
 - 1280 \overline{\gamma}
 - 411 \overline{\gamma}^2
 - 33 \overline{\gamma}^3\bigr)
 + \bigl(- \frac{11}{4} - 2 \overline{\gamma}\bigr) \overline{\beta}_{1}
 + \frac{1}{6} \bigl(-5
 + \overline{\gamma}\bigr) \overline{\delta}_{2}
 -  \frac{1}{3} \overline{\chi}_{1} \nonumber\\
&\quad + \Bigl(- \frac{11}{4} \bigl(2
 + \overline{\gamma}\bigr)^2
 + \overline{\delta}_{2}\Bigr) \ln\bigl(r'_{2}\bigr)
 + \Bigl(\frac{11}{4} \bigl(2
 + \overline{\gamma}\bigr)^2 -  \overline{\delta}_{2}\Bigr) \ln\bigl(r_{12}\bigr)\biggr)\biggr]\Biggr) + 1 \leftrightarrow 2\,.
\end{align}
\end{subequations}
In addition to the instantaneous terms, there is also a nonlocal contribution coming from the tail part of the 3PN Lagrangian. It reads,
\be\label{Gitail}
G^{i}_{\mathrm{tail}} = \frac{4G^{2}M}{3c^{6}\phi_{0}}m_{1}\left(1-2s_{1}\right)\int_{0}^{+\infty}\mathrm{d}\tau\ln\left(\frac{c\tau}{2r_{12}}\right)\left[I_{s}^{i\,(3)}(t-\tau)-I_{s}^{i\,(3)}(t+\tau)\right] + 1 \leftrightarrow 2 \,.
\ee
The center-of-mass frame is then defined by the equation
\be
\mathbf{G} = 0\,.
\ee
We solve this equation iteratively at each PN order, order-reducing the accelerations using the CM equations of motion. It results in the expression of the positions of the particles $\mathbf{y}_{A}$ expressed in the CM frame,
\begin{subequations}\label{y3PN}
\begin{align}
\mathbf{y}_{1} &= \left[\frac{m_{2}}{m}+\nu\mathcal{P}\right]\mathbf{x} + \nu\mathcal{Q}\mathbf{v} \,,\\
%%%%%%%%%%%%%%%%%%%%%%%%%%%%%%%%%%%%%%%%%%%%%%%%%%%%%%%%%%%%%%%%%
\mathbf{y}_{2} &= \left[-\frac{m_{1}}{m}+\nu\mathcal{P}\right]\mathbf{x} + \nu\mathcal{Q}\mathbf{v} \,,
\end{align}
\end{subequations}
where $\mathbf{x}=\mathbf{y}_{1}-\mathbf{y}_{2}$ and $\mathbf{v}=\mathbf{v}_{1}-\mathbf{v}_{2}$ are respectively the relative position and velocity. The coefficients $\mathcal{P}$ and $\mathcal{Q}$ are given by
\begin{subequations}\label{P}
\begin{align}
\mathcal{P}_{1\mathrm{PN}}={}&- \frac{1}{2} \frac{\alpha \tilde{G} m_{-} m}{r}
 + \frac{1}{2} m_{-} v^{2} \,,\\
%%%%%%%%%%%%%%%%%%%%%%%%%%%%%%%%%%%%%%%%%%%%%%%%%%%%%%%%%%%%%%%%%
\mathcal{P}_{2\mathrm{PN}}={}&\Bigl(2 \beta_{-} \nu
 + \bigl(\frac{7}{4}
 + \overline{\gamma}
 -  \frac{1}{2} \nu\bigr) m_{-}\Bigr) \frac{\alpha^2 \tilde{G}^2 m^2}{r^2}
 + \frac{\alpha \tilde{G}}{r} \Bigl(\dot{r}^2 \bigl(- \frac{1}{8}
 + \frac{3}{4} \nu\bigr) m_{-} m
 + \bigl(\frac{19}{8}
 + \frac{3}{2} \overline{\gamma} + \frac{3}{2} \nu\bigr) m_{-} m v^{2}\Bigr) \nonumber\\
&\quad + \bigl(\frac{3}{8}
 -  \frac{3}{2} \nu\bigr) m_{-} v^{4} \,,\\
%%%%%%%%%%%%%%%%%%%%%%%%%%%%%%%%%%%%%%%%%%%%%%%%%%%%%%%%%%%%%%%%%
\mathcal{P}_{3\mathrm{PN}}={}&\tilde{G}^3 \Biggl[ \frac{\alpha^2 m^3}{r^3} \biggl(- \frac{\delta_{-} \bigl(10 + \overline{\gamma}\bigr)}{6 \bigl(2 + \overline{\gamma}\bigr)}
 + \frac{\delta_{-} \bigl(10 + \overline{\gamma}\bigr) \nu}{6 + 3 \overline{\gamma}}
 + \Bigl(\frac{11}{24} \overline{\gamma} \bigl(2
 + \overline{\gamma}\bigr)
 -  \frac{\delta_{+} \bigl(10 + \overline{\gamma}\bigr)}{6 \bigl(2 + \overline{\gamma}\bigr)}\Bigr) m_{-}\biggr) \nonumber\\
&\  + \frac{\alpha^3 m^3}{r^3} \Biggl(\biggl[\frac{1}{3} \chi_{-}
 + \frac{1}{6} \delta_{-} \bigl(-5
 + \overline{\gamma}\bigr)
 + \beta_{-} \bigl(\frac{11}{4}
 + 2 \overline{\gamma}\bigr)
 + \Bigl(\beta_{-} \bigl(- \frac{27}{2}
 - 8 \overline{\gamma}\bigr)
 + \frac{1}{9} \delta_{-} \bigl(-20 - 3 \overline{\gamma}\bigr)\Bigr) \nu
 + 4 \beta_{-} \nu^2 \nonumber\\
&\quad + \biggl(- \frac{1}{3} \chi_{+}
 + \beta_{+} \bigl(- \frac{11}{4}
 - 2 \overline{\gamma}\bigr)
 + \frac{1}{6} \delta_{+} \bigl(-5
 + \overline{\gamma}\bigr)
 + \frac{1}{72} \bigl(-1160 - 1280 \overline{\gamma}
 - 411 \overline{\gamma}^2
 - 33 \overline{\gamma}^3\bigr) \nonumber\\
&\quad + \Bigl(- \frac{1}{2} \beta_{+}
 -  \frac{2}{3} \chi_{+}
 + \frac{1}{3} \delta_{+}
 -  \frac{4 \beta_{-}^2}{\overline{\gamma}}
 + \frac{4 \beta_{+}^2}{\overline{\gamma}}
 + \frac{1}{24} \bigl(3
 + 8 \overline{\gamma} + 2 \overline{\gamma}^2\bigr)\Bigr) \nu
 -  \frac{1}{2} \nu^2\biggr) m_{-}\biggr] + \biggl(- \delta_{+}
 + \frac{11}{4} \bigl(2
 + \overline{\gamma}\bigr)^2 \nonumber\\
&\quad + \Bigl(2 \delta_{+}
 -  \frac{11}{2} \bigl(2
 + \overline{\gamma}\bigr)^2\Bigr) \nu -  \delta_{-} m_{-}\biggr) \ln\bigl(\frac{r}{r'_{-}}\bigr)
 + \biggl(- \delta_{-}
 + 2 \delta_{-} \nu
 + \Bigl(- \delta_{+}
 + \frac{11}{4} \bigl(2 + \overline{\gamma}\bigr)^2\Bigr) m_{-}\biggr) \ln\bigl(\frac{r}{r'_{+}}\bigr)\Biggr)\Biggr] \nonumber\\
& + \frac{\alpha^2 \tilde{G}^2 m^2}{r^2} \biggl[\dot{r}^2 \biggl(- \frac{1}{6} \delta_{-}
 + \bigl(4 \beta_{-}
 -  \delta_{-}\bigr) \nu
 - 6 \beta_{-} \nu^2 + \Bigl(- \frac{1}{6} \delta_{+}
 -  \frac{1}{24} \bigl(2
 + \overline{\gamma}\bigr) \bigl(28
 + \overline{\gamma}\bigr)
 + \bigl(\frac{73}{8}
 - 4 \beta_{+}
 + 7 \overline{\gamma}\bigr) \nu \nonumber\\
&\quad + 4 \nu^2\Bigr) m_{-}\biggr) + \biggl(- \beta_{-} -  \frac{1}{6} \delta_{-}
 + \bigl(4 \beta_{-}
 + 3 \delta_{-}\bigr) \nu
 - 6 \beta_{-} \nu^2
 + \Bigl(\beta_{+}
 -  \frac{1}{6} \delta_{+}
 + \frac{1}{24} \bigl(202
 + 234 \overline{\gamma}
 + 71 \overline{\gamma}^2\bigr) \nonumber\\
&\quad + \bigl(- \frac{33}{8}
 - 2 \overline{\gamma}\bigr) \nu
 + 3 \nu^2\Bigr) m_{-}\biggr) v^{2}\biggr] \nonumber\\
& + \frac{\alpha \tilde{G} m}{r} \biggl(\dot{r}^4 \bigl(\frac{1}{16}
 -  \frac{5}{8} \nu
 + \frac{21}{16} \nu^2\bigr) m_{-} + \dot{r}^2 \Bigl(\frac{1}{16} \bigl(-5
 - 2 \overline{\gamma}\bigr)
 + \frac{1}{16} \bigl(21
 - 4 \overline{\gamma}\bigr) \nu
 -  \frac{11}{2} \nu^2\Bigr) m_{-} v^{2} \nonumber\\
&\quad + \Bigl(\frac{1}{16} \bigl(53
 + 30 \overline{\gamma}\bigr)
 + \bigl(-7 -  \frac{11}{2} \overline{\gamma}\bigr) \nu
 -  \frac{15}{2} \nu^2\Bigr) m_{-} v^{4}\biggr)
 + \bigl(\frac{5}{16}
 -  \frac{11}{4} \nu
 + 6 \nu^2\bigr) m_{-} v^{6} \,,
\end{align}
\end{subequations}
and
\begin{subequations}\label{Q}
\begin{align}
\mathcal{Q}_{2\mathrm{PN}}={}&\bigl(- \frac{7}{4}
 -  \overline{\gamma}\bigr) \dot{r} \alpha \tilde{G} m_{-} m \,,\\
%%%%%%%%%%%%%%%%%%%%%%%%%%%%%%%%%%%%%%%%%%%%%%%%%%%%%%%%%%%%%%%%%
\mathcal{Q}_{3\mathrm{PN}}={}&\dot{r} \biggl(- \frac{1}{3} \delta_{-}
 + \bigl(-2 \beta_{-}
 - 2 \delta_{-}\bigr) \nu
 + \Bigl(- \frac{1}{3} \delta_{+}
 + \frac{1}{24} \bigl(-235
 - 312 \overline{\gamma}
 - 98 \overline{\gamma}^2\bigr)
 -  \frac{3}{4} \bigl(7 + 4 \overline{\gamma}\bigr) \nu\Bigr) m_{-}\biggr) \frac{\alpha^2 \tilde{G}^2 m^2}{r} \nonumber\\
&\quad + \alpha \tilde{G} m \biggl(\dot{r}^3 \Bigl(\frac{1}{12} \bigl(5
 + 3 \overline{\gamma}\bigr)
 + \bigl(- \frac{19}{24}
 -  \frac{1}{2} \overline{\gamma}\bigr) \nu\Bigr) m_{-} + \dot{r} \Bigl(- \frac{15}{8}
 -  \overline{\gamma}
 + \bigl(\frac{21}{4}
 + 3 \overline{\gamma}\bigr) \nu\Bigr) m_{-} v^{2}\biggr) \,.
\end{align}\end{subequations}
As a consequence of the nonlocal tail term~\eqref{Gitail} in the center-of-mass position, there is an additionnal nonlocal contribution to the particle positions, given by
\be\label{y1itail}
y^{i}_{1\,\mathrm{tail}} = -\frac{4G^{2}}{3c^{6}\phi_{0}}\left[m_{1}\left(1-2s_{1}\right)+m_{2}\left(1-2s_{2}\right)\right]\int_{0}^{+\infty}\mathrm{d}\tau\ln\left(\frac{c\tau}{2r}\right)\left[I_{s}^{i\,(3)}(t-\tau)-I_{s}^{i\,(3)}(t+\tau)\right] \,.
\ee
As this is a pure 3PN contribution, it will not give an extra contribution to the relative acceleration.
The particle velocities $\mathbf{v}_{A}$ are obtained by differentiating the positions $\mathbf{y}_{A}$ with respect to time, and order-reducing the accelerations. Note that in addtion to the conservative contributions~\eqref{P}-\eqref{Q}, there are also some dissipative terms at $1.5$PN and $2.5$PN to be added to the particles' positions in the centre-of-mass frame. As we are only dealing with the conservative dynamics, we shall not display them here.

%%%%%%%%%%%%%%%%%%%%%%%%%%%%%%%%%%%%%%%%%%%%%%%%%%%%%%%
\subsection{Acceleration in the center-of-mass frame}\label{subsec:nonlocal}
%%%%%%%%%%%%%%%%%%%%%%%%%%%%%%%%%%%%%%%%%%%%%%%%%%%%%%%

Inserting the previous results~\eqref{y3PN}--\eqref{y1itail} into the equations of motion derived in Paper I, one gets the conservative relative acceleration $\mathbf{a}\equiv\mathbf{a}_{1}-\mathbf{a}_{2}$. The instantaneous part of the relative acceleration have the form
\be
\mathbf{a}=-\frac{\tilde{G}\alpha m}{r^{2}}\left[\left(1+\mathcal{A}\right)\mathbf{n}+\mathcal{B}\mathbf{v}\right]\,,
\ee
where the coefficients $\mathcal{A}$ and $\mathcal{B}$ are given by
\begin{subequations}\label{A}
\begin{align}
\mathcal{A}_{1\mathrm{PN}}={}&- \frac{3}{2} \dot{r}^2 \nu
 + \Bigl(-2 \beta_{+}
 - 2 \bigl(2
 + \overline{\gamma}\bigr)
 - 2 \nu
 + 2 \beta_{-} m_{-}\Bigr) \frac{\alpha \tilde{G} m}{r}
 + \bigl(1
 + \overline{\gamma}
 + 3 \nu\bigr) v^{2} \,,\\
%%%%%%%%%%%%%%%%%%%%%%%%%%%%%%%%%%%%%%%%%%%%%%%%%%%%%%%%%%%%%%%%%
\mathcal{A}_{2\mathrm{PN}}={}&\dot{r}^4 \bigl(\frac{15}{8} \nu
 -  \frac{45}{8} \nu^2\bigr)
 + \biggl(-2 \chi_{+}
 + \delta_{+}
 + 4 \beta_{+} \bigl(2
 + \overline{\gamma}\bigr)
 + \frac{9}{4} \bigl(2
 + \overline{\gamma}\bigr)^2
 + \Bigl(15 \beta_{+}
 + 4 \chi_{+} - 2 \delta_{+}
 + \frac{24 \beta_{-}^2}{\overline{\gamma}}
 -  \frac{24 \beta_{+}^2}{\overline{\gamma}} \nonumber\\
&\quad -  \frac{15 \delta_{-}^2}{\bigl(2 + \overline{\gamma}\bigr)^2}
 + \frac{15 \delta_{+}^2}{\bigl(2 + \overline{\gamma}\bigr)^2}
 -  \frac{1}{16} \bigl(2
 + \overline{\gamma}\bigr) \bigl(-174
 + 23 \overline{\gamma}\bigr)\Bigr) \nu + \Bigl(2 \chi_{-}
 + \delta_{-}
 - 4 \beta_{-} \bigl(2
 + \overline{\gamma}\bigr)
 - 7 \beta_{-} \nu\Bigr) m_{-}\biggr) \frac{\alpha^2 \tilde{G}^2 m^2}{r^2} \nonumber\\
&\  + \dot{r}^2 \Bigl(- \frac{3}{2} \bigl(3
 + \overline{\gamma}\bigr) \nu + 6 \nu^2\Bigr) v^{2}
 + \frac{\alpha \tilde{G}}{r} \biggl[\dot{r}^2 \biggl(-2 \delta_{+}
 -  \frac{1}{2} \bigl(2
 + \overline{\gamma}\bigr)^2
 + \Bigl(-25
 + 12 \beta_{+}
 - 18 \overline{\gamma}
 - 3 \overline{\gamma}^2 -  \frac{48 \delta_{-}^2}{\bigl(2 + \overline{\gamma}\bigr)^2} \nonumber\\
&\quad + \frac{48 \delta_{+}^2}{\bigl(2 + \overline{\gamma}\bigr)^2}\Bigr) \nu
 + \Bigl(-2
 + \frac{3}{2} \overline{\gamma}^2
 + \frac{24 \delta_{-}^2}{\bigl(2 + \overline{\gamma}\bigr)^2}
 -  \frac{24 \delta_{+}^2}{\bigl(2 + \overline{\gamma}\bigr)^2}\Bigr) \nu^2
 + \bigl(-2 \delta_{-} - 4 \beta_{-} \nu\bigr) m_{-}\biggr) m \nonumber\\
&\quad + \Bigl(2 \beta_{+}
 + \bigl(- \frac{13}{2}
 - 8 \beta_{+}
 - 2 \overline{\gamma}\bigr) \nu
 + 2 \nu^2
 + \bigl(-2 \beta_{-} + 6 \beta_{-} \nu\bigr) m_{-}\Bigr) m v^{2}\biggr]
 + \Bigl(\bigl(3
 + \overline{\gamma}\bigr) \nu
 - 4 \nu^2\Bigr) v^{4} \,,\\[7pt]
%%%%%%%%%%%%%%%%%%%%%%%%%%%%%%%%%%%%%%%%%%%%%%%%%%%%%%%%%%%%%%%%%
\mathcal{A}_{3\mathrm{PN}}^{(0)}={}&\dot{r}^6 \bigl(- \frac{35}{16} \nu
 + \frac{175}{16} \nu^2
 -  \frac{175}{16} \nu^3\bigr)
 + \dot{r}^4 \Bigl(\frac{15}{8} \bigl(4
 + \overline{\gamma}\bigr) \nu
 -  \frac{45}{8} \bigl(6
 + \overline{\gamma}\bigr) \nu^2
 + \frac{255}{8} \nu^3\Bigr) v^{2}\nonumber\\
& + \dot{r}^2 \Bigl(- \frac{3}{2} \bigl(5
 + 2 \overline{\gamma}\bigr) \nu
 + \bigl(\frac{237}{8}
 + 9 \overline{\gamma}\bigr) \nu^2
 -  \frac{45}{2} \nu^3\Bigr) v^{4}
 + \Bigl(\bigl(\frac{11}{4}
 + \overline{\gamma}\bigr) \nu
 + \bigl(- \frac{49}{4} - 3 \overline{\gamma}\bigr) \nu^2
 + 13 \nu^3\Bigr) v^{6} \,,\\[7pt]
%%%%%%%%%%%%%%%%%%%%%%%%%%%%%%%%%%%%%%%%%%%%%%%%%%%%%%%%%%%%%%%%%
\mathcal{A}_{3\mathrm{PN}}^{(1)}={}&\frac{\alpha}{r} \biggl[\dot{r}^4 \biggl(\Bigl(79
 + 18 \beta_{+}
 + 26 \delta_{+}
 + \frac{87}{2} \overline{\gamma}
 + \frac{35}{16} \overline{\gamma}^2
 -  \frac{69 \delta_{-}^2}{\bigl(2 + \overline{\gamma}\bigr)^2}
 + \frac{69 \delta_{+}^2}{\bigl(2 + \overline{\gamma}\bigr)^2}\Bigr) \nu
 + \Bigl(42 \beta_{+} -  \frac{192 \delta_{-}^2}{\bigl(2 + \overline{\gamma}\bigr)^2}
 + \frac{192 \delta_{+}^2}{\bigl(2 + \overline{\gamma}\bigr)^2} \nonumber\\
&\quad -  \frac{3}{2} \bigl(23
 + 28 \overline{\gamma}
 + 8 \overline{\gamma}^2\bigr)\Bigr) \nu^2
 + \Bigl(\frac{48 \delta_{-}^2}{\bigl(2 + \overline{\gamma}\bigr)^2}
 -  \frac{48 \delta_{+}^2}{\bigl(2 + \overline{\gamma}\bigr)^2} + 3 \bigl(-10
 + \overline{\gamma}^2\bigr)\Bigr) \nu^3
 + \Bigl(\bigl(18 \beta_{-}
 + 6 \delta_{-}\bigr) \nu
 - 18 \beta_{-} \nu^2\Bigr) m_{-}\biggr) m \nonumber\\
&\ + m v^{2} \dot{r}^2 \biggl(\Bigl(-121
 - 22 \beta_{+} - 30 \delta_{+}
 - 92 \overline{\gamma}
 - 16 \overline{\gamma}^2
 + \frac{88 \delta_{-}^2}{\bigl(2 + \overline{\gamma}\bigr)^2}
 -  \frac{88 \delta_{+}^2}{\bigl(2 + \overline{\gamma}\bigr)^2}\Bigr) \nu
 + \Bigl(16
 - 36 \beta_{+}
 + 26 \overline{\gamma}
 + \frac{19}{2} \overline{\gamma}^2 \nonumber\\
&\quad + \frac{152 \delta_{-}^2}{\bigl(2 + \overline{\gamma}\bigr)^2}
 -  \frac{152 \delta_{+}^2}{\bigl(2 + \overline{\gamma}\bigr)^2}\Bigr) \nu^2
 + \Bigl(20
 - 3 \overline{\gamma}^2
 -  \frac{48 \delta_{-}^2}{\bigl(2 + \overline{\gamma}\bigr)^2}
 + \frac{48 \delta_{+}^2}{\bigl(2 + \overline{\gamma}\bigr)^2}\Bigr) \nu^3 + \Bigl(\bigl(-14 \beta_{-}
 - 4 \delta_{-}\bigr) \nu
 + 16 \beta_{-} \nu^2\Bigr) m_{-}\biggr) \nonumber\\
&\ + m v^{4} \biggl(\Bigl(\frac{3}{2} \beta_{+}
 + 4 \delta_{+}
 -  \frac{14 \delta_{-}^2}{\bigl(2 + \overline{\gamma}\bigr)^2} + \frac{14 \delta_{+}^2}{\bigl(2 + \overline{\gamma}\bigr)^2}
 + \frac{3}{8} \bigl(50
 + 48 \overline{\gamma}
 + 11 \overline{\gamma}^2\bigr)\Bigr) \nu + \Bigl(8
 + 10 \beta_{+}
 -  \frac{5}{8} \overline{\gamma}^2 -  \frac{10 \delta_{-}^2}{\bigl(2 + \overline{\gamma}\bigr)^2} \nonumber\\
&\quad + \frac{10 \delta_{+}^2}{\bigl(2 + \overline{\gamma}\bigr)^2}\Bigr) \nu^2 - 10 \nu^3
 + \bigl(3 \beta_{-} \nu
 - 8 \beta_{-} \nu^2\bigr) m_{-}\biggr)\biggr] \,,\\
%%%%%%%%%%%%%%%%%%%%%%%%%%%%%%%%%%%%%%%%%%%%%%%%%%%%%%%%%%%%%%%%%
\mathcal{A}_{3\mathrm{PN}}^{(2)}={}&\frac{\alpha}{r^2} \biggl[\dot{r}^2 \biggl(\Bigl(- \frac{55}{8} \overline{\gamma} \bigl(2
 + \overline{\gamma}\bigr)
 + \frac{5 \delta_{+} \bigl(10 + \overline{\gamma}\bigr)}{2 \bigl(2 + \overline{\gamma}\bigr)}\Bigr) \nu
 + \frac{5 \delta_{-} \bigl(10 + \overline{\gamma}\bigr) \nu m_{-}}{2 \bigl(2 + \overline{\gamma}\bigr)}\biggr) m^2
 + \biggl(\Bigl(\frac{11}{8} \overline{\gamma} \bigl(2 + \overline{\gamma}\bigr)
 -  \frac{\delta_{+} \bigl(10 + \overline{\gamma}\bigr)}{4 + 2 \overline{\gamma}}\Bigr) \nu \nonumber\\
&\quad -  \frac{\delta_{-} \bigl(10 + \overline{\gamma}\bigr) \nu m_{-}}{4 + 2 \overline{\gamma}}\biggr) m^2 v^{2}\biggr]
 + \frac{\alpha^2}{r^2} \Biggl\{\dot{r}^2 \Biggl[\Biggl(\delta_{+} \bigl(1
 + \overline{\gamma}\bigr) + \frac{1}{4} \bigl(1
 + \overline{\gamma}\bigr) \bigl(2
 + \overline{\gamma}\bigr)^2
 + \biggl(15 \chi_{+}
 -  \frac{40 \beta_{-}^2}{\overline{\gamma}}
 + \frac{40 \beta_{+}^2}{\overline{\gamma}} \nonumber\\
&\quad -  \frac{120 \beta_{-} \delta_{-}}{\overline{\gamma}}
 + \frac{1}{96} \bigl(28228 + 32312 \overline{\gamma}
 + 11297 \overline{\gamma}^2
 + 1057 \overline{\gamma}^3\bigr)
 + \beta_{+} \Bigl(- \frac{120 \delta_{+}}{\overline{\gamma}}
 -  \frac{170 \delta_{-}^2}{\overline{\gamma} \bigl(2 + \overline{\gamma}\bigr)^2}
 + \frac{170 \delta_{+}^2}{\overline{\gamma} \bigl(2 + \overline{\gamma}\bigr)^2} \nonumber\\
&\quad + \frac{1}{8} \bigl(836
 + 571 \overline{\gamma}\bigr)\Bigr)
 -  \frac{15}{512} \bigl(2
 + \overline{\gamma}\bigr)^2 \bigl(-82
 + 7 \overline{\gamma}\bigr) \pi^2
 + \delta_{+}^2 \Bigl(\frac{367 + 323 \overline{\gamma}}{6 \bigl(2 + \overline{\gamma}\bigr)^2} -  \frac{105 \pi^2}{32 \bigl(2 + \overline{\gamma}\bigr)}\Bigr)
 + \delta_{-}^2 \Bigl(- \frac{367 + 323 \overline{\gamma}}{6 \bigl(2 + \overline{\gamma}\bigr)^2} \nonumber\\
&\quad + \frac{105 \pi^2}{32 \bigl(2 + \overline{\gamma}\bigr)}\Bigr)
 + \delta_{+} \Bigl(96
 + \frac{15}{2} \overline{\gamma}
 -  \frac{105}{64} \bigl(2 + \overline{\gamma}\bigr) \pi^2\Bigr)\biggr) \nu
 + \biggl(-10 \chi_{+}
 + 5 \delta_{+}
 -  \frac{60 \beta_{-}^2}{\overline{\gamma}}
 + \frac{60 \beta_{+}^2}{\overline{\gamma}}
 -  \frac{\delta_{-}^2 \bigl(255 + 37 \overline{\gamma}\bigr)}{2 \bigl(2 + \overline{\gamma}\bigr)^2} \nonumber\\
&\quad + \frac{\delta_{+}^2 \bigl(255 + 37 \overline{\gamma}\bigr)}{2 \bigl(2 + \overline{\gamma}\bigr)^2}
 + \frac{1}{32} \bigl(44
 + 32 \overline{\gamma}
 - 215 \overline{\gamma}^2
 - 37 \overline{\gamma}^3\bigr)
 + \beta_{+} \Bigl(- \frac{7}{2}
 -  \frac{65}{4} \overline{\gamma} -  \frac{260 \delta_{-}^2}{\overline{\gamma} \bigl(2 + \overline{\gamma}\bigr)^2}
 + \frac{260 \delta_{+}^2}{\overline{\gamma} \bigl(2 + \overline{\gamma}\bigr)^2}\Bigr)\biggr) \nu^2 \nonumber\\
&\quad + \Bigl(-7
 + 6 \overline{\gamma}^2
 + \frac{96 \delta_{-}^2}{\bigl(2 + \overline{\gamma}\bigr)^2}
 -  \frac{96 \delta_{+}^2}{\bigl(2 + \overline{\gamma}\bigr)^2}\Bigr) \nu^3 + \biggl[\delta_{-} \bigl(1
 + \overline{\gamma}\bigr)
 + \biggl(-15 \chi_{-}
 + \delta_{-} \bigl(- \frac{68}{3}
 -  \frac{5}{2} \overline{\gamma}\bigr)
 + \beta_{-} \Bigl(\frac{1}{8} \bigl(-580
 - 443 \overline{\gamma}\bigr) \nonumber\\
&\quad + \frac{170 \delta_{-}^2}{\overline{\gamma} \bigl(2 + \overline{\gamma}\bigr)^2}
 -  \frac{170 \delta_{+}^2}{\overline{\gamma} \bigl(2 + \overline{\gamma}\bigr)^2}\Bigr)\biggr) \nu
 + \beta_{-} \Bigl(\frac{1}{4} \bigl(-46
 - 25 \overline{\gamma}\bigr)
 -  \frac{100 \delta_{-}^2}{\overline{\gamma} \bigl(2 + \overline{\gamma}\bigr)^2} + \frac{100 \delta_{+}^2}{\overline{\gamma} \bigl(2 + \overline{\gamma}\bigr)^2}\Bigr) \nu^2\biggr] m_{-}\Biggr) m^2
 + \biggl(15 \delta_{-} \nu \nonumber\\
&\quad + \Bigl(15 \delta_{+}
 -  \frac{165}{4} \bigl(2 + \overline{\gamma}\bigr)^2\Bigr) \nu m_{-}\biggr) m^2 \ln\bigl(\frac{r}{r'_{-}}\bigr)
 + \biggl(\Bigl(15 \delta_{+}
 -  \frac{165}{4} \bigl(2
 + \overline{\gamma}\bigr)^2\Bigr) \nu
 + 15 \delta_{-} \nu m_{-}\biggr) m^2 \ln\bigl(\frac{r}{r'_{+}}\bigr)\Biggr]\nonumber\\
& + \Biggl[\Biggl(2 \chi_{+}
 + \biggl(-12 \chi_{+}
 -  \frac{20 \beta_{-}^2}{\overline{\gamma}}
 + \frac{20 \beta_{+}^2}{\overline{\gamma}}
 + \frac{24 \beta_{-} \delta_{-}}{\overline{\gamma}}
 + \frac{1}{96} \bigl(-4796
 - 6400 \overline{\gamma} - 2587 \overline{\gamma}^2
 - 281 \overline{\gamma}^3\bigr) + \beta_{+} \Bigl(-12 \nonumber\\
&\quad + \frac{24 \delta_{+}}{\overline{\gamma}}
 -  \frac{77}{8} \overline{\gamma}
 -  \frac{74 \delta_{-}^2}{\overline{\gamma} \bigl(2 + \overline{\gamma}\bigr)^2}
 + \frac{74 \delta_{+}^2}{\overline{\gamma} \bigl(2 + \overline{\gamma}\bigr)^2}\Bigr) + \frac{3}{512} \bigl(2
 + \overline{\gamma}\bigr)^2 \bigl(-82
 + 7 \overline{\gamma}\bigr) \pi^2 + \delta_{-}^2 \Bigl(\frac{353 + 139 \overline{\gamma}}{6 \bigl(2 + \overline{\gamma}\bigr)^2}
 -  \frac{21 \pi^2}{32 \bigl(2 + \overline{\gamma}\bigr)}\Bigr) \nonumber\\
&\quad + \delta_{+}^2 \Bigl(- \frac{353 + 139 \overline{\gamma}}{6 \bigl(2 + \overline{\gamma}\bigr)^2}
 + \frac{21 \pi^2}{32 \bigl(2 + \overline{\gamma}\bigr)}\Bigr)
 + \delta_{+} \Bigl(\frac{1}{2} \bigl(-33
 + \overline{\gamma}\bigr)
 + \frac{21}{64} \bigl(2
 + \overline{\gamma}\bigr) \pi^2\Bigr)\biggr) \nu + \biggl(12 \chi_{+}
 - 6 \delta_{+}
 + \frac{72 \beta_{-}^2}{\overline{\gamma}} \nonumber\\
&\quad -  \frac{72 \beta_{+}^2}{\overline{\gamma}}
 + \frac{\delta_{-}^2 \bigl(-251 - 113 \overline{\gamma}\bigr)}{\bigl(2 + \overline{\gamma}\bigr)^2}
 + \frac{\delta_{+}^2 \bigl(251 + 113 \overline{\gamma}\bigr)}{\bigl(2 + \overline{\gamma}\bigr)^2} -  \frac{1}{16} \overline{\gamma} \bigl(128
 + 275 \overline{\gamma}
 + 113 \overline{\gamma}^2\bigr)
 + \beta_{+} \Bigl(21
 + \frac{33}{2} \overline{\gamma} \nonumber\\
&\quad + \frac{264 \delta_{-}^2}{\overline{\gamma} \bigl(2 + \overline{\gamma}\bigr)^2}
 -  \frac{264 \delta_{+}^2}{\overline{\gamma} \bigl(2 + \overline{\gamma}\bigr)^2}\Bigr)\biggr) \nu^2 + \nu^3
 + \biggl[-2 \chi_{-}
 + \biggl(8 \chi_{-}
 + \frac{1}{6} \delta_{-} \bigl(41
 + 3 \overline{\gamma}\bigr)
 + \beta_{-} \Bigl(-8
 -  \frac{3}{8} \overline{\gamma}
 -  \frac{22 \delta_{-}^2}{\overline{\gamma} \bigl(2 + \overline{\gamma}\bigr)^2}\nonumber\\
& + \frac{22 \delta_{+}^2}{\overline{\gamma} \bigl(2 + \overline{\gamma}\bigr)^2}\Bigr)\biggr) \nu
 + \beta_{-} \Bigl(5
 -  \frac{3}{2} \overline{\gamma}
 -  \frac{24 \delta_{-}^2}{\overline{\gamma} \bigl(2 + \overline{\gamma}\bigr)^2}
 + \frac{24 \delta_{+}^2}{\overline{\gamma} \bigl(2 + \overline{\gamma}\bigr)^2}\Bigr) \nu^2\biggr] m_{-}\Biggr) m^2 + \biggl(-3 \delta_{-} \nu
 + \Bigl(-3 \delta_{+} \nonumber\\
&\quad + \frac{33}{4} \bigl(2
 + \overline{\gamma}\bigr)^2\Bigr) \nu m_{-}\biggr) m^2 \ln\bigl(\frac{r}{r'_{-}}\bigr)
 + \biggl(\Bigl(-3 \delta_{+}
 + \frac{33}{4} \bigl(2 + \overline{\gamma}\bigr)^2\Bigr) \nu
 - 3 \delta_{-} \nu m_{-}\biggr) m^2 \ln\bigl(\frac{r}{r'_{+}}\bigr)\Biggr] v^{2}\Biggr\} \,,\\[7pt]
%%%%%%%%%%%%%%%%%%%%%%%%%%%%%%%%%%%%%%%%%%%%%%%%%%%%%%%%%%%%%%%%%
\mathcal{A}_{3\mathrm{PN}}^{(3)}={}&\biggl(\Bigl(\frac{2 \delta_{+} \overline{\gamma}}{2 + \overline{\gamma}}
 -  \frac{11}{2} \overline{\gamma} \bigl(2
 + \overline{\gamma}\bigr)\Bigr) \nu
 + \frac{2 \delta_{-} \overline{\gamma} \nu m_{-}}{2 + \overline{\gamma}}\biggr) \frac{\alpha^2 m^3}{r^3}
 + \Biggl(\frac{8}{3} \beta_{-} \delta_{-}
 + 4 \chi_{+} \bigl(2
 + \overline{\gamma}\bigr) -  \frac{8}{3} \delta_{+} \bigl(2
 + \overline{\gamma}\bigr)
 - 2 \bigl(2
 + \overline{\gamma}\bigr)^3 \nonumber\\
&\quad + \beta_{+} \Bigl(- \frac{8}{3} \delta_{+}
 -  \frac{14}{3} \bigl(2
 + \overline{\gamma}\bigr)^2\Bigr)
 -  \frac{4}{3} \kappa_{+}
 + \biggl(\beta_{+}^2 \bigl(28 + \frac{136}{\overline{\gamma}}\bigr)
 + \chi_{+} \bigl(- \frac{4}{3}
 - 8 \overline{\gamma}\bigr)
 -  \frac{64 \beta_{+}^3}{\overline{\gamma}^2}
 + \frac{1}{144} \bigl(2
 + \overline{\gamma}\bigr)^2 \bigl(-3801 \nonumber\\
&\quad + 1357 \overline{\gamma}\bigr) + \beta_{-} \Bigl(\frac{32 \chi_{-}}{\overline{\gamma}} - \frac{32 \delta_{-} \bigl(2 + \overline{\gamma}\bigr)}{3 \overline{\gamma}}\Bigr)
 + \beta_{-}^2 \Bigl(\frac{64 \beta_{+}}{\overline{\gamma}^2}
 -  \frac{8 \bigl(17 + 7 \overline{\gamma}\bigr)}{\overline{\gamma}}\Bigr)
 + \beta_{+} \Bigl(- \frac{32 \chi_{+}}{\overline{\gamma}}
 + \frac{16 \delta_{+} \bigl(-4 + \overline{\gamma}\bigr)}{3 \overline{\gamma}} \nonumber\\
&\quad -  \frac{116 \delta_{-}^2 \bigl(8 + 3 \overline{\gamma}\bigr)}{\overline{\gamma} \bigl(2 + \overline{\gamma}\bigr)^2}
 + \frac{116 \delta_{+}^2 \bigl(8 + 3 \overline{\gamma}\bigr)}{\overline{\gamma} \bigl(2 + \overline{\gamma}\bigr)^2}
 -  \frac{5}{12} \bigl(2
 + \overline{\gamma}\bigr) \bigl(134
 + 49 \overline{\gamma}\bigr)\Bigr)
 + 4 \kappa_{+}
 -  \frac{1}{128} \bigl(2 + \overline{\gamma}\bigr)^2 \bigl(-82
 + 7 \overline{\gamma}\bigr) \pi^2 \nonumber\\
&\quad + \delta_{+}^2 \Bigl(- \frac{7 \bigl(129 + 55 \overline{\gamma}\bigr)}{9 \bigl(2 + \overline{\gamma}\bigr)^2}
 -  \frac{7 \pi^2}{8 \bigl(2 + \overline{\gamma}\bigr)}\Bigr)
 + \delta_{-}^2 \Bigl(\frac{7 \bigl(129 + 55 \overline{\gamma}\bigr)}{9 \bigl(2 + \overline{\gamma}\bigr)^2} + \frac{7 \pi^2}{8 \bigl(2 + \overline{\gamma}\bigr)}\Bigr)
 + \delta_{+} \Bigl(\frac{1}{9} \bigl(205
 + 39 \overline{\gamma}\bigr) \nonumber\\
&\quad -  \frac{7}{16} \bigl(2
 + \overline{\gamma}\bigr) \pi^2\Bigr)\biggr) \nu + \biggl(- \frac{40}{3} \chi_{+}
 + \frac{20}{3} \delta_{+} + \beta_{-}^2 \bigl(24
 -  \frac{80}{\overline{\gamma}}\bigr)
 + \frac{80 \beta_{+}^2}{\overline{\gamma}}
 + \frac{1}{24} \bigl(2
 + \overline{\gamma}\bigr)^2 \bigl(-213
 + 145 \overline{\gamma}\bigr) \nonumber\\
&\quad + \frac{2 \delta_{-}^2 \bigl(327 + 145 \overline{\gamma}\bigr)}{3 \bigl(2 + \overline{\gamma}\bigr)^2} -  \frac{2 \delta_{+}^2 \bigl(327 + 145 \overline{\gamma}\bigr)}{3 \bigl(2 + \overline{\gamma}\bigr)^2}
 + \beta_{+} \Bigl(-3 \bigl(2
 + \overline{\gamma}\bigr) \bigl(4
 + 7 \overline{\gamma}\bigr)
 -  \frac{48 \delta_{-}^2 \bigl(18 + 7 \overline{\gamma}\bigr)}{\overline{\gamma} \bigl(2 + \overline{\gamma}\bigr)^2} \nonumber\\
&\quad + \frac{48 \delta_{+}^2 \bigl(18 + 7 \overline{\gamma}\bigr)}{\overline{\gamma} \bigl(2 + \overline{\gamma}\bigr)^2}\Bigr)\biggr) \nu^2 + \biggl[- \frac{8}{3} \beta_{+} \delta_{-}
 - 4 \chi_{-} \bigl(2
 + \overline{\gamma}\bigr)
 -  \frac{8}{3} \delta_{-} \bigl(2
 + \overline{\gamma}\bigr)
 + \beta_{-} \Bigl(\frac{8}{3} \delta_{+} + \frac{14}{3} \bigl(2
 + \overline{\gamma}\bigr)^2\Bigr)
 + \frac{4}{3} \kappa_{-} \nonumber\\
&\quad + \biggl(- \frac{44}{3} \chi_{-}
 + \beta_{+} \bigl(\frac{32 \chi_{-}}{\overline{\gamma}}
 + \frac{32 \delta_{-}}{3 \overline{\gamma}}\bigr)
 + \delta_{-} \bigl(-7
 -  \overline{\gamma}\bigr) -  \frac{64 \beta_{-}^3}{\overline{\gamma}^2}
 + \beta_{-} \Bigl(28 \beta_{+}
 + \frac{64 \beta_{+}^2}{\overline{\gamma}^2}
 -  \frac{32 \chi_{+}}{\overline{\gamma}} -  \frac{16 \delta_{+} \bigl(-2 + \overline{\gamma}\bigr)}{3 \overline{\gamma}} \nonumber\\
&\quad + \frac{4 \delta_{-}^2 \bigl(176 + 65 \overline{\gamma}\bigr)}{\overline{\gamma} \bigl(2 + \overline{\gamma}\bigr)^2} -  \frac{4 \delta_{+}^2 \bigl(176 + 65 \overline{\gamma}\bigr)}{\overline{\gamma} \bigl(2 + \overline{\gamma}\bigr)^2}
 + \frac{1}{12} \bigl(2
 + \overline{\gamma}\bigr) \bigl(538
 + 179 \overline{\gamma}\bigr)\Bigr) -  \frac{4}{3} \kappa_{-}\biggr) \nu\biggr] m_{-}\Biggr) \frac{\alpha^3 m^3}{r^3} \,,
%%%%%%%%%%%%%%%%%%%%%%%%%%%%%%%%%%%%%%%%%%%%%%%%%%%%%%%%%%%%%%%%%
\end{align}\end{subequations}
and
\begin{subequations}\label{B}
\begin{align}
\mathcal{B}_{1\mathrm{PN}}={}&\dot{r} \Bigl(-2 \bigl(2 + \overline{\gamma}\bigr) + 2 \nu\Bigr) \,,\\
%%%%%%%%%%%%%%%%%%%%%%%%%%%%%%%%%%%%%%%%%%%%%%%%%%%%%%%%%%%%%%%%%
\mathcal{B}_{2\mathrm{PN}}={}&\dot{r}^3 \Bigl(\bigl(\frac{9}{2}
 + 3 \overline{\gamma}\bigr) \nu
 + 3 \nu^2\Bigr)
 + \dot{r} \biggl(2 \delta_{+}
 + \frac{1}{2} \bigl(2
 + \overline{\gamma}\bigr)^2
 + \Bigl(\frac{41}{2}
 - 8 \beta_{+}
 + 14 \overline{\gamma}
 + \frac{13}{8} \overline{\gamma}^2 \nonumber\\
&\quad + \frac{26 \delta_{-}^2}{\bigl(2 + \overline{\gamma}\bigr)^2}
 -  \frac{26 \delta_{+}^2}{\bigl(2 + \overline{\gamma}\bigr)^2}\Bigr) \nu
 + 4 \nu^2
 + \bigl(2 \delta_{-}
 + 4 \beta_{-} \nu\bigr) m_{-}\biggr) \frac{\alpha \tilde{G} m}{r}
 + \dot{r} \Bigl(\bigl(- \frac{15}{2}
 - 4 \overline{\gamma}\bigr) \nu - 2 \nu^2\Bigr) v^{2} \,,\\
%%%%%%%%%%%%%%%%%%%%%%%%%%%%%%%%%%%%%%%%%%%%%%%%%%%%%%%%%%%%%%%%%
\mathcal{B}_{3\mathrm{PN}}={}&\dot{r}^5 \Bigl(- \frac{15}{8} \bigl(3
 + 2 \overline{\gamma}\bigr) \nu
 + \bigl(15
 + \frac{45}{4} \overline{\gamma}\bigr) \nu^2
 + \frac{15}{4} \nu^3\Bigr) + \dot{r}^3 \Bigl(\frac{3}{2} \bigl(8
 + 5 \overline{\gamma}\bigr) \nu
 -  \frac{3}{4} \bigl(37
 + 28 \overline{\gamma}\bigr) \nu^2
 - 12 \nu^3\Bigr) v^{2} \nonumber\\
&\  + \dot{r} \Bigl(\bigl(- \frac{65}{8} -  \frac{9}{2} \overline{\gamma}\bigr) \nu
 + \bigl(19
 + 12 \overline{\gamma}\bigr) \nu^2
 + 6 \nu^3\Bigr) v^{4} + \frac{\tilde{G}^2 m^2}{r^2} \Biggl\{\alpha \dot{r} \biggl(\Bigl(\frac{11}{4} \overline{\gamma} \bigl(2
 + \overline{\gamma}\bigr) -  \frac{\delta_{+} \bigl(10 + \overline{\gamma}\bigr)}{2 + \overline{\gamma}}\Bigr) \nu \nonumber\\
&\quad -  \frac{\delta_{-} \bigl(10 + \overline{\gamma}\bigr) \nu m_{-}}{2 + \overline{\gamma}}\biggr)
 + \alpha^2 \dot{r} \Biggl[\Biggl(-2 \delta_{+} \bigl(2
 + \overline{\gamma}\bigr)
 -  \frac{1}{2} \bigl(2
 + \overline{\gamma}\bigr)^3 + \biggl(-8 \chi_{+}
 + \frac{24 \beta_{-}^2}{\overline{\gamma}}
 -  \frac{24 \beta_{+}^2}{\overline{\gamma}} + \frac{48 \beta_{-} \delta_{-}}{\overline{\gamma}} \nonumber\\
&\quad + \frac{1}{96} \bigl(-5500
 - 4928 \overline{\gamma}
 - 617 \overline{\gamma}^2
 + 221 \overline{\gamma}^3\bigr) + \beta_{+} \Bigl(\frac{48 \delta_{+}}{\overline{\gamma}}
 + \frac{92 \delta_{-}^2}{\overline{\gamma} \bigl(2 + \overline{\gamma}\bigr)^2}
 -  \frac{92 \delta_{+}^2}{\overline{\gamma} \bigl(2 + \overline{\gamma}\bigr)^2} - \frac{3}{4} \bigl(56
 + 43 \overline{\gamma}\bigr)\Bigr) \nonumber\\
&\quad + \frac{3}{256} \bigl(2
 + \overline{\gamma}\bigr)^2 \bigl(-82 + 7 \overline{\gamma}\bigr) \pi^2
 + \delta_{-}^2 \Bigl(\frac{871 + 485 \overline{\gamma}}{6 \bigl(2 + \overline{\gamma}\bigr)^2}
 -  \frac{21 \pi^2}{16 \bigl(2 + \overline{\gamma}\bigr)}\Bigr)
 + \delta_{+}^2 \Bigl(- \frac{871 + 485 \overline{\gamma}}{6 \bigl(2 + \overline{\gamma}\bigr)^2}
 + \frac{21 \pi^2}{16 \bigl(2 + \overline{\gamma}\bigr)}\Bigr) \nonumber\\
&\quad + \delta_{+} \Bigl(-34
 - 7 \overline{\gamma}
 + \frac{21}{32} \bigl(2
 + \overline{\gamma}\bigr) \pi^2\Bigr)\biggr) \nu
 + \biggl(25
 + 8 \chi_{+}
 - 4 \delta_{+}
 + \frac{48 \beta_{-}^2}{\overline{\gamma}}
 -  \frac{48 \beta_{+}^2}{\overline{\gamma}} + 16 \overline{\gamma}  - 12 \overline{\gamma}^2
 - 6 \overline{\gamma}^3 \nonumber\\
&\quad -  \frac{16 \delta_{-}^2 \bigl(11 + 6 \overline{\gamma}\bigr)}{\bigl(2 + \overline{\gamma}\bigr)^2}
 + \frac{16 \delta_{+}^2 \bigl(11 + 6 \overline{\gamma}\bigr)}{\bigl(2 + \overline{\gamma}\bigr)^2}
 + \beta_{+} \Bigl(-6
 + 15 \overline{\gamma} + \frac{240 \delta_{-}^2}{\overline{\gamma} \bigl(2 + \overline{\gamma}\bigr)^2}
 -  \frac{240 \delta_{+}^2}{\overline{\gamma} \bigl(2 + \overline{\gamma}\bigr)^2}\Bigr)\biggr) \nu^2
 + 8 \nu^3 \nonumber\\
&\quad + m_{-} \biggl[-2 \delta_{-} \bigl(2
 + \overline{\gamma}\bigr)
 + \biggl(8 \chi_{-}
 + \delta_{-} \bigl(\frac{14}{3}
 + \overline{\gamma}\bigr) + \beta_{-} \Bigl(42
 + \frac{129}{4} \overline{\gamma}
 -  \frac{92 \delta_{-}^2}{\overline{\gamma} \bigl(2 + \overline{\gamma}\bigr)^2}
 + \frac{92 \delta_{+}^2}{\overline{\gamma} \bigl(2 + \overline{\gamma}\bigr)^2}\Bigr)\biggr) \nu 
 + 14 \beta_{-} \nu^2\biggr]\Biggr) \nonumber\\
&\quad + \biggl(-6 \delta_{-} \nu + \Bigl(-6 \delta_{+}
 + \frac{33}{2} \bigl(2
 + \overline{\gamma}\bigr)^2\Bigr) \nu m_{-}\biggr) \ln\bigl(\frac{r}{r'_{-}}\bigr)
 + \biggl(\Bigl(-6 \delta_{+}
 + \frac{33}{2} \bigl(2
 + \overline{\gamma}\bigr)^2\Bigr) \nu
 - 6 \delta_{-} \nu m_{-}\biggr) \ln\bigl(\frac{r}{r'_{+}}\bigr)\Biggr]\Biggr\} \nonumber\\
&\  + \frac{\alpha \tilde{G} m}{r} \biggl[\dot{r}^3 \biggl(\Bigl(-12 \beta_{+}
 -  \frac{4}{3} \delta_{+}
 + \frac{42 \delta_{-}^2}{\bigl(2 + \overline{\gamma}\bigr)^2}
 -  \frac{42 \delta_{+}^2}{\bigl(2 + \overline{\gamma}\bigr)^2}
 + \frac{1}{24} \bigl(2
 + \overline{\gamma}\bigr) \bigl(658 + 727 \overline{\gamma}\bigr)\Bigr) \nu
 + \Bigl(-16 \beta_{+}
 + \frac{88 \delta_{-}^2}{\bigl(2 + \overline{\gamma}\bigr)^2} \nonumber\\
&\quad -  \frac{88 \delta_{+}^2}{\bigl(2 + \overline{\gamma}\bigr)^2}
 + \frac{1}{2} \bigl(59
 + 56 \overline{\gamma}
 + 11 \overline{\gamma}^2\bigr)\Bigr) \nu^2 + 18 \nu^3
 + \Bigl(\bigl(-12 \beta_{-}
 - 8 \delta_{-}\bigr) \nu
 + 8 \beta_{-} \nu^2\Bigr) m_{-}\biggr) \nonumber\\
&\quad + \dot{r} \biggl(\Bigl(-15
 + 12 \beta_{+}
 + 4 \delta_{+}
 - 39 \overline{\gamma} -  \frac{125}{8} \overline{\gamma}^2
 -  \frac{42 \delta_{-}^2}{\bigl(2 + \overline{\gamma}\bigr)^2}
 + \frac{42 \delta_{+}^2}{\bigl(2 + \overline{\gamma}\bigr)^2}\Bigr) \nu
 + \Bigl(-27
 + 8 \beta_{+} \nonumber\\
&\quad - 20 \overline{\gamma}
 -  \frac{25}{8} \overline{\gamma}^2
 -  \frac{50 \delta_{-}^2}{\bigl(2 + \overline{\gamma}\bigr)^2} + \frac{50 \delta_{+}^2}{\bigl(2 + \overline{\gamma}\bigr)^2}\Bigr) \nu^2
 - 10 \nu^3
 + \Bigl(\bigl(6 \beta_{-}
 + 6 \delta_{-}\bigr) \nu
 - 4 \beta_{-} \nu^2\Bigr) m_{-}\biggr) v^{2}\biggr] \,.
\end{align}\end{subequations}
The conservative tail part of the relative acceleration reads,
\begin{align}\label{aitail}\nn
a_{\mathrm{tail}}^{i} =\ & \frac{8G^{2}M}{3c^{6}\phi_{0}}\left(s_{1}-s_{2}\right)\int_{0}^{+\infty}\ud\tau\,\ln\left(\frac{c\tau}{2r}\right)\left[I_{s}^{i\,(5)}(t-\tau)-I_{s}^{i\,(5)}(t+\tau)\right]\\
& + \frac{8\tilde{G}^{3}\alpha^{3}M^{3}\nu}{3c^{6}r^{4}}\left(2\overline{\delta}_{+}+\frac{\overline{\gamma}(2+\overline{\gamma})}{2}\right)\left[\left(v^{2}-8(nv)^2-\frac{3\tilde{G}\alpha M}{2r}\right)n^{i}+2(nv)v^{i}\right]\,,
\end{align}
where the scalar dipole moment $I^{i}_{s}$ is, in the center of mass frame,
\be
I_{s}^{i} = \frac{2 m\nu(s_{1}-s_{2})}{\phi_{0}(3+2\omega_{0})}\,x^{i} \,.
\ee
The contribution~\eqref{aitail} only corresponds to the conservative part of the tail effects. One can see it by analysing the time-symmetric structure of the integrand, $f(t-\tau)-f(t+\tau)$. Together with this term, there is a time-antisymmetric contribution that corresponds to the dissipative part of the tail term. It reads,
\begin{align}\label{aitaildiss}\nn
a_{\mathrm{tail,\,diss}}^{i} =\ & \frac{8G^{2}M}{3c^{6}\phi_{0}}\left(s_{1}-s_{2}\right)\int_{0}^{+\infty}\ud\tau\,\ln\left(\frac{c\tau}{2r}\right)\left[I_{s}^{i\,(5)}(t-\tau)+I_{s}^{i\,(5)}(t+\tau)\right]\,.
\end{align}
Note that we can use any regularisation scale in this expression as it will cancel out from the two terms in the integral. Such a dissipative contribution to the equations of motion will have to be taken into account when computing the energy and angular momentum rate of loss. However, as in this paper we are only interested into the conservative dynamics, we will not consider the consequences of the additional dissipative term.

%%%%%%%%%%%%%%%%%%%%%%%%%%%%%%%%%%%%%%%%%%%%%%%%%%%%%%%
\section{Conserved integrals in the center-of-mass frame}\label{sec:invariants}
%%%%%%%%%%%%%%%%%%%%%%%%%%%%%%%%%%%%%%%%%%%%%%%%%%%%%%%

%%%%%%%%%%%%%%%%%%%%%%%%%%%%%%%%%%%%%%%%%%%%%%%%%%%%%%%
\subsection{Local part}\label{subsec:local}
%%%%%%%%%%%%%%%%%%%%%%%%%%%%%%%%%%%%%%%%%%%%%%%%%%%%%%%

The center-of-mass conserved energy and angular momentum are composed of an instantaneous part and a nonlocal part originating from the tail contribution. We start here by displaying the local contribution, before explaining the subtle incorporation of the tail terms in the next section.  It has been obtained by writing the combination $\sum_{A}m_{A}\mathbf{v}_{A}\cdot\mathbf{a}_{A\,\mathrm{inst}}$, where $\mathbf{a}_{A\,\mathrm{inst}}$ is the instantaneous 3PN acceleration, as a total time derivative $-\frac{\mathrm{d}E^{\mathrm{inst}}_{0}}{\mathrm{d}t}$. The complete instantaneous energy is then made of a kinetic and a potential pieces, $E^{\mathrm{inst}}=\sum_{A}\frac{1}{2}m_{A}v_{A}^{2}+E^{\mathrm{inst}}_{0}$. We have also obtained the conserved quantities by an alternative Lagrangian method, as described in~\cite{deAndrade:2000gf}, and we have checked that the two methods give the same results. Using the center-of-mass coordinates~\eqref{P}-\eqref{y1itail} derived in the previous section, we can write these quantities in the CM frame. Defining the reduced CM energy $\mathcal{E}\equiv\frac{E}{\mu}$ and the reduced CM angular momentum $\mathcal{J}\equiv\frac{\vert\mathbf{J}\vert}{\vert\mathbf{J}_{N}\vert}$, where $\mathbf{J}_{N}=\mu\mathbf{x}\times\mathbf{v}$ is the Newtonian angular momentum, we have
\begin{subequations}\label{Einst}\begin{align}
\mathcal{E}_{\mathrm{N}}={}&- \frac{\alpha \tilde{G} m}{r} + \frac{1}{2} v^{2} \,,\\[7pt]
%%%%%%%%%%%%%%%%%%%%%%%%%%%%%%%%%%%%%%%%%%%%%%%%%%%%%%%%%%%%%%%%%
\mathcal{E}_{1\mathrm{PN}}={}&\bigl(\frac{1}{2} + \beta_{+}
 -  \beta_{-} m_{-}\bigr) \frac{\alpha^2 \tilde{G}^2 m^2}{r^2}
 + \frac{\alpha \tilde{G} m}{r} \Bigl(\frac{1}{2} \dot{r}^2 \nu
 + \bigl(\frac{3}{2}
 + \overline{\gamma}
 + \frac{1}{2} \nu\bigr) v^{2}\Bigr)
 + \bigl(\frac{3}{8}
 -  \frac{9}{8} \nu\bigr) v^{4} \,,\\[7pt]
%%%%%%%%%%%%%%%%%%%%%%%%%%%%%%%%%%%%%%%%%%%%%%%%%%%%%%%%%%%%%%%%%
\mathcal{E}_{2\mathrm{PN}}={}& \frac{\alpha^3 \tilde{G}^3 m^3}{r^3}\biggl(- \beta_{+}
 + \frac{2}{3} \chi_{+}
 -  \frac{1}{3} \delta_{+}
 + \frac{1}{12} \bigl(-6
 - 4 \overline{\gamma}
 -  \overline{\gamma}^2\bigr)
 + \Bigl(-2 \beta_{+}
 -  \frac{4}{3} \chi_{+}
 + \frac{2}{3} \delta_{+}
 -  \frac{8 \beta_{-}^2}{\overline{\gamma}} + \frac{8 \beta_{+}^2}{\overline{\gamma}} \nonumber\\
&\quad + \frac{1}{12} \bigl(-45
 - 16 \overline{\gamma} + 2 \overline{\gamma}^2\bigr)\Bigr) \nu
 + \bigl(\beta_{-}
 -  \frac{2}{3} \chi_{-}
 -  \frac{1}{3} \delta_{-}\bigr) m_{-}\biggr) \nonumber\\
&\  + \frac{\alpha^2 \tilde{G}^2 m^2}{r^2} \biggl(\dot{r}^2 \Bigl(\frac{1}{2} \delta_{+}
 + \frac{1}{8} \bigl(2
 + \overline{\gamma}\bigr)^2
 + \bigl(\frac{69}{8}
 - 3 \beta_{+}
 + 6 \overline{\gamma}\bigr) \nu
 + \frac{3}{2} \nu^2
 + \bigl(\frac{1}{2} \delta_{-} + \beta_{-} \nu\bigr) m_{-}\Bigr)
 + \Bigl(\frac{1}{2} \beta_{+} \nonumber\\
&\quad -  \frac{1}{2} \delta_{+}
 + \frac{1}{8} \bigl(14
 + 20 \overline{\gamma}
 + 7 \overline{\gamma}^2\bigr)
 + \bigl(- \frac{55}{8}
 + \frac{1}{2} \beta_{+}
 - 4 \overline{\gamma}\bigr) \nu + \frac{1}{2} \nu^2
 + \bigl(- \frac{1}{2} \beta_{-}
 -  \frac{1}{2} \delta_{-}
 + \frac{1}{2} \beta_{-} \nu\bigr) m_{-}\Bigr) v^{2}\biggr) \nonumber\\
&\  + \frac{\alpha \tilde{G} m}{r} \biggl(\dot{r}^4 \bigl(- \frac{3}{8} \nu
 + \frac{9}{8} \nu^2\bigr)
 + \dot{r}^2 \Bigl(\frac{1}{4} \bigl(1- 2 \overline{\gamma}\bigr) \nu
 -  \frac{15}{4} \nu^2\Bigr) v^{2}
 + \Bigl(\frac{3}{8} \bigl(7
 + 4 \overline{\gamma}\bigr) \nonumber\\
&\quad + \bigl(- \frac{23}{8}
 -  \frac{5}{2} \overline{\gamma}\bigr) \nu
 -  \frac{27}{8} \nu^2\Bigr) v^{4}\biggr)
 + \bigl(\frac{5}{16}
 -  \frac{35}{16} \nu + \frac{65}{16} \nu^2\bigr) v^{6} \,,\\[7pt]
%%%%%%%%%%%%%%%%%%%%%%%%%%%%%%%%%%%%%%%%%%%%%%%%%%%%%%%%%%%%%%%%%
\mathcal{E}_{3\mathrm{PN}}^{(0)}={}&\bigl(\frac{35}{128}
 -  \frac{413}{128} \nu
 + \frac{833}{64} \nu^2
 -  \frac{2261}{128} \nu^3\bigr) v^{8} \,,\\[7pt]
%%%%%%%%%%%%%%%%%%%%%%%%%%%%%%%%%%%%%%%%%%%%%%%%%%%%%%%%%%%%%%%%%
\mathcal{E}_{3\mathrm{PN}}^{(1)}={}&\frac{\alpha m}{r} \biggl(\dot{r}^6 \bigl(\frac{5}{16} \nu
 -  \frac{25}{16} \nu^2
 + \frac{25}{16} \nu^3\bigr)
 + \dot{r}^4 \Bigl(\frac{3}{16} \bigl(-3
 + 2 \overline{\gamma}\bigr) \nu
 + \bigl(\frac{21}{4}
 -  \frac{9}{8} \overline{\gamma}\bigr) \nu^2
 -  \frac{165}{16} \nu^3\Bigr) v^{2}+ \dot{r}^2 \Bigl(- \frac{7}{16} \bigl(3
 + 4 \overline{\gamma}\bigr) \nu \nonumber\\
&\quad + \frac{1}{16} \bigl(-75
 + 76 \overline{\gamma}\bigr) \nu^2
 + \frac{375}{16} \nu^3\Bigr) v^{4} + \Bigl(\frac{5}{16} \bigl(11
 + 6 \overline{\gamma}\bigr) + \bigl(- \frac{215}{16}
 -  \frac{69}{8} \overline{\gamma}\bigr) \nu
 + \frac{1}{8} \bigl(58
 + 91 \overline{\gamma}\bigr) \nu^2
 + \frac{325}{16} \nu^3\Bigr) v^{6}\biggr) \,,\\
%%%%%%%%%%%%%%%%%%%%%%%%%%%%%%%%%%%%%%%%%%%%%%%%%%%%%%%%%%%%%%%%%
\mathcal{E}_{3\mathrm{PN}}^{(2)}={}&\frac{\alpha^2 m^2}{r^2} \biggl[\dot{r}^4 \biggl(\Bigl(-3 \beta_{+}
 -  \frac{13}{3} \delta_{+}
 + \frac{1}{48} \bigl(-731
 - 408 \overline{\gamma}
 - 52 \overline{\gamma}^2\bigr)\Bigr) \nu
 + \bigl(\frac{41}{4}
 - 7 \beta_{+} + 10 \overline{\gamma}\bigr) \nu^2 + 6 \nu^3 \nonumber\\
&\quad + \Bigl(\bigl(-3 \beta_{-}
 -  \delta_{-}\bigr) \nu
 + 3 \beta_{-} \nu^2\Bigr) m_{-}\biggr)
 + \dot{r}^2 \biggl(\frac{3}{4} \delta_{+}
 + \frac{3}{16} \bigl(2
 + \overline{\gamma}\bigr)^2 + \Bigl(\beta_{+}
 + \frac{5}{4} \delta_{+}
 + \frac{1}{16} \bigl(248
 - 24 \overline{\gamma}
 - 75 \overline{\gamma}^2\bigr)\Bigr) \nu \nonumber\\
&\quad + \bigl(- \frac{815}{16}
 + \frac{39}{2} \beta_{+}
 - 40 \overline{\gamma}\bigr) \nu^2
 -  \frac{81}{4} \nu^3 + \Bigl(\frac{3}{4} \delta_{-}
 + \bigl(5 \beta_{-}
 -  \frac{1}{4} \delta_{-}\bigr) \nu
 -  \frac{15}{2} \beta_{-} \nu^2\Bigr) m_{-}\biggr) v^{2} \nonumber\\
&\quad + \biggl(\frac{3}{8} \beta_{+}
 -  \frac{3}{4} \delta_{+}
 + \frac{3}{16} \bigl(45
 + 52 \overline{\gamma} + 15 \overline{\gamma}^2\bigr)
 + \Bigl(- \frac{19}{8} \beta_{+}
 + \frac{7}{4} \delta_{+}
 + \frac{1}{16} \bigl(-194
 - 120 \overline{\gamma}
 - 17 \overline{\gamma}^2\bigr)\Bigr) \nu \nonumber\\
&\quad + \bigl(\frac{203}{8}
 -  \frac{9}{8} \beta_{+} + 13 \overline{\gamma}\bigr) \nu^2
 -  \frac{27}{4} \nu^3
 + \Bigl(- \frac{3}{8} \beta_{-}
 -  \frac{3}{4} \delta_{-}
 + \bigl(\frac{13}{8} \beta_{-}
 + \frac{5}{4} \delta_{-}\bigr) \nu
 -  \frac{27}{8} \beta_{-} \nu^2\Bigr) m_{-}\biggr) v^{4}\biggr] \,,\\[7pt]
%%%%%%%%%%%%%%%%%%%%%%%%%%%%%%%%%%%%%%%%%%%%%%%%%%%%%%%%%%%%%%%%%
\mathcal{E}_{3\mathrm{PN}}^{(3)}={}&\frac{\alpha^2 m^3}{r^3} \biggl[\dot{r}^2 \biggl(\Bigl(\frac{11}{8} \overline{\gamma} \bigl(2
 + \overline{\gamma}\bigr)
 -  \frac{\delta_{+} \bigl(10 + \overline{\gamma}\bigr)}{4 + 2 \overline{\gamma}}\Bigr) \nu
 -  \frac{\delta_{-} \bigl(10 + \overline{\gamma}\bigr) \nu m_{-}}{4 + 2 \overline{\gamma}}\biggr)
 + \biggl(\Bigl(- \frac{11}{24} \overline{\gamma} \bigl(2 + \overline{\gamma}\bigr)
 + \frac{\delta_{+} \bigl(10 + \overline{\gamma}\bigr)}{12 + 6 \overline{\gamma}}\Bigr) \nu \nonumber\\
&\quad + \frac{\delta_{-} \bigl(10 + \overline{\gamma}\bigr) \nu m_{-}}{12 + 6 \overline{\gamma}}\biggr) v^{2}\biggr]
 + \frac{\alpha^3 m^3}{r^3} \Biggl(\dot{r}^2 \biggl[\delta_{+} \bigl(\frac{3}{2}
 + \overline{\gamma}\bigr)
 + \frac{1}{8} \bigl(2 + \overline{\gamma}\bigr)^2 \bigl(3
 + 2 \overline{\gamma}\bigr)
 + \biggl(-3 \chi_{+} + \beta_{+} \bigl(-31
 + \frac{24 \delta_{+}}{\overline{\gamma}} \nonumber\\
&\quad - 24 \overline{\gamma}\bigr)
 + \frac{8 \beta_{-}^2}{\overline{\gamma}}
 -  \frac{8 \beta_{+}^2}{\overline{\gamma}}
 + \frac{24 \beta_{-} \delta_{-}}{\overline{\gamma}} + \frac{1}{24} \bigl(-405
 - 370 \overline{\gamma}
 - 185 \overline{\gamma}^2
 - 69 \overline{\gamma}^3\bigr) + \frac{3}{256} \bigl(-164
 - 150 \overline{\gamma} - 20 \overline{\gamma}^2 \nonumber\\
&\quad + 7 \overline{\gamma}^3\bigr) \pi^2 + \delta_{+} \Bigl(\frac{1}{6} \bigl(-101
 - 9 \overline{\gamma}\bigr)
 + \frac{21}{64} \bigl(2
 + \overline{\gamma}\bigr) \pi^2\Bigr)\biggr) \nu + \bigl(\frac{51}{4}
 - 5 \beta_{+}
 + 2 \chi_{+}
 -  \delta_{+}
 + \frac{12 \beta_{-}^2}{\overline{\gamma}} -  \frac{12 \beta_{+}^2}{\overline{\gamma}} \nonumber\\
&\quad + 9 \overline{\gamma}
 -  \frac{1}{4} \overline{\gamma}^2\bigr) \nu^2
 + \frac{7}{2} \nu^3
 + \biggl(\delta_{-} \bigl(\frac{3}{2}
 + \overline{\gamma}\bigr)
 + \Bigl(3 \chi_{-}
 + \frac{1}{6} \delta_{-} \bigl(29
 + 3 \overline{\gamma}\bigr)
 + \beta_{-} \bigl(19 + 16 \overline{\gamma}\bigr)\Bigr) \nu
 + 5 \beta_{-} \nu^2\biggr) m_{-} \nonumber\\
&\quad + \biggl(-3 \delta_{-} \nu + \Bigl(-3 \delta_{+}
 + \frac{33}{4} \bigl(2
 + \overline{\gamma}\bigr)^2\Bigr) \nu m_{-}\biggr) \ln\bigl(\frac{r}{r'_{-}}\bigr) + \biggl(\Bigl(-3 \delta_{+}
 + \frac{33}{4} \bigl(2
 + \overline{\gamma}\bigr)^2\Bigr) \nu
 - 3 \delta_{-} \nu m_{-}\biggr) \ln\bigl(\frac{r}{r'_{+}}\bigr)\biggr] \nonumber\\
&\quad + \biggl[\frac{1}{3} \chi_{+}
 + \delta_{+} \bigl(-1
 -  \frac{2}{3} \overline{\gamma}\bigr) + \beta_{+} \bigl(\frac{3}{2}
 + \overline{\gamma}\bigr)
 + \frac{1}{12} \bigl(15
 + 34 \overline{\gamma}
 + 25 \overline{\gamma}^2
 + 6 \overline{\gamma}^3\bigr)
 + \biggl(- \frac{1}{3} \chi_{+}
 -  \frac{8 \beta_{-}^2}{\overline{\gamma}} + \frac{8 \beta_{+}^2}{\overline{\gamma}} \nonumber\\
&\quad -  \frac{8 \beta_{-} \delta_{-}}{\overline{\gamma}} + \beta_{+} \bigl(13
 -  \frac{8 \delta_{+}}{\overline{\gamma}} + 10 \overline{\gamma}\bigr)
 + \frac{1}{72} \bigl(-1395
 - 1856 \overline{\gamma}
 - 565 \overline{\gamma}^2
 + 33 \overline{\gamma}^3\bigr)
 + \frac{1}{256} \bigl(164 + 150 \overline{\gamma}
 + 20 \overline{\gamma}^2 \nonumber\\
&\quad - 7 \overline{\gamma}^3\bigr) \pi^2 + \delta_{+} \Bigl(\frac{95}{18}
 + \frac{7}{6} \overline{\gamma}
 -  \frac{7}{64} \bigl(2
 + \overline{\gamma}\bigr) \pi^2\Bigr)\biggr) \nu
 + \Bigl(\beta_{+}
 + \frac{2}{3} \chi_{+} -  \frac{1}{3} \delta_{+}
 + \frac{4 \beta_{-}^2}{\overline{\gamma}} -  \frac{4 \beta_{+}^2}{\overline{\gamma}}
 + \frac{1}{12} \bigl(-63
 - 40 \overline{\gamma} \nonumber\\
&\quad -  \overline{\gamma}^2\bigr)\Bigr) \nu^2
 + \frac{1}{2} \nu^3
 + \biggl(- \frac{1}{3} \chi_{-}
 + \beta_{-} \bigl(- \frac{3}{2} -  \overline{\gamma}\bigr)
 + \delta_{-} \bigl(-1
 -  \frac{2}{3} \overline{\gamma}\bigr) + \Bigl(- \frac{1}{3} \chi_{-}
 + \frac{1}{18} \delta_{-} \bigl(-5
 - 3 \overline{\gamma}\bigr)
 - 4 \beta_{-} \bigl(3
 + 2 \overline{\gamma}\bigr)\Bigr) \nu \nonumber\\
&\quad + \beta_{-} \nu^2\biggr) m_{-} + \biggl(\delta_{-} \nu
 + \Bigl(\delta_{+}
 -  \frac{11}{4} \bigl(2
 + \overline{\gamma}\bigr)^2\Bigr) \nu m_{-}\biggr) \ln\bigl(\frac{r}{r'_{-}}\bigr)
 + \biggl(\Bigl(\delta_{+}
 -  \frac{11}{4} \bigl(2 + \overline{\gamma}\bigr)^2\Bigr) \nu
 + \delta_{-} \nu m_{-}\biggr) \ln\bigl(\frac{r}{r'_{+}}\bigr)\biggr] v^{2}\Biggr) \,,\\[7pt]
%%%%%%%%%%%%%%%%%%%%%%%%%%%%%%%%%%%%%%%%%%%%%%%%%%%%%%%%%%%%%%%%%
\mathcal{E}_{3\mathrm{PN}}^{(4)}={}&\biggl(\Bigl(\frac{11}{12} \overline{\gamma} \bigl(2
 + \overline{\gamma}\bigr)
 + \frac{\delta_{+} \bigl(5 -  \overline{\gamma}\bigr)}{6 + 3 \overline{\gamma}}\Bigr) \nu
 + \frac{\delta_{-} \bigl(5 -  \overline{\gamma}\bigr) \nu m_{-}}{6 + 3 \overline{\gamma}}\biggr) \frac{\alpha^3 m^4}{r^4}
 + \frac{\alpha^4 m^4}{r^4} \Biggl(\frac{1}{2} \beta_{-}^2
 + \frac{1}{2} \beta_{+}^2 -  \frac{2}{3} \chi_{+}
 -  \frac{2}{3} \beta_{-} \delta_{-} + \frac{1}{3} \delta_{+} \nonumber\\
&\quad + \frac{1}{24} \bigl(9
 + 8 \overline{\gamma}
 + 2 \overline{\gamma}^2\bigr)
 + \beta_{+} \Bigl(\frac{2}{3} \delta_{+}
 + \frac{1}{6} \bigl(7
 + 4 \overline{\gamma}
 + \overline{\gamma}^2\bigr)\Bigr) + \frac{1}{3} \kappa_{+}
 + \biggl(\beta_{+}^2 \bigl(\frac{5}{2}
 -  \frac{16}{\overline{\gamma}}\bigr) + \beta_{-}^2 \bigl(- \frac{3}{2}
 -  \frac{16 \beta_{+}}{\overline{\gamma}^2}
 + \frac{16}{\overline{\gamma}}\bigr) \nonumber\\
&\quad + \beta_{-} \Bigl(- \frac{8 \chi_{-}}{\overline{\gamma}} + \frac{8 \delta_{-} \bigl(-1 + \overline{\gamma}\bigr)}{3 \overline{\gamma}}\Bigr)
 + \frac{16 \beta_{+}^3}{\overline{\gamma}^2}
 + \frac{1}{12} \delta_{+} \bigl(16
 + \overline{\gamma}\bigr) + \frac{1}{144} \bigl(3846
 + 3196 \overline{\gamma}
 + 588 \overline{\gamma}^2 - 33 \overline{\gamma}^3\bigr) \nonumber\\
&\quad + \beta_{+} \Bigl(\frac{8 \chi_{+}}{\overline{\gamma}}
 -  \frac{4 \delta_{+} \bigl(2 + \overline{\gamma}\bigr)}{3 \overline{\gamma}}
 + \frac{1}{12} \bigl(227
 + 80 \overline{\gamma}
 - 4 \overline{\gamma}^2\bigr)\Bigr)
 -  \kappa_{+}\biggr) \nu
 + \biggl[\frac{2}{3} \chi_{-} + \frac{1}{3} \delta_{-}
 + \frac{2}{3} \beta_{+} \delta_{-}
 + \beta_{-} \Bigl(- \beta_{+} \nonumber\\
&\quad -  \frac{2}{3} \delta_{+} + \frac{1}{6} \bigl(-7
 - 4 \overline{\gamma}
 -  \overline{\gamma}^2\bigr)\Bigr)
 -  \frac{1}{3} \kappa_{-}
 + \biggl(\frac{4}{3} \chi_{-} + \beta_{+} \bigl(- \frac{8 \chi_{-}}{\overline{\gamma}}
 -  \frac{8 \delta_{-}}{3 \overline{\gamma}}\bigr)
 + \delta_{-} \bigl(- \frac{7}{9}
 + \frac{1}{12} \overline{\gamma}\bigr)
 + \frac{16 \beta_{-}^3}{\overline{\gamma}^2} \nonumber\\
&\quad + \beta_{-} \Bigl(-3 \beta_{+}
 -  \frac{16 \beta_{+}^2}{\overline{\gamma}^2}
 + \frac{8 \chi_{+}}{\overline{\gamma}} + \frac{4 \delta_{+} \bigl(-2 + \overline{\gamma}\bigr)}{3 \overline{\gamma}}
 + \frac{1}{12} \bigl(-191
 - 80 \overline{\gamma}
 + 4 \overline{\gamma}^2\bigr)\Bigr)
 + \frac{1}{3} \kappa_{-}\biggr) \nu\biggr] m_{-}
 + \biggl(\delta_{-} \nu \nonumber\\
&\quad + \Bigl(\delta_{+} -  \frac{11}{4} \bigl(2
 + \overline{\gamma}\bigr)^2\Bigr) \nu m_{-}\biggr) \ln\bigl(\frac{r}{r'_{-}}\bigr)
 + \biggl(\Bigl(\delta_{+}
 -  \frac{11}{4} \bigl(2
 + \overline{\gamma}\bigr)^2\Bigr) \nu
 + \delta_{-} \nu m_{-}\biggr) \ln\bigl(\frac{r}{r'_{+}}\bigr)\Biggr) \,,
\end{align}\end{subequations}
and
\begin{subequations}\label{J}\begin{align}
\mathcal{J}_{\mathrm{N}}={}& 1 \,,\\
%%%%%%%%%%%%%%%%%%%%%%%%%%%%%%%%%%%%%%%%%%%%%%%%%%%%%%%%%%%%%%%%%
\mathcal{J}_{1\mathrm{PN}}={}& \bigl(3
 + \nu
 + 2 \overline{\gamma}\bigr) \frac{\alpha \tilde{G} m}{r}
 + \bigl(\frac{1}{2}
 -  \frac{3}{2} \nu\bigr) v^{2} \,,\\ 
%%%%%%%%%%%%%%%%%%%%%%%%%%%%%%%%%%%%%%%%%%%%%%%%%%%%%%%%%%%%%%%%%
\mathcal{J}_{2\mathrm{PN}}={}& \frac{\alpha^2 \tilde{G}^2 m^2}{r^2} \Bigl(\frac{7}{2}
 + \beta_{+}
 -  \delta_{+}
 + \nu^2
 + \nu \bigl(- \frac{41}{4}
 + \beta_{+}
 - 6 \overline{\gamma}\bigr)
 + 5 \overline{\gamma}
 + \frac{7}{4} \overline{\gamma}^2
 + \bigl(- \beta_{-}
 -  \delta_{-} + \beta_{-} \nu\bigr) m_{-}\Bigr) \nonumber\\
&\  + \frac{\alpha \tilde{G} m}{r} \biggl(\dot{r}^2 \Bigl(- \frac{5}{2} \nu^2
 + \nu \bigl(-1
 -  \overline{\gamma}\bigr)\Bigr)
 + \Bigl(\frac{7}{2}
 -  \frac{9}{2} \nu^2
 + \nu \bigl(-5 - 4 \overline{\gamma}\bigr)
 + 2 \overline{\gamma}\Bigr) v^{2}\biggr)
 + \bigl(\frac{3}{8}
 -  \frac{21}{8} \nu
 + \frac{39}{8} \nu^2\bigr) v^{4} \,,\\ 
%%%%%%%%%%%%%%%%%%%%%%%%%%%%%%%%%%%%%%%%%%%%%%%%%%%%%%%%%%%%%%%%%
\mathcal{J}_{3\mathrm{PN}}={}&\tilde{G}^3 \Biggl( \frac{\alpha^2 m^3}{r^3} \biggl(\nu \Bigl(- \frac{11}{12} \overline{\gamma} \bigl(2
 + \overline{\gamma}\bigr)
 + \delta_{+} \bigl(10
 + \overline{\gamma}\bigr) \frac{1}{6 + 3 \overline{\gamma}}\Bigr)
 + \delta_{-} \nu \bigl(10
 + \overline{\gamma}\bigr) \frac{1}{6 + 3 \overline{\gamma}} m_{-}\biggr) \nonumber\\
&\  + \frac{\alpha^3 m^3}{r^3} \biggl[\frac{5}{2}
 + \frac{2}{3} \chi_{+}
 + \nu^3
 + \delta_{+} \bigl(-2
 -  \frac{4}{3} \overline{\gamma}\bigr)
 + \frac{17}{3} \overline{\gamma}
 + \beta_{+} \bigl(3
 + 2 \overline{\gamma}\bigr)
 + \nu^2 \bigl(-7
 + 2 \beta_{+} + \frac{4}{3} \chi_{+}
 -  \frac{2}{3} \delta_{+} \nonumber\\
&\quad + 8  \frac{\beta_{-}^2}{\overline{\gamma}} - 8 \frac{\beta_{+}^2 }{\overline{\gamma}}
 -  \frac{14}{3} \overline{\gamma}
 -  \frac{1}{6} \overline{\gamma}^2\bigr)
 + \frac{25}{6} \overline{\gamma}^2
 + \overline{\gamma}^3
 + \nu \biggl(- \frac{2}{3} \chi_{+} - 16  \frac{\beta_{-}^2}{\overline{\gamma}}
 + 16  \frac{\beta_{+}^2}{\overline{\gamma}}
 - 16  \frac{\beta_{-} \delta_{-}}{\overline{\gamma}} \nonumber\\
&\quad + \delta_{+} \Bigl(\frac{122}{9}
 + \frac{7}{3} \overline{\gamma}
 -  \frac{7}{32} \pi^2 \bigl(2
 + \overline{\gamma}\bigr)\Bigr) + \beta_{+} \Bigl(-16  \frac{\delta_{+}}{\overline{\gamma}}
 + 4 \bigl(5
 + 4 \overline{\gamma}\bigr)\Bigr)
 + \frac{1}{128} \pi^2 \bigl(164
 + 150 \overline{\gamma}
 + 20 \overline{\gamma}^2
 - 7 \overline{\gamma}^3\bigr) \nonumber\\
&\quad + \frac{1}{72} \bigl(-657 - 1156 \overline{\gamma}
 - 356 \overline{\gamma}^2
 + 66 \overline{\gamma}^3\bigr)\biggr)
 + \biggl(- \frac{2}{3} \chi_{-}
 + 2 \beta_{-} \nu^2
 + \beta_{-} \bigl(-3
 - 2 \overline{\gamma}\bigr)
 + \delta_{-} \bigl(-2 -  \frac{4}{3} \overline{\gamma}\bigr) \nonumber\\
&\quad + \nu \Bigl(- \frac{2}{3} \chi_{-}
 + \frac{1}{9} \delta_{-} \bigl(-14
 - 3 \overline{\gamma}\bigr)
 - 6 \beta_{-} \bigl(3
 + 2 \overline{\gamma}\bigr)\Bigr)\biggr) m_{-}
 + \biggl(2 \delta_{-} \nu
 + \nu \Bigl(2 \delta_{+} -  \frac{11}{2} \bigl(2
 + \overline{\gamma}\bigr)^2\Bigr) m_{-}\biggr) \ln\bigl(\frac{r}{r'_{-}}\bigr) \nonumber\\
&\quad + \biggl(\nu \Bigl(2 \delta_{+}
 -  \frac{11}{2} \bigl(2
 + \overline{\gamma}\bigr)^2\Bigr)
 + 2 \delta_{-} \nu m_{-}\biggr) \ln\bigl(\frac{r}{r'_{+}}\bigr)\biggr]\Biggr) \nonumber\\
&\ + \frac{\alpha^2 \tilde{G}^2 m^2}{r^2} \biggl[\dot{r}^2 \biggl(\frac{1}{2} \delta_{+}
 -  \frac{27}{2} \nu^3
 + \nu^2 \bigl(- \frac{317}{8}
 + 13 \beta_{+}
 - 30 \overline{\gamma}\bigr)
 + \nu \Bigl(- \frac{7}{6} \delta_{+}
 + \frac{1}{24} \bigl(-287 - 624 \overline{\gamma}
 - 247 \overline{\gamma}^2\bigr)\Bigr) \nonumber\\
&\quad + \frac{1}{8} \bigl(2
 + \overline{\gamma}\bigr)^2
 + \Bigl(\frac{1}{2} \delta_{-}
 + \bigl(4 \beta_{-}
 + \frac{1}{2} \delta_{-}\bigr) \nu
 - 5 \beta_{-} \nu^2\Bigr) m_{-}\biggr) + \biggl(\frac{1}{2} \beta_{+}
 -  \delta_{+}
 - 9 \nu^3 + \nu^2 \bigl(\frac{105}{4}
 -  \frac{3}{2} \beta_{+}
 + 13 \overline{\gamma}\bigr) \nonumber\\
&\quad + \nu \Bigl(- \frac{7}{2} \beta_{+}
 + \frac{4}{3} \delta_{+}
 + \frac{1}{6} \bigl(-161 - 141 \overline{\gamma}
 - 34 \overline{\gamma}^2\bigr)\Bigr)
 + \frac{1}{4} \bigl(45
 + 52 \overline{\gamma}
 + 15 \overline{\gamma}^2\bigr)
 + \Bigl(- \frac{1}{2} \beta_{-}
 -  \delta_{-} \nonumber\\
&\quad + \bigl(\frac{5}{2} \beta_{-}
 + 2 \delta_{-}\bigr) \nu -  \frac{9}{2} \beta_{-} \nu^2\Bigr) m_{-}\biggr) v^{2}\biggr] \nonumber\\
&\  + \frac{\alpha \tilde{G} m}{r} \biggl(\dot{r}^4 \Bigl(- \frac{33}{8} \nu^3
 + \frac{3}{4} \nu \bigl(1
 + \overline{\gamma}\bigr)
 -  \frac{3}{4} \nu^2 \bigl(1
 + 3 \overline{\gamma}\bigr)\Bigr) + \dot{r}^2 \Bigl(\frac{75}{4} \nu^3
 + \nu \bigl(-3
 -  \frac{5}{2} \overline{\gamma}\bigr) + \nu^2 \bigl(\frac{7}{4}
 + 7 \overline{\gamma}\bigr)\Bigr) v^{2} \nonumber\\
&\quad + \Bigl(\frac{195}{8} \nu^3
 + \frac{1}{4} \nu \bigl(-71
 - 45 \overline{\gamma}\bigr) + \frac{3}{8} \bigl(11
 + 6 \overline{\gamma}\bigr)
 + \frac{1}{4} \nu^2 \bigl(53
 + 65 \overline{\gamma}\bigr)\Bigr) v^{4}\biggr) + \bigl(\frac{5}{16}
 -  \frac{59}{16} \nu
 + \frac{119}{8} \nu^2
 -  \frac{323}{16} \nu^3\bigr) v^{6} \,.
\end{align}\end{subequations}
%

%%%%%%%%%%%%%%%%%%%%%%%%%%%%%%%%%%%%%%%%%%%%%%%%%%%%%%%
\subsection{Non-local contribution}\label{subsec:nonlocal}
%%%%%%%%%%%%%%%%%%%%%%%%%%%%%%%%%%%%%%%%%%%%%%%%%%%%%%%

Due to the non-locality of the tail terms, one has to be careful when deriving their contribution to the energy and angular momentum. Following the same procedure as for the instantaneous terms, but with the tail part of the acceleration, $\mathbf{a}_{A\,\mathrm{tail}}$, one obtains
\be\label{dEdt-E0H1}
\sum_{A}m_{A}\mathbf{v}_{A}\cdot\mathbf{a}_{A\,\mathrm{tail}} = -\frac{\mathrm{d}E^{\mathrm{tail}}_{0}}{\mathrm{d}t} + H_{1}^{\mathrm{tail}}\,,
\ee
where
\be\label{H1tail}
H^{\mathrm{tail}}_{1} = \frac{2G^{2}M}{3c^{6}}(3+2\omega_{0})\left[I_{s}^{i\,(3)}\mathcal{T}_{s}^{i\,(2)}-I_{s}^{i\,(2)}\mathcal{T}_{s}^{i\,(3)}\right]\,,
\ee
The quantity $E^{\mathrm{tail}}_{0}$ consists in the first part of the tail contribution to the energy. It reads
\be\label{E0tail}
E^{\mathrm{tail}}_{0} = -\frac{4G^{2}M}{3c^{6}}(3+2\omega_{0})\left[I_{s}^{i\,(1)}\mathcal{T}_{s}^{i\,(3)}-\frac{1}{2}I_{s}^{i\,(2)}\mathcal{T}_{s}^{i\,(2)}-2(\ln r_{12})^{(1)}I_{s}^{i\,(1)}I_{s}^{i\,(2)}\right]\,,
\ee
where $\mathcal{T}_{s}^{i\,(n)}\equiv\int_{0}^{+\infty}\mathrm{d}\tau\ln\left(\frac{c\tau}{2r_{12}}\right)\left[I_{s}^{i\,(n+1)}(t-\tau)-I_{s}^{i\,(n+1)}(t+\tau)\right]$. The local term in Eq.~\eqref{E0tail} comes from the derivation of the regularisation scale $r_{12}$. Due to the presence of the second term $H_{1}^{\mathrm{tail}}$ in Eq.~\eqref{dEdt-E0H1}, finding the complete tail contribution to the energy is more complicated. A similar problem arises at 4PN in general relativity and it has been solved in~\cite{Bernard:2016wrg} by resorting to Fourier series. In the following we will follow the same procedure in order to determine the additional contribution to the energy. We start by modifying the kernel function in the expression for $\mathcal{T}_{s}^{i\,(n)}$,
\be
\mathcal{T}_{s}^{i\,(n)}=\int_{0}^{+\infty}\mathrm{d}\tau\ln\left(\frac{c\tau}{2r_{12}}\right)\left[I_{s}^{i\,(n+1)}(t-\tau)-I_{s}^{i\,(n+1)}(t+\tau)\right]\mathrm{e}^{-\epsilon\vert\tau\vert}\,,
\ee
where $\mathrm{e}^{-\epsilon\vert\tau\vert}$, with $\epsilon>0$, is a cut-off function. At the end of the calculation, we will let $\epsilon$ tends to zero to get a finite result. Thanks to this regulator, we can perform a Taylor expansion of $\mathcal{T}_{s}^{i\,(n)}$ where all the individual integrals are convergent, and get
\be
\mathcal{T}_{s}^{i\,(n)}=\sum_{p=1}^{+\infty}I_{s}^{i\,(2p+n)}(t)\int_{-\infty}^{+\infty}\frac{\mathrm{d}\tau}{\vert\tau\vert}\frac{\tau^{2p}}{(2p)!}\mathrm{e}^{-\epsilon\vert\tau\vert}\,.
\ee
Inserting it into the expression for $H_{1}^{\mathrm{tail}}$, we rewrite the later as a total time derivative $H_{1}^{\mathrm{tail}}=-\frac{\mathrm{d}E_{1}^{\mathrm{tail}}}{\mathrm{d}t}$, with
\be\label{E1tailseries}
E^{\mathrm{tail}}_{1} = \frac{2G^{2}M}{3c^{6}}(3+2\omega_{0})\sum_{n=1}^{+\infty}\left[I_{s}^{i\,(2)}I_{s}^{i\,(2n+2)}-2\sum_{p=0}^{n-2}(-1)^{p}I_{s}^{i\,(3+p)}I_{s}^{i\,(2n+1-p)}+(-1)^{n}I_{s}^{i\,(n+2)}I_{s}^{i\,(n+2)}\right]\int_{-\infty}^{+\infty}\frac{\mathrm{d}\tau}{\vert\tau\vert}\frac{\tau^{2n}}{(2n)!}\mathrm{e}^{-\epsilon\vert\tau\vert}\,.
\ee
However this expression is still in the form of an infinite Taylor expansion. In order to obtain a non-perturbative expression, we decompose the scalar dipole moment $I_{s}^{i}$ in Fourier series,
\be
I_{s}^{i}(t)=\sum_{p=-\infty}^{+\infty}\mathop{\mathcal{I}}_{p}{}_{\!i}\,\mathrm{e}^{ip\ell}
% \quad\Longleftrightarrow\quad \mathop{\mathcal{I}}_{p}{}_{\!i}=\int_{0}^{2\pi}\frac{\mathrm{d}\ell}{2\pi}I_{s}^{i}\mathrm{e}^{-ip\ell}
\,,
\ee
where $\ell=n(t-t_{0})$ is the mean anomaly and $n$ is the radial frequency. Such an expansion is possible because we only need $I_{s}^{i}$ at Newtonian order where there is a unique frequency $\omega=n$. Inserting it into Eq.~\eqref{E1tailseries}, we get
\be\label{E1tail}
E^{\mathrm{tail}}_{1} = -\frac{4G^{2}M\omega^{4}}{3c^{6}}(3+2\omega_{0})\left[\sum_{p=-\infty}^{+\infty}\left\vert\mathop{\mathcal{I}}_{p}{}_{\!i}\right\vert^{2}p^{4}+\frac{1}{2}\sum_{p+q\neq 0}\mathop{\mathcal{I}}_{p}{}_{\!i}\mathop{\mathcal{I}}_{q}{}_{\!i}\,p^{2}q^{2}\frac{p-q}{p+q}\ln\left\vert\frac{p}{q}\right\vert\mathrm{e}^{i(p+q)\ell}\right]\,.
\ee
The second term in this equation is a AC contribution, oscillating with time with zero average value, and it will vanish for circular orbits. The first term in the right hand side of Eq.~\eqref{E1tail} is a constant DC contribution, and turns out to be crucial in the computation of the total energy, as it does not vanish for circular orbits. Note that the DC term also admits a closed-form formulation,
\be\label{E1tailDC}
E^{\mathrm{tail}}_{1,\mathrm{DC}} = -\frac{4G^{2}M}{3c^{6}}(3+2\omega_{0})\left\langle I_{s}^{i\,(2)}I_{s}^{i\,(2)}\right\rangle\,,
\ee
where the brackets stands for averaging over one orbital period. Finally, the total energy is given by
\be\label{Etotal}
E = E^{\mathrm{inst}}_{0}+E^{\mathrm{tail}}_{0}+E^{\mathrm{tail}}_{1}\,,
\ee
and is conserved
\be
\frac{\mathrm{d}E}{\mathrm{d}t} = 0\,.
\ee

With a similar reasonning we compute the tail contribution to the integral of angular momentum. First we write
\be\label{dJdt-J0K1}
\sum_{A}m_{A}\epsilon_{ijk}\mathbf{y}_{A}^{j}\cdot\mathbf{a}^{k}_{A\,\mathrm{tail}} = -\frac{\mathrm{d}J^{i}_{0,\,\mathrm{tail}}}{\mathrm{d}t} + K_{1,\,\mathrm{tail}}^{i}\,,
\ee
with
\be\label{K1tail}
K^{i}_{1,\,\mathrm{tail}}\equiv \frac{4G^{2}M}{3c^{6}}(3+2\omega_{0})\epsilon_{ijk}I_{s}^{j\,(2)}\mathcal{T}_{s}^{k\,(2)}\,,
\ee
and
\be\label{J0tail}
J^{i}_{0,\,\mathrm{tail}} = -\frac{4G^{2}M}{3c^{6}}(3+2\omega_{0})\epsilon_{ijk}\left[I_{s}^{j}\mathcal{T}_{s}^{k\,(3)}-I_{s}^{j\,(1)}\mathcal{T}_{s}^{k\,(2)}-2(\ln r_{12})^{(1)}I_{s}^{j}I_{s}^{k\,(2)}\right]\,.
\ee
Using the decomposition in Fourier series of the scalar dipole moment, we rewrite $K^{i}_{1,\,\mathrm{tail}}$ as a total time derivative,
\be
K^{i}_{1,\,\mathrm{tail}}=-\frac{\mathrm{d}J^{i}_{1,\,\mathrm{tail}}}{\mathrm{d}t}\,,
\ee
with
\be\label{J1tail}
J^{i}_{1,\,\mathrm{tail}}=\frac{4G^{2}M\omega^{3}}{3c^{6}}(3+2\omega_{0})\epsilon_{ijk}\left[\sum_{p=-\infty}^{+\infty}i\mathop{\mathcal{I}}_{p}{}_{\!j}\mathop{\mathcal{I}}_{-p}{}_{\!k}p^{3}+\sum_{p+q\neq 0}i\mathop{\mathcal{I}}_{p}{}_{\!j}\mathop{\mathcal{I}}_{q}{}_{\!k}\,\frac{p^{2}q^{2}}{p+q}\ln\left\vert\frac{p}{q}\right\vert\mathrm{e}^{i(p+q)\ell}\right]\,.
\ee
Once again, the first term in this equation is a constant DC contribution that does not vanishes for circular orbit. It admits the closed-form expansion
\be\label{J1tailDC}
J^{i}_{1,\,\mathrm{tail},\,\mathrm{DC}} = \frac{4G^{2}M}{3c^{6}}(3+2\omega_{0})\left\langle \epsilon_{ijk}\,I_{s}^{j\,(2)}I_{s}^{k\,(1)}\right\rangle\,.
\ee
The second term in Eq.~\eqref{J1tail} is a time-varying AC contribution and is zero for circular orbit. Finally, the total angular momentum is
\be
J^{i}=J^{i}_{0,\,\mathrm{inst}}+J^{i}_{0,\,\mathrm{tail}}+J^{i}_{1,\,\mathrm{tail}}\,,
\ee
and verifies
\be
\frac{\mathrm{d}J^{i}}{\mathrm{d}t}=0\,.
\ee
%

%%%%%%%%%%%%%%%%%%%%%%%%%%%%%%%%%%%%%%%%%%%%%%%%%%%%%%%
\section{Reduction to circular orbits}\label{sec:CircOrbit}
%%%%%%%%%%%%%%%%%%%%%%%%%%%%%%%%%%%%%%%%%%%%%%%%%%%%%%%

%%%%%%%%%%%%%%%%%%%%%%%%%%%%%%%%%%%%%%%%%%%%%%%%%%%%%%%
\subsection{Conserved energy for circular orbits}
%%%%%%%%%%%%%%%%%%%%%%%%%%%%%%%%%%%%%%%%%%%%%%%%%%%%%%%

From the equations of motion in the center-of-mass frame, we then reduce the dynamics to the case of circular orbits. We have $\dot{r}=(nv)=0$, and the acceleration is purely radial,
\be
\mathbf{a}=-\omega^{2}\mathbf{x}\,,
\ee
where $\omega$ is the orbital frequency, that can be expressed as a function of the radial parameter $r$ only. Again we split our calculation between the instantaneous and tail parts. For the local part of the dynamics, the orbital frequency is obtained by putting $\dot{r}=0$ and $v^{2}=r^{2}\omega^{2}$ in the CM dynamics. We get,
\begin{align}\label{omegaInst}
\omega^{2}_\mathrm{3PN,\,inst}={}&\frac{\alpha \tilde{G} m}{r^3} \Biggl\{1
 + \bigl(-3
 - 2 \beta_{+}
 + \nu
 -  \overline{\gamma}
 + 2 \beta_{-} m_{-}\bigr) \gamma
 + \biggl(6
 + 8 \beta_{+}
 - 2 \chi_{+}
 + \delta_{+}
 + \nu^2
 + \bigl(5 + 2 \beta_{+}\bigr) \overline{\gamma}
 + \frac{5}{4} \overline{\gamma}^2 \nonumber\\
&\quad + \Bigl(2 \chi_{-}
 + \delta_{-}
 - 2 \beta_{-} \bigl(4
 + \overline{\gamma}\bigr)\Bigr) m_{-}
 + \nu \bigl(\frac{41}{4}
 + \beta_{+}
 + 4 \chi_{+}
 - 2 \delta_{+} + 24 \beta_{-}^2 \frac{1}{\overline{\gamma}}
 - 24 \beta_{+}^2 \frac{1}{\overline{\gamma}}
 + 5 \overline{\gamma}
 -  \frac{1}{2} \overline{\gamma}^2 \nonumber\\
&\quad + 5 \beta_{-} m_{-}\bigr)\biggr) \gamma^2
 + \Biggl[\nu^3
 + \nu^2 \Bigl(14 \beta_{+}
 - 64 \beta_{+}^2 \frac{1}{\overline{\gamma}} + 8 \beta_{-}^2 \frac{1}{\overline{\gamma}} \bigl(8
 + 3 \overline{\gamma}\bigr)
 + \frac{1}{6} \bigl(57
 + 64 \chi_{+}
 - 32 \delta_{+}
 - 2 \overline{\gamma}
 - 8 \overline{\gamma}^2\bigr) \nonumber\\
&\quad + 6 \beta_{-} m_{-}\Bigr) -  \frac{1}{8} \nu \frac{1}{\alpha} \frac{1}{2 + \overline{\gamma}} \Bigl(\delta_{+} \bigl(40
 - 12 \overline{\gamma}\bigr)
 + 132 \overline{\gamma}^2
 + 33 \overline{\gamma}^3
 + 40 \delta_{-} m_{-}
 - 12 \overline{\gamma} \bigl(-11
 + \delta_{-} m_{-}\bigr)\Bigr) \nonumber\\
&\quad + \frac{1}{12} \biggl(-4 \Bigl(30
 + 12 \beta_{+}^2
 - 24 \chi_{+}
 - 8 \beta_{-} \delta_{-}
 + 13 \delta_{+}
 + \beta_{+} \bigl(50
 + 8 \delta_{+}\bigr)
 + 4 \kappa_{+}\Bigr)
 - 4 \bigl(39 + 32 \beta_{+}
 - 6 \chi_{+} \nonumber\\
&\quad + 5 \delta_{+}\bigr) \overline{\gamma}
 -  \bigl(69
 + 32 \beta_{+}\bigr) \overline{\gamma}^2
 - 9 \overline{\gamma}^3
 + 4 \Bigl(-24 \chi_{-}
 - 13 \delta_{-}
 - 8 \beta_{+} \delta_{-} + \beta_{-} \bigl(50
 + 24 \beta_{+}
 + 8 \delta_{+}\bigr)
 + 4 \kappa_{-}\Bigr) m_{-} \nonumber\\
&\quad + 4 \bigl(32 \beta_{-}
 - 6 \chi_{-}
 - 5 \delta_{-}\bigr) \overline{\gamma} m_{-}
 + 32 \beta_{-} \overline{\gamma}^2 m_{-} - 48 \beta_{-}^2 m_{-}^2\biggr)
 + \nu \Biggl(- \frac{1}{256} \pi^2 \bigl(2
 + \overline{\gamma}\bigr) \bigl(-82
 + 28 \delta_{+}
 - 34 \overline{\gamma}
 + 7 \overline{\gamma}^2\bigr) \nonumber\\
&\quad + \frac{1}{72\overline{\gamma}^{2}} \biggl[-4608 \beta_{+}^3
 + 288 \beta_{+}^2 \overline{\gamma} \bigl(33
 + 5 \overline{\gamma}\bigr)
 - 4608 \beta_{-}^3 m_{-}
 + 12 \beta_{+} \overline{\gamma} \Bigl(-192 \chi_{+}
 - 421 \overline{\gamma} + 16 \delta_{+} \bigl(1
 + 2 \overline{\gamma}\bigr)
 - 172 \overline{\gamma}^2 \nonumber\\
&\quad + 8 \overline{\gamma}^3
 + 192 \chi_{-} m_{-}
 + 64 \delta_{-} m_{-}\Bigr) + \overline{\gamma}^2 \Bigl(-7509
 + 288 \kappa_{+} - 6106 \overline{\gamma}
 - 48 \chi_{+} \bigl(23
 + 6 \overline{\gamma}\bigr)
 + 4 \delta_{+} \bigl(203
 + 51 \overline{\gamma}\bigr) \nonumber\\
&\quad - 889 \overline{\gamma}^2 + 135 \overline{\gamma}^3
 - 48 \chi_{-} m_{-} + 204 \delta_{-} m_{-}
 - 96 \kappa_{-} m_{-}
 - 36 \delta_{-} \overline{\gamma} m_{-}\Bigr)
 + 48 \beta_{-} \biggl(48 \chi_{-} \overline{\gamma}
 + 4 \delta_{-} \bigl(1
 - 4 \overline{\gamma}\bigr) \overline{\gamma} \nonumber\\
&\quad + 96 \beta_{+}^2 m_{-}
 + \Bigl(-48 \chi_{+}
 - 8 \delta_{+} \bigl(-2
 + \overline{\gamma}\bigr)
 + \overline{\gamma} \bigl(67
 + 37 \overline{\gamma}
 - 2 \overline{\gamma}^2\bigr)\Bigr) \overline{\gamma} m_{-}\biggr) + 288 \beta_{-}^2 \biggl(16 \beta_{+}
 + \overline{\gamma} \Bigl(-33 \nonumber\\
&\quad + \overline{\gamma} \bigl(-8
 + 3 m_{-}^2\bigr)\Bigr)\biggr)\biggr]
 + \biggl(-3 \delta_{-}
 -  \frac{3}{4} \Bigl(4 \delta_{+}
 - 11 \bigl(2 + \overline{\gamma}\bigr)^2\Bigr) m_{-}\biggr) \ln\bigl(\frac{r}{r'_{-}}\bigr) \nonumber\\
&\quad + \bigl(33
 - 3 \delta_{+}
 + 33 \overline{\gamma}
 + \frac{33}{4} \overline{\gamma}^2
 - 3 \delta_{-} m_{-}\bigr) \ln\bigl(\frac{r}{r'_{+}}\bigr)\Biggr)\Biggr] \gamma^3\Biggr\} \,,
\end{align}
where we have defined the post-Newtonian parameter $\gamma\equiv\frac{\tilde{G}\alpha m}{c^{2}r}$. The tail contribution to the orbital frequency is obtained directly from Eq.~\eqref{aitail}. We have $a_{\mathrm{tail}}^{i}=-\omega_{\mathrm{tail}}^{2}x^{i}$, with
\be\label{omegaTail}
\omega_{\mathrm{tail}}^{2}=\frac{4\tilde{G}\alpha m}{3r^{3}}\gamma^{3}\nu\left(2\overline{\delta}_{+}+\frac{\overline{\gamma}(2+\overline{\gamma})}{2}\right)\left[\ln\left(4\gamma\right)+2\gamma_{\mathrm{E}}+1\right]\,,
\ee
where $\gamma_{\mathrm{E}}$ is the Euler constant. Note that these two expressions are not yet gauge invariant, as they depend on the separation $r$ which is a harmonic coordinate quantity. As a consequence they still depend logarithmically on the regularisation constants $r_{+}$ and $r_{-}$. Inverting Eqs.~\eqref{omegaInst}-\eqref{omegaTail}, we get $\gamma$ as a function of $\omega$, see App.~\ref{sec:App2}. Next we use these expressions to reduce the energy to circular orbits. There are several contributions coming from the instantaneous terms in the center-of-mass frame presented in Eq.~\eqref{Einst}, together with the tail terms given by Eqs.~\eqref{E0tail} and~\eqref{E1tail}. Note that the contribution coming form $E_{1}^{\mathrm{tail}}$ is crucial in the process of obtaining a complete result. In a first stage, we obtain the energy as a function of the separation $r$, or rather $\gamma=\frac{\tilde{G}\alpha m}{c^{2}r}$. Then, using Eqs.~\eqref{gammaInst}-\eqref{gammaTail}, we rewrite the energy in a  gauge-invariant way, using the PN parameter $x\equiv\left(\frac{\tilde{G}\alpha m \omega}{c^{3}}\right)^{2/3}$. Finally, we get
\begin{align}\label{Ecirc3PN}
E_\mathrm{3PN}={}& -\frac{1}{2}m\nu c^{2}x\,\Biggl\{1
 + \Bigl(- \frac{1}{12} \nu
 + \frac{1}{12} \bigl(-9
 + 8 \beta_{+}
 - 8 \overline{\gamma}
 - 8 \beta_{-} m_{-}\bigr)\Bigr) x
 + \biggl(- \frac{1}{24} \nu^2
 + \frac{1}{24} \nu \frac{1}{\overline{\gamma}} \Bigl(-384 \beta_{-}^2 + 384 \beta_{+}^2 \nonumber\\
&\quad - 152 \beta_{+} \overline{\gamma} + \overline{\gamma} \bigl(57
 - 64 \chi_{+}
 + 32 \delta_{+}
 + 88 \overline{\gamma}
 + 8 \overline{\gamma}^2\bigr)
 + 8 \beta_{-} \overline{\gamma} m_{-}\Bigr) + \frac{1}{24} \Bigl(-81
 + 32 \beta_{+}^2
 + 32 \chi_{+}
 + 8 \delta_{+}
 - 112 \overline{\gamma} \nonumber\\
&\quad - 38 \overline{\gamma}^2
 - 24 \beta_{-} m_{-}
 - 32 \chi_{-} m_{-} + 8 \delta_{-} m_{-}
 + 8 \beta_{+} \bigl(3
 + 4 \overline{\gamma}
 - 8 \beta_{-} m_{-}\bigr)
 - 32 \beta_{-} \overline{\gamma} m_{-}
 + 32 \beta_{-}^2 m_{-}^2\Bigr)\biggr) x^2 \nonumber\\
&\quad + \Biggl(- \frac{35}{5184} \nu^3 + \frac{5}{12} \nu \frac{\overline{\gamma}}{\alpha} \frac{1}{2 + \overline{\gamma}} \bigl(44
 - 4 \delta_{+}
 + 44 \overline{\gamma}
 + 11 \overline{\gamma}^2
 - 4 \delta_{-} m_{-}\bigr)
 + \frac{1}{864} \nu^2 \frac{1}{\overline{\gamma}} \biggl(-2304 \beta_{-}^2 \bigl(-5 + 12 \overline{\gamma}\bigr) \nonumber\\
&\quad - 5 \Bigl(2304 \beta_{+}^2 - 856 \beta_{+} \overline{\gamma}
 + \overline{\gamma} \bigl(279
 - 384 \chi_{+}
 + 192 \delta_{+}
 + 472 \overline{\gamma}
 + 48 \overline{\gamma}^2\bigr)\Bigr) + 40 \beta_{-} \overline{\gamma} m_{-}\biggr)
 + \nu \Biggl[\frac{5}{384} \pi^2 \bigl(2
 + \overline{\gamma}\bigr) \bigl(-82 \nonumber\\
&\quad + 28 \delta_{+} - 34 \overline{\gamma}
 + 7 \overline{\gamma}^2\bigr) + \frac{1}{1728} \frac{1}{\overline{\gamma}^{2}} \Biggl(92160 \beta_{-}^3 \bigl(1
 + \overline{\gamma}\bigr) m_{-}
 - 80 \beta_{-} \biggl[576 \chi_{-} \overline{\gamma}
 - 192 \delta_{-} \overline{\gamma} \bigl(2
 + \overline{\gamma}\bigr) + \biggl(1152 \beta_{+}^2 \bigl(1
 + \overline{\gamma}\bigr) \nonumber\\
&\quad + \overline{\gamma} \Bigl(192 \delta_{+}
 + \bigl(21
 - 160 \overline{\gamma}\bigr) \overline{\gamma}
 - 192 \chi_{+} \bigl(3
 + \overline{\gamma}\bigr)\Bigr) - 56 \beta_{+} \overline{\gamma}^2\biggr) m_{-}\biggr]
 + 5 \biggl(18432 \beta_{+}^3 \bigl(1
 + \overline{\gamma}\bigr) + 7616 \beta_{+}^2 \overline{\gamma}^2 \nonumber\\
&\quad - 16 \beta_{+} \overline{\gamma} \Bigl(-384 \delta_{+} + 192 \chi_{+} \bigl(-3
 + \overline{\gamma}\bigr)
 + 249 \overline{\gamma}
 + 448 \overline{\gamma}^2
 + 576 \chi_{-} m_{-}
 + 192 \delta_{-} m_{-}\Bigr) + \overline{\gamma}^2 \Bigl(20667 - 1152 \kappa_{+} \nonumber\\
&\quad + 20368 \overline{\gamma}
 - 384 \chi_{+} \bigl(13
 + 4 \overline{\gamma}\bigr)
 - 32 \delta_{+} \bigl(151
 + 15 \overline{\gamma}\bigr)
 + 4488 \overline{\gamma}^2
 - 600 \overline{\gamma}^3 + 2688 \chi_{-} m_{-}
 + 96 \delta_{-} m_{-} + 384 \kappa_{-} m_{-} \nonumber\\
&\quad + 288 \delta_{-} \overline{\gamma} m_{-}\Bigr)\biggr)
 - 64 \beta_{-}^2 \Bigl(1440 \beta_{+} \bigl(1
 + \overline{\gamma}\bigr) + \overline{\gamma}^2 \bigl(351
 + 206 m_{-}^2\bigr)\Bigr)\Biggr)\Biggr]
 + \frac{1}{5184} \biggl[5 \Bigl(-10935
 + 3584 \beta_{+}^3
 + 2016 \delta_{+} \nonumber\\
&\quad + 1152 \kappa_{+} - 19440 \overline{\gamma}
 + 1152 \delta_{+} \overline{\gamma}
 + 1152 \chi_{+} \bigl(3
 + 2 \overline{\gamma}\bigr)
 + 192 \beta_{+}^2 \bigl(27
 + 16 \overline{\gamma}\bigr)
 - 11304 \overline{\gamma}^2 - 2144 \overline{\gamma}^3
 - 3456 \chi_{-} m_{-} \nonumber\\
&\quad + 2016 \delta_{-} m_{-} - 1152 \kappa_{-} m_{-} + 48 \beta_{+} \bigl(-33
 + 96 \chi_{+}
 + 8 \overline{\gamma}^2 - 96 \chi_{-} m_{-}\bigr)
 - 2304 \chi_{-} \overline{\gamma} m_{-}
 + 1152 \delta_{-} \overline{\gamma} m_{-}\Bigr) \nonumber\\
&\quad + 240 \beta_{-} \biggl(m_{-} \Bigl(33
 - 224 \beta_{+}^2
 - 96 \chi_{+} - 8 \beta_{+} \bigl(27
 + 16 \overline{\gamma}\bigr)
 - 8 \overline{\gamma}^2
 + 96 \chi_{-} m_{-}\Bigr)
 + 48 \delta_{-} \bigl(-1
 + m_{-}^2\bigr)\biggr) \nonumber\\
&\quad + 192 \beta_{-}^2 \Bigl(-27 + 2 \bigl(81
 + 140 \beta_{+}
 + 40 \overline{\gamma}\bigr) m_{-}^2\Bigr)
 - 17920 \beta_{-}^3 m_{-}^3\biggr]\Biggr) x^3\Biggr\}\,,
\end{align}
together with the tail part
\be
E_{3\mathrm{PN}}^{\mathrm{tail}} = -\frac{1}{2}m\nu c^{2}x\,\cdot\,\frac{20}{9}\nu\left(2\overline{\delta}_{+}+\frac{\overline{\gamma}(2+\overline{\gamma})}{2}\right)\left[\ln\left(4x\right)+2\gamma_{\mathrm{E}}+\frac{2}{5}\right]\,x^{3}\,.
\ee
As expected for a gauge invariant quantity, the gauge dependent regularisation constants $r'_{-}$ and $r'_{+}$ are absent from this final result. We have also checked that the 2PN part of Eq.~\eqref{Ecirc3PN} is in full agreement with the result of~\cite{Sennett:2016klh}.

%%%%%%%%%%%%%%%%%%%%%%%%%%%%%%%%%%%%%%%%%%%%%%%%%%%%%%%
\subsection{Conserved angular momentum for circular orbits}
%%%%%%%%%%%%%%%%%%%%%%%%%%%%%%%%%%%%%%%%%%%%%%%%%%%%%%%

Similarly, one can reduce the angular momentum to circular orbits. One gets, for the instantaneous contribution,
\begin{align}
J_\mathrm{3PN}={}& \frac{\tilde{G}\alpha m^{2}\nu}{c\sqrt{x}}\Biggr\{ 1
 + \Bigl(\frac{1}{6} \nu
 + \frac{1}{6} \bigl(9
 - 8 \beta_{+}
 + 8 \overline{\gamma}
 + 8 \beta_{-} m_{-}\bigr)\Bigr) x
 + \biggl(\frac{1}{24} \nu^2
 -  \frac{1}{24} \nu \frac{1}{\overline{\gamma}} \Bigl(-384 \beta_{-}^2
 + 384 \beta_{+}^2 - 152 \beta_{+} \overline{\gamma} \nonumber\\
&\quad + \overline{\gamma} \bigl(57
 - 64 \chi_{+}
 + 32 \delta_{+}
 + 88 \overline{\gamma}
 + 8 \overline{\gamma}^2\bigr)
 + 8 \beta_{-} \overline{\gamma} m_{-}\Bigr)
 + \frac{1}{24} \Bigl(81
 - 32 \beta_{+}^2 - 32 \chi_{+}
 - 8 \delta_{+}
 + 112 \overline{\gamma}
 + 38 \overline{\gamma}^2 \nonumber\\
&\quad + 24 \beta_{-} m_{-}
 + 32 \chi_{-} m_{-}
 - 8 \delta_{-} m_{-}
 - 8 \beta_{+} \bigl(3
 + 4 \overline{\gamma} - 8 \beta_{-} m_{-}\bigr)
 + 32 \beta_{-} \overline{\gamma} m_{-}
 - 32 \beta_{-}^2 m_{-}^2\Bigr)\biggr) x^2 \nonumber\\
&\quad + \Biggl(\frac{135}{16}
 -  \frac{224}{81} \beta_{+}^3
 + \frac{16}{9} \beta_{-} \delta_{-}
 -  \frac{14}{9} \delta_{+} -  \frac{8}{9} \kappa_{+}
 + \frac{7}{1296} \nu^3
 + 15 \overline{\gamma}
 -  \frac{8}{9} \delta_{+} \overline{\gamma}
 + \frac{157}{18} \overline{\gamma}^2
 + \frac{134}{81} \overline{\gamma}^3 -  \frac{11}{9} \beta_{-} m_{-} \nonumber\\
&\quad + \frac{8}{3} \chi_{-} m_{-} -  \frac{14}{9} \delta_{-} m_{-}
 + \frac{8}{9} \kappa_{-} m_{-}
 -  \frac{4}{27} \beta_{+}^2 \bigl(27
 + 16 \overline{\gamma}
 - 56 \beta_{-} m_{-}\bigr)
 -  \frac{8}{9} \chi_{+} \bigl(3
 + 2 \overline{\gamma}
 - 4 \beta_{-} m_{-}\bigr) \nonumber\\
&\quad -  \frac{1}{3} \nu \frac{\overline{\gamma}}{\alpha} \frac{1}{2 + \overline{\gamma}} \bigl(44
 - 4 \delta_{+}
 + 44 \overline{\gamma}
 + 11 \overline{\gamma}^2
 - 4 \delta_{-} m_{-}\bigr)
 + \frac{16}{9} \chi_{-} \overline{\gamma} m_{-}
 -  \frac{8}{9} \delta_{-} \overline{\gamma} m_{-} + \frac{1}{216} \nu^2 \frac{1}{\overline{\gamma}} \Bigl(2304 \beta_{+}^2
 - 856 \beta_{+} \overline{\gamma} \nonumber\\
&\quad + 1152 \beta_{-}^2 \bigl(-2
 + 3 \overline{\gamma}\bigr)
 + \overline{\gamma} \bigl(279
 - 384 \chi_{+}
 + 192 \delta_{+} + 472 \overline{\gamma}
 + 48 \overline{\gamma}^2\bigr)
 - 8 \beta_{-} \overline{\gamma} m_{-}\Bigr)
 + \frac{8}{27} \beta_{-} \overline{\gamma}^2 m_{-}
 - 4 \beta_{-}^2 m_{-}^2 \nonumber\\
&\quad -  \frac{32}{9} \beta_{-} \chi_{-} m_{-}^2 -  \frac{16}{9} \beta_{-} \delta_{-} m_{-}^2
 -  \frac{1}{27} \beta_{+} \bigl(-33
 + 96 \chi_{+}
 + 8 \overline{\gamma}^2
 - 216 \beta_{-} m_{-}
 - 96 \chi_{-} m_{-} - 128 \beta_{-} \overline{\gamma} m_{-} \nonumber\\
&\quad + 224 \beta_{-}^2 m_{-}^2\bigr) + \nu \Biggl[- \frac{1}{96} \pi^2 \bigl(2
 + \overline{\gamma}\bigr) \bigl(-82
 + 28 \delta_{+}
 - 34 \overline{\gamma}
 + 7 \overline{\gamma}^2\bigr) + \frac{1}{432} \frac{1}{\overline{\gamma}^{2}} \Biggl(-18432 \beta_{+}^3 \bigl(1
 + \overline{\gamma}\bigr)
 - 7616 \beta_{+}^2 \overline{\gamma}^2 \nonumber\\
&\quad - 18432 \beta_{-}^3 \bigl(1
 + \overline{\gamma}\bigr) m_{-} + 16 \beta_{+} \overline{\gamma} \Bigl(-384 \delta_{+}
 + 192 \chi_{+} \bigl(-3
 + \overline{\gamma}\bigr)
 + 249 \overline{\gamma}
 + 448 \overline{\gamma}^2
 + 576 \chi_{-} m_{-}
 + 192 \delta_{-} m_{-}\Bigr) \nonumber\\
&\quad + 16 \beta_{-} \biggl[576 \chi_{-} \overline{\gamma}  - 192 \delta_{-} \overline{\gamma} \bigl(2
 + \overline{\gamma}\bigr) + \biggl(1152 \beta_{+}^2 \bigl(1
 + \overline{\gamma}\bigr)
 + \overline{\gamma} \Bigl(192 \delta_{+}
 + \bigl(21 - 160 \overline{\gamma}\bigr) \overline{\gamma}
 - 192 \chi_{+} \bigl(3
 + \overline{\gamma}\bigr)\Bigr) \nonumber\\
&\quad - 56 \beta_{+} \overline{\gamma}^2\biggr) m_{-}\biggr] + \overline{\gamma}^2 \Bigl(-20667
 + 1152 \kappa_{+}
 - 20368 \overline{\gamma} + 384 \chi_{+} \bigl(13
 + 4 \overline{\gamma}\bigr)
 + 32 \delta_{+} \bigl(151
 + 15 \overline{\gamma}\bigr)
 - 4488 \overline{\gamma}^2 \nonumber\\
&\quad + 600 \overline{\gamma}^3
 - 2688 \chi_{-} m_{-} - 96 \delta_{-} m_{-}
 - 384 \kappa_{-} m_{-}
 - 288 \delta_{-} \overline{\gamma} m_{-}\Bigr)
 + 64 \beta_{-}^2 \Bigl(288 \beta_{+} \bigl(1
 + \overline{\gamma}\bigr)
 + \overline{\gamma}^2 \bigl(108 + 25 m_{-}^2\bigr)\Bigr)\Biggr)\Biggr] \nonumber\\
&\quad -  \frac{64}{27} \beta_{-}^2 \overline{\gamma} m_{-}^2
 + \frac{224}{81} \beta_{-}^3 m_{-}^3\Biggr) x^3 \Biggr\}\,.
\end{align}
The contribution from the nonlocal tail part is computed by reducing Eqs.~\eqref{J0tail}-\eqref{J1tail} to circular orbits. It gives,
\be
J_{3\mathrm{PN}}^{\mathrm{tail}} = \frac{\tilde{G}\alpha m^{2}\nu}{c\sqrt{x}}\,\cdot\,\left(-\frac{10}{9}\right)\nu\left(2\overline{\delta}_{+}+\frac{\overline{\gamma}(2+\overline{\gamma})}{2}\right)\left[\ln\left(4x\right)+2\gamma_{\mathrm{E}}-\frac{1}{5}\right]\,x^{3}\,.
\ee
%

%%%%%%%%%%%%%%%%%%%%%%%%%%%%%%%%%%%%%%%%%%%%%%%%%%%%%%%
\section{Conclusions}\label{sec:conclusions}
%%%%%%%%%%%%%%%%%%%%%%%%%%%%%%%%%%%%%%%%%%%%%%%%%%%%%%%

In Paper I~\cite{Bernard:2018hta}, we computed the harmonic coordinates equations of motion at 3PN order in scalar-tensor theories. In the present paper, we have complemented this work by deriving the conserved integrals of motion at the same 3PN order and their reduction to the center-of-mass frame. We have also performed the reduction of the conserved energy and angular momentum to the case of circular orbits. This result, which ends the computation of the 3PN conservative dynamics in ST theories, will be directly used to compute the energy flux and scalar waveform at 2PN order~\cite{Heffernan:2018}.

One can note some specific effects that were not expected, originating in the presence of the non-local term in the action~\eqref{a1itail}. As it is the case at 4PN order in GR~\cite{Bernard:2016wrg}, this term gives rise to additional contributions in the energy and angular momentum, see Eqs.~\eqref{E1tail} and \eqref{J1tail}. Note also that contrary to the GR 4PN case, there is also an additional contribution in the center-of-mass position~\eqref{y1itail}. However this term does not have any further consequence in the dynamics.

Looking in more details at the final result one can see, as expected from our previous result of Paper~1, that in the case of a neutron star-black hole binary the conserved energy depends only on one parameter. As a consequence, the observations would not allow to distinguish between Brans-Dicke (with a constant scalar function, $\omega=\omega_0$) and the general scalar-tensor theories. Such a degeneracy might be lifted by including the tidal effecs. Indeed, in scalar-tensor theories the finite-size effects are expected to start contributing to the dynamics at 3PN order~\cite{Damour:1998jk}\footnote{The tidal effects may even start at a lower order (1PN) due to the dynamical scalarisation phenomenon that could be responsible for the large value of some coefficients in the expansion of the mass w.r.t. the scalar field~\cite{Esposito-Farese:2011cha} .}. Thus, if one wants to capture all the effects in the dynamics, the tidal effects should be properly included in the 3PN dynamics and waveforms. This will be the object of a future work~\cite{Bernard:2019}.

\acknowledgments 

The author thanks L. Blanchet and A. Heffernan for useful discussions.
The author acknowledges financial support provided under the European Union's H2020 ERC Consolidator Grant ``Matter and strong-field gravity: New frontiers in Einstein's theory'' grant agreement no. MaGRaTh-646597.
%The author thankfully acknowledges the computer resources, technical expertise and assistance provided by CENTRA/IST. Computations were performed at the cluster “Baltasar-Sete-S\' ois” and supported by the H2020 ERC Consolidator Grant "Matter and strong field gravity: New frontiers in Einstein's theory" grant agreement no. MaGRaTh-646597."
This research was supported in part by Perimeter Institute for Theoretical Physics. Research at Perimeter Institute is supported by the Government of Canada through Innovation, Science and Economic Development Canada and by the Province of Ontario through the Ministry of Research, Innovation and Science.​

%%%%%%%%%%%%%%%%%%%%%%%%%%%%%%%%%%%%%%%%%%%%%%%%%%%%%%%%%%%%%%%%%%%
%%%%%%%%%%%%%%%%%%%%%%%%%%%%%%%%%%%%%%%%%%%%%%%%%%%%%%%%%%%%%%%%%%%

\appendix

%%%%%%%%%%%%%%%%%%%%%%%%%%%%%%%%%%%%%%%%%%%%%%%%%%%%%%%%
\section{The 3PN scalar-tensor Lagrangian in harmonic coordinates}\label{sec:App1}
%%%%%%%%%%%%%%%%%%%%%%%%%%%%%%%%%%%%%%%%%%%%%%%%%%%%%%%%

In this Appendix, we give the complete 3PN scalar-tensor Lagrangian in harmonic coordinates. The equations of motion displayed in paper I can be directly derived from this Lagrangian.
{\allowdisplaybreaks
\begin{subequations}\label{L3PN}
\begin{align}
L_\text{N}={}&\frac{1}{2} \frac{\alpha \tilde{G} m_{1} m_{2}}{r_{12}}
 + \frac{1}{2} m_{1} v_1^{2} + 1 \leftrightarrow 2\,,\\[7pt]
%%%%%%%%%%%%%%%%%%%%%%%%%%%%%%%%%%%%%%%%%%%%%%%%%%%%%%%%%%%%%%%%%
L_\text{1PN}={}&\bigl(- \frac{1}{2}
 -  \overline{\beta}_{2}\bigr) \frac{\alpha^2 \tilde{G}^2 m_{1}^2 m_{2}}{r_{12}^2}
 + \frac{\alpha \tilde{G} m_{1} m_{2}}{r_{12}} \Bigl(- \frac{1}{4} (n_{12} v_1) (n_{12} v_2)
 + \bigl(- \frac{7}{4}
 -  \overline{\gamma}\bigr) (v_1 v_2)
 + \bigl(\frac{3}{2} + \overline{\gamma}\bigr) v_1^{2}\Bigr)
 + \frac{1}{8} m_{1} v_1^{4} + 1 \leftrightarrow 2\,,\\[7pt]
%%%%%%%%%%%%%%%%%%%%%%%%%%%%%%%%%%%%%%%%%%%%%%%%%%%%%%%%%%%%%%%%%
L_\text{2PN}={}&\frac{\alpha^3 \tilde{G}^3}{r_{12}^3} \biggl(\Bigl(\frac{19}{8}
 + \overline{\beta}_{1} \bigl(2
 - 4 \overline{\beta}_{2} \frac{1}{\overline{\gamma}}\bigr)
 + \overline{\gamma}\Bigr) m_{1}^2 m_{2}^2
 + m_{1}^3 m_{2} \Bigl(\overline{\beta}_{2}
 + \frac{1}{3} \overline{\delta}_{1}
 + \frac{1}{12} \bigl(6
 + 4 \overline{\gamma}
 + \overline{\gamma}^2\bigr) -  \frac{2}{3} \overline{\chi}_{2}\Bigr)\biggr) \nonumber\\
& + \frac{\alpha^2 \tilde{G}^2}{r_{12}^2} \biggl[m_{1}^2 m_{2} \biggl(\Bigl(- \overline{\beta}_{2}
 + \frac{1}{2} \overline{\delta}_{1}
 + \frac{1}{8} \bigl(28
 + 20 \overline{\gamma}
 + \overline{\gamma}^2\bigr)\Bigr) (n_{12} v_1)^2
 + \bigl(- \frac{7}{2} + 2 \overline{\beta}_{2}
 -  \overline{\delta}_{1}
 - 3 \overline{\gamma}
 -  \frac{1}{4} \overline{\gamma}^2\bigr) (n_{12} v_1) (n_{12} v_2) \nonumber\\
&\quad + \bigl(- \frac{7}{4}
 -  \overline{\beta}_{2}
 + \overline{\delta}_{1}
 - 4 \overline{\gamma}
 -  \frac{7}{4} \overline{\gamma}^2\bigr) (v_1 v_2) + \Bigl(\overline{\beta}_{2}
 -  \frac{1}{2} \overline{\delta}_{1}
 + \frac{1}{8} \bigl(2
 + 12 \overline{\gamma}
 + 7 \overline{\gamma}^2\bigr)\Bigr) v_1^{2}\biggr) \nonumber\\
&\  + m_{1} m_{2}^2 \biggl(\Bigl(\frac{1}{2} \overline{\delta}_{2}
 + \frac{1}{8} \bigl(2
 + \overline{\gamma}\bigr)^2\Bigr) (n_{12} v_1)^2 + \Bigl(\frac{1}{2} \overline{\beta}_{1}
 -  \frac{1}{2} \overline{\delta}_{2}
 + \frac{1}{8} \bigl(14
 + 20 \overline{\gamma}
 + 7 \overline{\gamma}^2\bigr)\Bigr) v_1^{2}\biggr)\biggr] \nonumber\\
& + \alpha \tilde{G} \biggl[m_{1} m_{2} \Bigl(\bigl(\frac{7}{4} + \overline{\gamma}\bigr) (a_2 v_1) (n_{12} v_1) + \frac{1}{8} (a_2 n_{12}) (n_{12} v_1)^2 + \frac{1}{8} \bigl(-7
 - 4 \overline{\gamma}\bigr) (a_2 n_{12}) v_1^{2}\Bigr) 
 + \frac{m_{1} m_{2}}{r_{12}} \biggl(\frac{3}{16} (n_{12} v_1)^2 (n_{12} v_2)^2 \nonumber\\
&\quad + \frac{1}{4} \bigl(3
 + 2 \overline{\gamma}\bigr) (n_{12} v_1) (n_{12} v_2) (v_1 v_2) + \frac{1}{8} (v_1 v_2)^2 + \frac{1}{8} \bigl(7
 + 4 \overline{\gamma}\bigr) v_1^{4}
 + v_1^{2} \Bigl(\frac{1}{8} \bigl(-7
 - 4 \overline{\gamma}\bigr) (n_{12} v_2)^2
 + \bigl(-2
 -  \overline{\gamma}\bigr) (v_1 v_2) \nonumber\\
&\quad + \frac{1}{16} \bigl(15 + 8 \overline{\gamma}\bigr) v_2^{2}\Bigr)\biggr)\biggr]
 + \frac{1}{16} m_{1} v_1^{6} + 1 \leftrightarrow 2\,.
\end{align}
\end{subequations}}\noindent
Note that, as in GR, acceleration terms start appearing at 2PN order. The 3PN part of the harmonic coordinate Lagrangian in ST theories is,
{\allowdisplaybreaks
\begin{subequations}\label{result4PN}
\begin{align}
L_\text{3PN}^{(0)}={}&\frac{5}{128} m_{1} v_1^{8} + 1 \leftrightarrow 2\,,\\[7pt]
%%%%%%%%%%%%%%%%%%%%%%%%%%%%%%%%%%%%%%%%%%%%%%%%%%%%%%%%%%%%%%%%%
L_\text{3PN}^{(1)}={}&\alpha m_{1} m_{2} \Biggl[- \frac{1}{16} (a_2 n_{12}) (n_{12} v_1)^4
 + (n_{12} v_1)^3 \Bigl(\frac{1}{12} \bigl(-5
 - 3 \overline{\gamma}\bigr) (a_1 v_2)
 + \frac{1}{6} \bigl(-11\ - 5 \overline{\gamma}\bigr) (a_2 v_2)
 + \bigl(\frac{7}{4}
 + \overline{\gamma}\bigr) (a_2 n_{12}) (n_{12} v_2)\Bigr) \nonumber\\
&\quad + 5 \bigl(2
 + \overline{\gamma}\bigr) (a_1 v_1) (n_{12} v_2) (v_1 v_2)
 + \bigl(\frac{7}{4} + \overline{\gamma}\bigr) (a_1 n_{12}) (v_1 v_2)^2
 + (a_2 v_1) \biggl(\frac{1}{6} \bigl(-5
 - 3 \overline{\gamma}\bigr) (n_{12} v_1)^3 \nonumber\\
&\quad + \bigl(- \frac{11}{2} - 3 \overline{\gamma}\bigr) (n_{12} v_1)^2 (n_{12} v_2)
 + 3 \bigl(2
 + \overline{\gamma}\bigr) (n_{12} v_2) v_1^{2}
 + (n_{12} v_1) \Bigl(\bigl(\frac{13}{4}
 + 2 \overline{\gamma}\bigr) (v_1 v_2)
 + \bigl(\frac{15}{4} + 2 \overline{\gamma}\bigr) v_1^{2}\Bigr)\biggr) \nonumber\\
&\quad + (a_2 n_{12}) \Bigl(3 \bigl(2
 + \overline{\gamma}\bigr) (v_1 v_2) v_1^{2}
 + \frac{1}{8} \bigl(-11
 - 6 \overline{\gamma}\bigr) v_1^{4}\Bigr)
 + (n_{12} v_1)^2 \biggl(\frac{1}{8} \bigl(5 + 2 \overline{\gamma}\bigr) (a_1 v_1) (n_{12} v_2) \nonumber\\
&\quad + \frac{1}{8} \bigl(-41
 - 22 \overline{\gamma}\bigr) (a_1 v_2) (n_{12} v_2)
 + \frac{3}{16} \bigl(15 + 8 \overline{\gamma}\bigr) (a_1 n_{12}) (n_{12} v_2)^2
 + (a_2 n_{12}) \Bigl(\bigl(- \frac{11}{2}
 - 3 \overline{\gamma}\bigr) (v_1 v_2)
 + \frac{1}{8} \bigl(5
 + 2 \overline{\gamma}\bigr) v_1^{2}\Bigr) \nonumber\\
&\quad + (a_1 n_{12}) \Bigl(\frac{1}{4} \bigl(-5
 - 3 \overline{\gamma}\bigr) (v_1 v_2)
 + \frac{1}{8} \bigl(-27
 - 13 \overline{\gamma}\bigr) v_2^{2}\Bigr)\biggr)
 + v_1^{2} \biggl(\frac{1}{4} \bigl(-11 - 6 \overline{\gamma}\bigr) (a_1 v_1) (n_{12} v_2) \nonumber\\
&\quad + \frac{5}{2} \bigl(2
 + \overline{\gamma}\bigr) (a_1 v_2) (n_{12} v_2)
 -  \frac{7}{8} \bigl(2
 + \overline{\gamma}\bigr) (a_1 n_{12}) (n_{12} v_2)^2 + (a_1 n_{12}) \Bigl(\bigl(\frac{15}{8}
 + \overline{\gamma}\bigr) (v_1 v_2)
 + \bigl(\frac{97}{16}
 + 3 \overline{\gamma}\bigr) v_2^{2}\Bigr)\biggr) \nonumber\\
&\quad + (n_{12} v_1) \biggl(\bigl(\frac{7}{2}
 + 2 \overline{\gamma}\bigr) (a_1 v_2) (v_1 v_2) + \frac{1}{4} \bigl(-41
 - 22 \overline{\gamma}\bigr) (a_1 n_{12}) (n_{12} v_2) (v_1 v_2) -  \frac{3}{2} \bigl(2
 + \overline{\gamma}\bigr) (a_2 n_{12}) (n_{12} v_2) v_1^{2} \nonumber\\
&\quad + \Bigl(\bigl(\frac{15}{8} + \overline{\gamma}\bigr) (a_1 v_2)
 + \bigl(\frac{41}{4}
 + 5 \overline{\gamma}\bigr) (a_2 v_2) + \frac{1}{8} \bigl(5
 + 2 \overline{\gamma}\bigr) (a_1 n_{12}) (n_{12} v_2)\Bigr) v_1^{2}
 + (a_1 v_1) \Bigl(- \frac{7}{4} \bigl(2 + \overline{\gamma}\bigr) (n_{12} v_2)^2 \nonumber\\
&\quad + \bigl(\frac{15}{4}
 + 2 \overline{\gamma}\bigr) (v_1 v_2) + \bigl(\frac{97}{8}
 + 6 \overline{\gamma}\bigr) v_2^{2}\Bigr)\biggr) + \frac{1}{r_{12}} \Biggl(\bigl(- \frac{27}{16}
 -  \overline{\gamma}\bigr) (v_1 v_2)^3 + (n_{12} v_1)^3 \Bigl(\frac{5}{32} \bigl(29
 + 16 \overline{\gamma}\bigr) (n_{12} v_2)^3 \nonumber\\
&\quad + \bigl(16
 + \frac{17}{2} \overline{\gamma}\bigr) (n_{12} v_2) (v_1 v_2)
 -  \frac{3}{16} \bigl(5 + 2 \overline{\gamma}\bigr) (n_{12} v_2) v_1^{2}\Bigr)
 + \frac{1}{16} \bigl(11
 + 6 \overline{\gamma}\bigr) v_1^{6}
 + v_1^{4} \Bigl(\bigl(- \frac{23}{8}
 -  \frac{3}{2} \overline{\gamma}\bigr) (n_{12} v_2)^2 \nonumber\\
&\quad + \frac{1}{16} \bigl(-5 - 2 \overline{\gamma}\bigr) (v_1 v_2)
 + \frac{15}{4} \bigl(2
 + \overline{\gamma}\bigr) v_2^{2}\Bigr)
 + (n_{12} v_1)^4 \Bigl(- \frac{5}{16} \bigl(15
 + 8 \overline{\gamma}\bigr) (n_{12} v_2)^2
 + \frac{1}{4} \bigl(5 + 3 \overline{\gamma}\bigr) (v_1 v_2) \nonumber\\
&\quad + \frac{1}{8} \bigl(27
 + 13 \overline{\gamma}\bigr) v_2^{2}\Bigr)
 + v_1^{2} \Bigl(\bigl(16
 + \frac{65}{8} \overline{\gamma}\bigr) (n_{12} v_2)^2 (v_1 v_2)
 + \bigl(\frac{47}{8} + 3 \overline{\gamma}\bigr) (v_1 v_2)^2
 + \bigl(- \frac{387}{32}
 - 6 \overline{\gamma}\bigr) (v_1 v_2) v_2^{2}\Bigr) \nonumber\\
&\quad + (n_{12} v_1)^2 \biggl(- \frac{3}{32} \bigl(199 + 108 \overline{\gamma}\bigr) (n_{12} v_2)^2 (v_1 v_2)
 + \bigl(- \frac{45}{8}
 - 3 \overline{\gamma}\bigr) (v_1 v_2)^2
 + v_1^{2} \Bigl(\frac{9}{2} \bigl(2
 + \overline{\gamma}\bigr) (n_{12} v_2)^2 \nonumber\\
&\quad + \frac{1}{16} \bigl(-45 - 26 \overline{\gamma}\bigr) (v_1 v_2) + \frac{1}{4} \bigl(-39
 - 19 \overline{\gamma}\bigr) v_2^{2}\Bigr)\biggr)
 + (n_{12} v_1) \biggl[\bigl(\frac{197}{16}
 + \frac{27}{4} \overline{\gamma}\bigr) (n_{12} v_2) (v_1 v_2)^2 + \bigl(\frac{21}{16}
 + \frac{5}{8} \overline{\gamma}\bigr) (n_{12} v_2) v_1^{4} \nonumber\\
&\quad + v_1^{2} \biggl(- \frac{39}{8} \bigl(2
 + \overline{\gamma}\bigr) (n_{12} v_2)^3
 + (n_{12} v_2) \Bigl(\bigl(- \frac{35}{2}
 - 9 \overline{\gamma}\bigr) (v_1 v_2) + \bigl(\frac{283}{32}
 + \frac{35}{8} \overline{\gamma}\bigr) v_2^{2}\Bigr)\biggr)\biggr]\Biggr)\Biggr] + 1 \leftrightarrow 2\,,\\[7pt]
%%%%%%%%%%%%%%%%%%%%%%%%%%%%%%%%%%%%%%%%%%%%%%%%%%%%%%%%%%%%%%%%%
L_\text{3PN}^{(2)}={}&\alpha^2 \Biggl[m_{1} m_{2}^2 \biggl[\frac{1}{r_{12}} \biggl(\Bigl(\frac{1}{3} \overline{\delta}_{2}
 + \frac{1}{24} \bigl(235
 + 312 \overline{\gamma}
 + 98 \overline{\gamma}^2\bigr)\Bigr) (a_2 v_1) (n_{12} v_1)
 + \Bigl(\frac{2}{3} \overline{\delta}_{2}
 + \frac{1}{12} \bigl(34 + 21 \overline{\gamma}
 + 2 \overline{\gamma}^2\bigr)\Bigr) (a_2 n_{12}) (n_{12} v_1)^2 \nonumber\\
&\quad + \Bigl(- \frac{1}{2} \overline{\beta}_{1}
 -  \frac{1}{3} \overline{\delta}_{2}
 + \frac{1}{12} \bigl(-80
 - 87 \overline{\gamma} - 25 \overline{\gamma}^2\bigr)\Bigr) (a_2 n_{12}) v_1^{2}\biggr)
 + \dfrac{1}{r_{12}^{2}} \biggl(\Bigl(\frac{1}{4} \overline{\delta}_{2}
 + \frac{1}{16} \bigl(2
 + \overline{\gamma}\bigr)^2\Bigr) (n_{12} v_1)^2 v_1^{2} \nonumber\\
&\quad + \Bigl(- \frac{1}{2} \overline{\delta}_{2}
 -  \frac{1}{8} \bigl(2 + \overline{\gamma}\bigr)^2\Bigr) (n_{12} v_1) (n_{12} v_2) v_1^{2}
 + \Bigl(\frac{1}{2} \overline{\delta}_{2}
 + \frac{1}{8} \bigl(-76
 - 84 \overline{\gamma}
 - 23 \overline{\gamma}^2\bigr)\Bigr) (v_1 v_2) v_1^{2} \nonumber\\
&\quad + \Bigl(\frac{1}{8} \overline{\beta}_{1} -  \frac{1}{4} \overline{\delta}_{2}
 + \frac{1}{16} \bigl(45
 + 52 \overline{\gamma}
 + 15 \overline{\gamma}^2\bigr)\Bigr) v_1^{4}\biggr)\biggr] \nonumber\\
&\  + m_{1}^2 m_{2} \Biggl(\frac{1}{r_{12}} \biggl[\bigl(- \frac{235}{24}
 + \overline{\beta}_{2}
 -  \frac{4}{3} \overline{\delta}_{1} - 10 \overline{\gamma}
 -  \frac{7}{3} \overline{\gamma}^2\bigr) (a_2 v_1) (n_{12} v_1) + \Bigl(- \overline{\beta}_{2}
 -  \frac{1}{6} \overline{\delta}_{1}
 + \frac{1}{24} \bigl(-29
 - 6 \overline{\gamma}
 -  \overline{\gamma}^2\bigr)\Bigr) (a_2 n_{12}) (n_{12} v_1)^2 \nonumber\\
&\quad + \Bigl(- \frac{1}{6} \overline{\delta}_{1}
 + \frac{1}{48} \bigl(-235
 - 312 \overline{\gamma}
 - 98 \overline{\gamma}^2\bigr)\Bigr) (a_2 n_{12}) v_1^{2} + (a_1 n_{12}) \biggl(\Bigl(-2 \overline{\delta}_{1}
 + \frac{1}{8} \bigl(-185 - 196 \overline{\gamma}
 - 52 \overline{\gamma}^2\bigr)\Bigr) (v_1 v_2) \nonumber\\
&\quad + \Bigl(\overline{\delta}_{1}
 + \frac{1}{16} \bigl(185
 + 196 \overline{\gamma}
 + 52 \overline{\gamma}^2\bigr)\Bigr) v_1^{2}\biggr)\biggr] + \dfrac{1}{r_{12}^{2}} \biggl[\Bigl(- \overline{\beta}_{2} -  \frac{5}{9} \overline{\delta}_{1}
 + \frac{1}{36} \bigl(26
 + 24 \overline{\gamma}
 - 5 \overline{\gamma}^2\bigr)\Bigr) (n_{12} v_1)^4 \nonumber\\
&\quad + \Bigl(4 \overline{\beta}_{2}
 + \frac{20}{9} \overline{\delta}_{1}
 + \frac{1}{18} \bigl(83
 + 24 \overline{\gamma} + 10 \overline{\gamma}^2\bigr)\Bigr) (n_{12} v_1)^3 (n_{12} v_2) + \Bigl(- \overline{\beta}_{2}
 + \frac{5}{6} \overline{\delta}_{1}
 + \frac{1}{24} \bigl(463
 + 492 \overline{\gamma}
 + 125 \overline{\gamma}^2\bigr)\Bigr) (v_1 v_2)^2 \nonumber\\
&\quad + (n_{12} v_1) \biggl(\Bigl(4 \overline{\beta}_{2}
 -  \frac{4}{3} \overline{\delta}_{1}
 + \frac{1}{6} \bigl(-97
 - 111 \overline{\gamma}
 - 26 \overline{\gamma}^2\bigr)\Bigr) (n_{12} v_2) (v_1 v_2)
 + \Bigl(-2 \overline{\beta}_{2}
 + \frac{1}{3} \overline{\delta}_{1} \nonumber\\
&\quad + \frac{1}{12} \bigl(179
 + 246 \overline{\gamma}
 + 73 \overline{\gamma}^2\bigr)\Bigr) (n_{12} v_2) v_1^{2}\biggr) + \bigl(\frac{373}{48}
 + \frac{1}{6} \overline{\delta}_{1}
 + 9 \overline{\gamma}
 + \frac{61}{24} \overline{\gamma}^2\bigr) v_1^{4} + (n_{12} v_1)^2 \biggl(\bigl(- \frac{35}{6}
 - 4 \overline{\beta}_{2}
 -  \frac{5}{3} \overline{\delta}_{1} \nonumber\\
&\quad - 2 \overline{\gamma}
 -  \frac{5}{12} \overline{\gamma}^2\bigr) (n_{12} v_2)^2 + \Bigl(-3 \overline{\beta}_{2}
 + \frac{5}{6} \overline{\delta}_{1} + \frac{1}{24} \bigl(529
 + 600 \overline{\gamma}
 + 149 \overline{\gamma}^2\bigr)\Bigr) (v_1 v_2)
 + \Bigl(\overline{\beta}_{2}
 -  \frac{1}{6} \overline{\delta}_{1}
 + \frac{1}{24} \bigl(-245
 - 282 \overline{\gamma} \nonumber\\
&\quad - 73 \overline{\gamma}^2\bigr)\Bigr) v_1^{2} + \bigl(- \frac{7}{24}
 + \frac{1}{2} \overline{\beta}_{2}
 + \frac{11}{12} \overline{\delta}_{1}
 - 2 \overline{\gamma}
 -  \frac{37}{48} \overline{\gamma}^2\bigr) v_2^{2}\biggr)
 + v_1^{2} \biggl(\Bigl(- \frac{1}{3} \overline{\delta}_{1}
 + \frac{1}{24} \bigl(-235 - 312 \overline{\gamma}
 - 98 \overline{\gamma}^2\bigr)\Bigr) (n_{12} v_2)^2 \nonumber\\
&\quad + \Bigl(\overline{\beta}_{2}
 -  \frac{7}{6} \overline{\delta}_{1}
 + \frac{1}{24} \bigl(-719
 - 816 \overline{\gamma}
 - 223 \overline{\gamma}^2\bigr)\Bigr) (v_1 v_2) + \Bigl(- \frac{1}{12} \overline{\delta}_{1}
 + \frac{1}{48} \bigl(463
 + 564 \overline{\gamma}
 + 167 \overline{\gamma}^2\bigr)\Bigr) v_2^{2}\biggr)\biggr]\Biggr)\Biggr] + 1 \leftrightarrow 2\,,\\[7pt]
%%%%%%%%%%%%%%%%%%%%%%%%%%%%%%%%%%%%%%%%%%%%%%%%%%%%%%%%%%%%%%%%%
L_\text{3PN}^{(3)}={}&\frac{\alpha^2 m_{1}^3 m_{2}}{r_{12}^3} \biggl(\Bigl(\frac{11}{8} \overline{\gamma} \bigl(2
 + \overline{\gamma}\bigr)
 -  \overline{\delta}_{1} \bigl(10
 + \overline{\gamma}\bigr) \frac{1}{4 + 2 \overline{\gamma}}\Bigr) (n_{12} v_1)^2
 + \Bigl(- \frac{11}{8} \overline{\gamma} \bigl(2
 + \overline{\gamma}\bigr)
 + \overline{\delta}_{1} \bigl(10 + \overline{\gamma}\bigr) \frac{1}{4 + 2 \overline{\gamma}}\Bigr) (n_{12} v_1) (n_{12} v_2) \nonumber\\
&\quad + \Bigl(\frac{11}{24} \overline{\gamma} \bigl(2
 + \overline{\gamma}\bigr)
 -  \frac{1}{6} \overline{\delta}_{1} \bigl(10
 + \overline{\gamma}\bigr) \frac{1}{2 + \overline{\gamma}}\Bigr) (v_1 v_2) + \Bigl(- \frac{11}{24} \overline{\gamma} \bigl(2
 + \overline{\gamma}\bigr)
 + \overline{\delta}_{1} \bigl(10
 + \overline{\gamma}\bigr) \frac{1}{12 + 6 \overline{\gamma}}\Bigr) v_1^{2}\biggr) \nonumber\\
& + \alpha^3 \Biggl(\frac{m_{1}^2 m_{2}^2}{r_{12}^3} \biggl[\biggl(- \frac{34}{3} \overline{\delta}_{1}
 - 5 \overline{\delta}_{2} + \overline{\beta}_{1} \bigl(- \frac{63}{4}
 - 8 \overline{\beta}_{2} \frac{1}{\overline{\gamma}}
 + 12 \overline{\delta}_{1} \frac{1}{\overline{\gamma}}
 - 10 \overline{\gamma}\bigr)
 + \overline{\beta}_{2} \bigl(-6
 + 12 \overline{\delta}_{2} \frac{1}{\overline{\gamma}}
 - 4 \overline{\gamma}\bigr) + \frac{1}{24} \bigl(383
 + 454 \overline{\gamma} \nonumber\\
&\quad + 94 \overline{\gamma}^2
 - 24 \overline{\gamma}^3\bigr)
 + \pi^2 \Bigl(\frac{21}{128} \overline{\delta}_{1} \bigl(2
 + \overline{\gamma}\bigr)
 + \frac{21}{128} \overline{\delta}_{2} \bigl(2
 + \overline{\gamma}\bigr)
 + \frac{3}{256} \bigl(-164
 - 150 \overline{\gamma} - 20 \overline{\gamma}^2
 + 7 \overline{\gamma}^3\bigr)\Bigr)\biggr) (n_{12} v_1)^2 \nonumber\\
&\quad + \biggl(\frac{49}{3} \overline{\delta}_{1}
 + \overline{\beta}_{1} \bigl(\frac{75}{4}
 + 14 \overline{\beta}_{2} \frac{1}{\overline{\gamma}}
 - 24 \overline{\delta}_{1} \frac{1}{\overline{\gamma}}
 + 14 \overline{\gamma}\bigr) + \frac{1}{48} \bigl(-889
 - 956 \overline{\gamma}
 - 188 \overline{\gamma}^2
 + 48 \overline{\gamma}^3\bigr) + \pi^2 \Bigl(- \frac{21}{64} \overline{\delta}_{1} \bigl(2
 + \overline{\gamma}\bigr) \nonumber\\
&\quad -  \frac{3}{256} \bigl(-164
 - 150 \overline{\gamma} - 20 \overline{\gamma}^2
 + 7 \overline{\gamma}^3\bigr)\Bigr)\biggr) (n_{12} v_1) (n_{12} v_2)
 + \biggl(- \frac{49}{9} \overline{\delta}_{1}
 + \overline{\beta}_{1} \bigl(- \frac{33}{4}
 - 6 \overline{\beta}_{2} \frac{1}{\overline{\gamma}} + 8 \overline{\delta}_{1} \frac{1}{\overline{\gamma}}
 - 6 \overline{\gamma}\bigr) \nonumber\\
&\quad + \frac{1}{144} \bigl(439
 + 344 \overline{\gamma}
 - 196 \overline{\gamma}^2
 - 144 \overline{\gamma}^3\bigr)
 + \pi^2 \Bigl(\frac{7}{64} \overline{\delta}_{1} \bigl(2
 + \overline{\gamma}\bigr)
 + \frac{1}{256} \bigl(-164
 - 150 \overline{\gamma} - 20 \overline{\gamma}^2
 + 7 \overline{\gamma}^3\bigr)\Bigr)\biggr) (v_1 v_2) \nonumber\\
&\quad + \biggl(- \frac{305}{72}
 + \frac{34}{9} \overline{\delta}_{1}
 + \frac{5}{3} \overline{\delta}_{2}
 -  \frac{26}{9} \overline{\gamma}
 + \overline{\beta}_{1} \bigl(\frac{21}{4}
 + 8 \overline{\beta}_{2} \frac{1}{\overline{\gamma}} - 4 \overline{\delta}_{1} \frac{1}{\overline{\gamma}}
 + 4 \overline{\gamma}\bigr)
 + \overline{\beta}_{2} \Bigl(-4 \overline{\delta}_{2} \frac{1}{\overline{\gamma}}
 + 2 \bigl(1
 + \overline{\gamma}\bigr)\Bigr)
 + \frac{49}{36} \overline{\gamma}^2 \nonumber\\
&\quad + \pi^2 \Bigl(- \frac{7}{128} \overline{\delta}_{1} \bigl(2
 + \overline{\gamma}\bigr) -  \frac{7}{128} \overline{\delta}_{2} \bigl(2
 + \overline{\gamma}\bigr)
 + \frac{1}{256} \bigl(164
 + 150 \overline{\gamma}
 + 20 \overline{\gamma}^2
 - 7 \overline{\gamma}^3\bigr)\Bigr)
 + \overline{\gamma}^3\biggr) v_1^{2}\biggr] + \frac{m_{1} m_{2}^3}{r_{12}^3} \biggl(\Bigl(\overline{\delta}_{2} \bigl(\frac{3}{2} + \overline{\gamma}\bigr) \nonumber\\
&\quad + \frac{1}{8} \bigl(3
 + 2 \overline{\gamma}\bigr) \bigl(2
 + \overline{\gamma}\bigr)^2\Bigr) (n_{12} v_1)^2
 + \Bigl(\overline{\delta}_{2} \bigl(-1
 -  \frac{2}{3} \overline{\gamma}\bigr)
 + \overline{\beta}_{1} \bigl(\frac{3}{2}
 + \overline{\gamma}\bigr)
 + \frac{1}{12} \bigl(15 + 34 \overline{\gamma}
 + 25 \overline{\gamma}^2
 + 6 \overline{\gamma}^3\bigr)
 + \frac{1}{3} \overline{\chi}_{1}\Bigr) v_1^{2}\biggr) \nonumber\\
&\  + \frac{m_{1}^3 m_{2}}{r_{12}^3} \biggl[\biggl(\frac{1}{2} \overline{\delta}_{1} \bigl(1
 + 3 \overline{\gamma}\bigr)
 -  \frac{3}{4} \overline{\beta}_{2} \bigl(9
 + 8 \overline{\gamma}\bigr) + \frac{1}{24} \bigl(-1370
 - 1472 \overline{\gamma}
 - 441 \overline{\gamma}^2
 - 27 \overline{\gamma}^3\bigr)
 - 2 \overline{\chi}_{2} \nonumber\\
&\quad + \Bigl(3 \overline{\delta}_{1}
 -  \frac{33}{4} \bigl(2
 + \overline{\gamma}\bigr)^2\Bigr) \ln\bigl(r'_{1}\bigr) + \Bigl(-3 \overline{\delta}_{1}
 + \frac{33}{4} \bigl(2
 + \overline{\gamma}\bigr)^2\Bigr) \ln\bigl(r_{12}\bigr)\biggr) (n_{12} v_1)^2
 + \biggl(- \frac{5}{2} \overline{\delta}_{1} \bigl(1
 + \overline{\gamma}\bigr)
 + \overline{\beta}_{2} \bigl(\frac{21}{4}
 + 6 \overline{\gamma}\bigr) \nonumber\\
&\quad + \frac{1}{24} \bigl(1316
 + 1400 \overline{\gamma}
 + 405 \overline{\gamma}^2
 + 21 \overline{\gamma}^3\bigr)
 + 3 \overline{\chi}_{2}
 + \Bigl(-3 \overline{\delta}_{1}
 + \frac{33}{4} \bigl(2
 + \overline{\gamma}\bigr)^2\Bigr) \ln\bigl(r'_{1}\bigr) \nonumber\\
&\quad + \Bigl(3 \overline{\delta}_{1}
 -  \frac{33}{4} \bigl(2
 + \overline{\gamma}\bigr)^2\Bigr) \ln\bigl(r_{12}\bigr)\biggr) (n_{12} v_1) (n_{12} v_2)
 + \biggl(\overline{\beta}_{2} \bigl(- \frac{23}{4}
 - 4 \overline{\gamma}\bigr)
 + \frac{1}{6} \overline{\delta}_{1} \bigl(7 + 9 \overline{\gamma}\bigr)
 + \frac{1}{72} \bigl(-1340
 - 1688 \overline{\gamma} \nonumber\\
&\quad - 711 \overline{\gamma}^2
 - 105 \overline{\gamma}^3\bigr)
 -  \overline{\chi}_{2}
 + \Bigl(\overline{\delta}_{1}
 -  \frac{11}{4} \bigl(2 + \overline{\gamma}\bigr)^2\Bigr) \ln\bigl(r'_{1}\bigr)
 + \Bigl(- \overline{\delta}_{1}
 + \frac{11}{4} \bigl(2
 + \overline{\gamma}\bigr)^2\Bigr) \ln\bigl(r_{12}\bigr)\biggr) (v_1 v_2) \nonumber\\
&\quad + \biggl(\frac{1}{6} \overline{\delta}_{1} \bigl(-2
 - 5 \overline{\gamma}\bigr) + \overline{\beta}_{2} \bigl(\frac{15}{4}
 + 3 \overline{\gamma}\bigr)
 + \frac{1}{72} \bigl(1232
 + 1472 \overline{\gamma}
 + 558 \overline{\gamma}^2
 + 69 \overline{\gamma}^3\bigr) + \overline{\chi}_{2} \nonumber\\
&\quad + \Bigl(- \overline{\delta}_{1}
 + \frac{11}{4} \bigl(2 + \overline{\gamma}\bigr)^2\Bigr) \ln\bigl(r'_{1}\bigr)
 + \Bigl(\overline{\delta}_{1}
 -  \frac{11}{4} \bigl(2
 + \overline{\gamma}\bigr)^2\Bigr) \ln\bigl(r_{12}\bigr)\biggr) v_1^{2}\biggr]\Biggr) + 1 \leftrightarrow 2\,,\\[7pt]
%%%%%%%%%%%%%%%%%%%%%%%%%%%%%%%%%%%%%%%%%%%%%%%%%%%%%%%%%%%%%%%%%
L_\text{3PN}^{(4)}={}&\Bigl(- \frac{11}{12} \overline{\gamma} \bigl(2
 + \overline{\gamma}\bigr)
 + \frac{1}{3} \overline{\delta}_{1} \bigl(-5
 + \overline{\gamma}\bigr) \frac{1}{2 + \overline{\gamma}}\Bigr) \frac{\alpha^3 m_{1}^3 m_{2}^2}{r_{12}^4}
 + \frac{\alpha^4}{r_{12}^4} \biggl[m_{1}^4 m_{2} \biggl(- \frac{1}{2} \bigl(\overline{\beta}_{2}\bigr)^2
 -  \frac{1}{3} \overline{\delta}_{1} + \overline{\beta}_{2} \Bigl(- \frac{2}{3} \overline{\delta}_{1}
 + \frac{1}{6} \bigl(-7
 - 4 \overline{\gamma}
 -  \overline{\gamma}^2\bigr)\Bigr) \nonumber\\
&\quad + \frac{1}{24} \bigl(-9
 - 8 \overline{\gamma}
 - 2 \overline{\gamma}^2\bigr)
 -  \frac{1}{3} \overline{\kappa}_{2}
 + \frac{2}{3} \overline{\chi}_{2}\biggr) + m_{1}^3 m_{2}^2 \biggl(-2 \bigl(\overline{\beta}_{2}\bigr)^2
 -  \frac{25}{18} \overline{\delta}_{2}
 + \overline{\beta}_{2} \bigl(- \frac{79}{4}
 - 8 \overline{\gamma}\bigr)
 + \frac{1}{36} \overline{\delta}_{1} \bigl(-34
 - 3 \overline{\gamma}\bigr) \nonumber\\
&\quad + \frac{1}{144} \bigl(-4008 - 3340 \overline{\gamma}
 - 624 \overline{\gamma}^2
 + 33 \overline{\gamma}^3\bigr)
 + 2 \overline{\chi}_{2}
 + \overline{\beta}_{1} \Bigl(-16 \bigl(\overline{\beta}_{2}\bigr)^2 \dfrac{1}{\overline{\gamma}^{2}}
 + \overline{\beta}_{2} \bigl(-2
 + 16 \frac{1}{\overline{\gamma}}\bigr) -  \frac{2}{3} \overline{\delta}_{1} \frac{1}{\overline{\gamma}} \bigl(-4
 + \overline{\gamma}\bigr) \nonumber\\
&\quad + \frac{1}{6} \bigl(-16
 - 4 \overline{\gamma}
 -  \overline{\gamma}^2\bigr)
 - 8 \frac{1}{\overline{\gamma}} \overline{\chi}_{2}\Bigr)
 + \Bigl(\overline{\delta}_{1}
 -  \frac{11}{4} \bigl(2
 + \overline{\gamma}\bigr)^2\Bigr) \ln\bigl(r'_{1}\bigr) + \Bigl(- \overline{\delta}_{1}
 + \frac{11}{4} \bigl(2
 + \overline{\gamma}\bigr)^2\Bigr) \ln\bigl(r_{12}\bigr)\biggr)\biggr] + 1 \leftrightarrow 2\,.
\end{align}
\end{subequations}}\noindent
Finally, the instantaneous Lagrangian has to be completed with the nonlocal tail contribution,
\be\label{L3PNtail}
L^{\mathrm{tail}} = \frac{2G^{2}M}{3c^{6}\phi_{0}}\left(3+2\omega_{0}\right)\,I_{\mathrm{s},i}^{(2)}(t)\int_{0}^{+\infty}\ud\tau\,\ln\left(\frac{c\tau}{2r_{12}}\right)\,\left(I_{\mathrm{s},i}^{(3)}(t-\tau)-I_{\mathrm{s},i}^{(3)}(t+\tau)\right) \,.
\ee
%

%%%%%%%%%%%%%%%%%%%%%%%%%%%%%%%%%%%%%%%%%%%%%%%%%%%%%%%
\section{Expression of $\gamma$ as a function of $\omega$}\label{sec:App2}
%%%%%%%%%%%%%%%%%%%%%%%%%%%%%%%%%%%%%%%%%%%%%%%%%%%%%%%%

In this appendix, we display the expression of $\gamma=\frac{\tilde{G}\alpha m}{c^2 r}$ as a function of the orbital frequency $\omega$. Using the PN variable $x\equiv\left(\frac{\tilde{G}\alpha m \omega}{c^{3}}\right)^{2/3}$, the instantaneous part reads
\begin{align}\label{gammaInst}
\gamma_\mathrm{3PN}={}&x
 + \Bigl(- \frac{1}{3} \nu
 + \frac{1}{3} \bigl(3
 + 2 \beta_{+}
 + \overline{\gamma}
 - 2 \beta_{-} m_{-}\bigr)\Bigr) x^2
 + \biggl(\frac{1}{12} \nu \frac{1}{\overline{\gamma}} \Bigl(-96 \beta_{-}^2
 + 96 \beta_{+}^2
 - 20 \beta_{+} \overline{\gamma} + \overline{\gamma} \bigl(-65
 - 16 \chi_{+}
 + 8 \delta_{+} \nonumber\\
&\quad - 28 \overline{\gamma}
 + 2 \overline{\gamma}^2\bigr)
 - 4 \beta_{-} \overline{\gamma} m_{-}\Bigr)
 + \frac{1}{12} \Bigl(12
 + 16 \beta_{+}^2
 + 8 \chi_{+} - 4 \delta_{+}
 + 4 \overline{\gamma}
 -  \overline{\gamma}^2
 - 16 \beta_{-} m_{-}
 - 8 \chi_{-} m_{-}
 - 4 \delta_{-} m_{-} \nonumber\\
&\quad + 8 \beta_{+} \bigl(2
 + \overline{\gamma}
 - 4 \beta_{-} m_{-}\bigr) - 8 \beta_{-} \overline{\gamma} m_{-}
 + 16 \beta_{-}^2 m_{-}^2\Bigr)\biggr) x^3
 + \Biggl\{\frac{1}{81} \nu^3
 + \frac{1}{108} \nu^2 \frac{1}{\overline{\gamma}} \Bigl(288 \beta_{+}^2
 - 308 \beta_{+} \overline{\gamma} \nonumber\\
&\quad - 288 \beta_{-}^2 \bigl(1 + 3 \overline{\gamma}\bigr)
 + \overline{\gamma} \bigl(687
 - 48 \chi_{+}
 + 24 \delta_{+}
 + 488 \overline{\gamma}
 + 6 \overline{\gamma}^2\bigr)
 + 92 \beta_{-} \overline{\gamma} m_{-}\Bigr) + \nu \frac{1}{\alpha} \frac{1}{48 + 24 \overline{\gamma}} \bigl(40 \delta_{+}
 + 132 \overline{\gamma} \nonumber\\
&\quad - 12 \delta_{+} \overline{\gamma} + 132 \overline{\gamma}^2
 + 33 \overline{\gamma}^3
 + 40 \delta_{-} m_{-}
 - 12 \delta_{-} \overline{\gamma} m_{-}\bigr) + \frac{1}{162} \biggl[162
 + 144 \beta_{+}
 + 720 \beta_{+}^2
 + 560 \beta_{+}^3
 + 324 \chi_{+} \nonumber\\
&\quad + 504 \beta_{+} \chi_{+}
 - 144 \beta_{-} \delta_{-} - 144 \delta_{+} - 108 \beta_{+} \delta_{+}
 + 72 \kappa_{+}
 - 54 \overline{\gamma}
 + 72 \beta_{+} \overline{\gamma}
 + 336 \beta_{+}^2 \overline{\gamma}
 + 144 \chi_{+} \overline{\gamma}
 - 36 \delta_{+} \overline{\gamma}
 - 162 \overline{\gamma}^2 \nonumber\\
&\quad - 3 \beta_{+} \overline{\gamma}^2
 - 47 \overline{\gamma}^3
 - 3 \biggl(12 \Bigl(2 \kappa_{-}
 + \delta_{-} \bigl(4
 + 3 \beta_{+}
 + \overline{\gamma}\bigr)
 + \chi_{-} \bigl(9
 + 14 \beta_{+}
 + 4 \overline{\gamma}\bigr)\Bigr) + \beta_{-} \Bigl(48
 + 560 \beta_{+}^2
 + 168 \chi_{+}
 - 36 \delta_{+} \nonumber\\
&\quad + 24 \overline{\gamma}
 + 32 \beta_{+} \bigl(15
 + 7 \overline{\gamma}\bigr)
 -  \overline{\gamma}^2\Bigr)\biggr) m_{-} + 12 \beta_{-} \Bigl(21 \bigl(2 \chi_{-}
 + \delta_{-}\bigr)
 + 4 \beta_{-} \bigl(15
 + 35 \beta_{+}
 + 7 \overline{\gamma}\bigr)\Bigr) m_{-}^2
 - 560 \beta_{-}^3 m_{-}^3\biggr] \nonumber\\
&\quad + \nu \Biggl[\frac{1}{768} \pi^2 \bigl(2
 + \overline{\gamma}\bigr) \bigl(-82
 + 28 \delta_{+}
 - 34 \overline{\gamma}
 + 7 \overline{\gamma}^2\bigr)
 + \frac{1}{216} \frac{1}{\overline{\gamma}^{2}} \Biggl(1152 \beta_{+}^3 \bigl(4
 + 7 \overline{\gamma}\bigr) + 16 \beta_{+}^2 \overline{\gamma} \bigl(162
 + 71 \overline{\gamma}\bigr) \nonumber\\
&\quad + 1152 \beta_{-}^3 \bigl(4
 + 7 \overline{\gamma}\bigr) m_{-}
 - 8 \beta_{+} \overline{\gamma} \Bigl(\delta_{+} \bigl(24
 - 36 \overline{\gamma}\bigr)
 + 114 \overline{\gamma} + 24 \chi_{+} \bigl(-12
 + 7 \overline{\gamma}\bigr)
 + 71 \overline{\gamma}^2
 - 9 \overline{\gamma}^3 + 288 \chi_{-} m_{-} \nonumber\\
&\quad + 96 \delta_{-} m_{-}\Bigr)
 - 4 \beta_{-} \biggl[576 \chi_{-} \overline{\gamma} - 48 \delta_{-} \overline{\gamma} \bigl(-1
 + 4 \overline{\gamma}\bigr)
 + \biggl(288 \beta_{+}^2 \bigl(4
 + 7 \overline{\gamma}\bigr) - 224 \beta_{+} \overline{\gamma}^2
 + \overline{\gamma} \Bigl(24 \delta_{+} \bigl(8
 + 3 \overline{\gamma}\bigr) \nonumber\\
&\quad - 48 \chi_{+} \bigl(12
 + 7 \overline{\gamma}\bigr)
 + \overline{\gamma} \bigl(69
 + 38 \overline{\gamma}
 + 18 \overline{\gamma}^2\bigr)\Bigr)\biggr) m_{-}\biggr]
 + \overline{\gamma}^2 \Bigl(831
 - 288 \kappa_{+} + 1024 \overline{\gamma} - 96 \chi_{+} \bigl(13
 + 4 \overline{\gamma}\bigr) \nonumber\\
&\quad + 4 \delta_{+} \bigl(91
 + 33 \overline{\gamma}\bigr)
 + 231 \overline{\gamma}^2
 - 51 \overline{\gamma}^3
 + 384 \chi_{-} m_{-}
 - 36 \delta_{-} m_{-} + 96 \kappa_{-} m_{-}
 + 36 \delta_{-} \overline{\gamma} m_{-}\Bigr) - 16 \beta_{-}^2 \biggl(72 \beta_{+} \bigl(4
 + 7 \overline{\gamma}\bigr) \nonumber\\
&\quad + \overline{\gamma} \Bigl(162
 + \overline{\gamma} \bigl(108 + 19 m_{-}^2\bigr)\Bigr)\biggr)\Biggr)
 + \biggl(- \delta_{-} + \frac{1}{4} \Bigl(-4 \delta_{+}
 + 11 \bigl(2
 + \overline{\gamma}\bigr)^2\Bigr) m_{-}\biggr) \ln\bigl(\frac{r'_{-}}{r}\bigr) \nonumber\\
&\quad + \bigl(11
 -  \delta_{+} + 11 \overline{\gamma}
 + \frac{11}{4} \overline{\gamma}^2
 -  \delta_{-} m_{-}\bigr) \ln\bigl(\frac{r'_{+}}{r}\bigr)\Biggr]\Biggr\} x^4\,,
\end{align} 
while the tail contribution is
\be
\label{gammaTail}
\gamma_{\mathrm{tail}}=-\frac{4\nu}{9}\left(2\overline{\delta}_{+}+\frac{\overline{\gamma}(2+\overline{\gamma})}{2}\right)\left[\ln\left(4x\right)+2\gamma_{\mathrm{E}}+1\right]\,x^{3} \,.
\ee
The presence of the logarithmic dependence in the constant $r_{+}$ and $r_{-}$ is due to the fact that the relative distance $r$ is not a gauge invariant quantity.

%%%%%%%%%%%%%%%%%%%%%%%%%%%%%%%%%%%%%%%%%%%%%%%%%%%%%%%%%%%%%%%%%%%
%%%%%%%%%%%%%%%%%%%%%%%%%%%%%%%%%%%%%%%%%%%%%%%%%%%%%%%%%%%%%%%%%%%

\bibliography{ListeRef_ST}

\end{document}